\documentclass[11pt]{article}
\usepackage[utf8]{inputenc}
\usepackage{setspace}
\usepackage{url}
\usepackage{amsmath}
\usepackage{amssymb}
\usepackage{amsfonts}
\usepackage[dvipsnames]{xcolor}
\usepackage[colorlinks=true, linkcolor=NavyBlue, citecolor=NavyBlue, filecolor=NavyBlue, urlcolor=NavyBlue]{hyperref}
\usepackage{bbm}
\usepackage{graphicx}
\usepackage{float}
\usepackage{subcaption}
\usepackage{authblk}

\usepackage{multirow}
\usepackage{booktabs}

\usepackage[backgroundcolor=white, linecolor=red, bordercolor=red]{todonotes}
\definecolor{boldcolor}{rgb}{0.5, 0., 1}
\definecolor{boldcolor2}{rgb}{0., 0.5, 1}

\usepackage{amsthm}

\newtheorem{proposition}{Proposition}

\newtheorem{assumption}{Assumption}

\theoremstyle{definition}
\newtheorem{definition}{Definition}

\usepackage{geometry}
\newcommand\blfootnote[1]{%
  \begingroup
  \renewcommand\thefootnote{}\footnote{#1}%
  \addtocounter{footnote}{-1}%
  \endgroup
}
\renewcommand{\footnoterule}{\vfill\kern -3pt \hrule width 0.4\columnwidth \kern 2.6pt}
\usepackage{natbib}
\usepackage{mathpazo}

\title{\textbf{Explaining the Macroeconomic Inertia Puzzle}\blfootnote{
Northwestern University. Email: \href{michaelcai@u.northwestern.edu}{michaelcai@u.northwestern.edu}\\
I am deeply grateful to my advisors---George-Marios Angeletos, Matthias Doepke, and Matthew Rognlie---for their encouragement and guidance.
I would also like to thank Patrick Adams, Bence Bard{\'o}czy, Kwok Yan Chiu, Lawrence Christiano, Keshav Dogra, Martin Eichenbaum, Wouter den Haan, Stephan Hobler, Ethan Ilzetzki, Yuriy Gorodnichenko,
Joao Guerreiro, Fergal Hanks, Jonathon Hazell, Zhen Huo, Diego K{\"a}nzig, Chen Lian, Xiaojie Liu, Evan Majic, Pooya Molavi, Jane Olmstead-Rumsey, Giorgio Primiceri,
Ricardo Reis, Aditya Soenarjo, Amilcar Velez, Rongxuan (Dalton) Zhang and participants of the Northwestern Macro Lunch
and Society for Economic Dynamics 2025 meeting for insightful discussion and helpful comments.
I gratefully acknowledge the financial support of the Alfred P. Sloan Foundation Pre-Doctoral Fellowship in Behavioral Macroeconomics awarded by the NBER.
}}
\author{Michael Cai}
\date{July 2025}

\begin{document}

\maketitle

\vspace{-0.5cm}

\begin{center}
  \href{www.michaelcai.com/files/MichaelCai_JMP_Inertia.pdf}{Click here for the most recent version}
\end{center}

\abstract{
Benchmark macroeconomic models require additional frictions to explain the 
sluggish response of aggregate variables to sudden shocks or changes in policy. 
I show that standard heterogeneous-agent (HA) models---the \cite{blanchard1985olg} perpetual youth
and \cite{bewley1986stationary} incomplete markets models---are consistent with 
aggregate consumption inertia without the use of habit preferences or any specific 
model of expectation underreaction to dampen the responsiveness of 
consumption-savings decisions.
I instead replicate observed consumption inertia in standard HA models 
by directly substituting survey expectations of income and interest rates for 
agents' expectations.
I propose a new theory of macroeconomic inertia that rationalizes the observed 
extrapolation bias in survey expectations by embedding an unobserved components 
model of expectations into a tractable HA general equilibrium environment.
Inertia results when expectations imperfectly account for the equilibrium amplification 
of shocks, which is large in HA economies. 
This imperfect inference causes expectations to gradually unanchor as agents repeatedly 
misattribute large responses of equilibrium outcomes simply to larger shocks.
This theory also illustrates a novel drawback to inertial monetary policy rules 
and the delayed financing of fiscal deficits: 
Policy regimes that act more gradually experience longer transmission lags due to their 
decreased effectiveness at anchoring expectations.
}

\pagebreak

\section{Introduction}
Macroeconomic variables often display a sluggish response to sudden 
shocks or policy changes\footnote{These include monetary policy shocks (\citealt{romer2004new}), 
productivity shocks (\citealt{kurmann2021revisions}), government spending 
shocks (\citealt{ramey2011identifying}), oil price shocks (\citealt{kanzig2021macroeconomic}),
and ``max-share'' shocks that explain the majority of business cycle 
fluctuations in a large panel of macroeconomic aggregates 
(\citealt{angeletos2020business}).}. 
Understanding the sources of this macroeconomic inertia is an important 
challenge for business cycle research for two reasons.
First, long lags in monetary and fiscal policy transmission 
make them unreliable tools for stabilizing business cycle fluctuations.
Second, the textbook New Keynesian model, 
the benchmark model for policy analysis, cannot rationalize these slow responses
and instead predicts that policy actions should have immediate effects.
Theories of adjustment frictions and bounded rationality 
have been proposed to limit the agents' foresight and the responsiveness of their decisions,
two features preventing the textbook model from generating inertia. 
However, no consensus has emerged among these theories.

This paper focuses on explaining the observed inertia in aggregate consumer spending 
because it plays a central role in facilitating policy transmission.
I first adopt a semi-structural approach to analyze 
consumption-savings models without imposing restrictive 
assumptions on expectation formation or altering preferences. 
My goal is to identify whether the basic structure of these models is consistent 
with aggregate consumption inertia without taking a stand on how expectations are formed. 
To do so, I first compute the empirical impulse responses of survey expectations and 
realizations of income and interest rates to an identified shock. 
I then evaluate whether model-implied consumption, conditional on these 
measured income and interest rate paths, matches the empirical consumption impulse response.

My first main result finds that standard heterogeneous-agent models\footnote{
Perpetual youth, overlapping generations 
(\citealt{blanchard1985olg}, \citealt{yaari1965uncertain}) and the 
standard incomplete markets model with idiosyncratic risk and 
borrowing constraints (\citealt{bewley1986stationary}, \citealt{imrohorouglu1989cost}, 
\citealt{huggett1993risk}, \citealt{aiyagari1994uninsured}). 
} 
can match observed aggregate consumption inertia as long as agents' average expectations 
of income and interest rates align with average survey expectations of these variables.
In contrast, the standard representative-agent model implies an overly muted 
consumption response that fails to match observed consumption because 
of agents' low marginal propensities to consume (MPCs).

The consistency of heterogeneous-agent models with aggregate consumption inertia 
does not explain the underlying causes of inertia.
Additionally, patterns of extrapolation bias that I document in survey expectations 
are not easily explained by many existing models of expectation formation.
I therefore proceed to propose a new theory of macroeconomic inertia that 
rationalizes expectations data by embedding an unobserved components model of 
expectations into a tractable heterogeneous-agent equilibrium environment.


My second main result explains why high MPCs in 
heterogeneous-agent models contribute to the emergence of inertia.
Agents learn by observing equilibrium output (income) but cannot perfectly 
separate the contributions of the direct effect of an exogenous demand shock from the 
ensuing Keynesian amplification, which is large when the average MPC is large.
As agents overweight the direct impact of the shock and optimistically consume more, 
equilibrium output is further magnified, which partially confirms their initial beliefs. 
Inertia arises due to the repeated misattribution and delayed amplification that results 
from this feedback loop.
As a consequence of imperfect learning, policy regimes which delay action and rely on 
far-horizon reasoning through equilibrium outcomes become less effective.

\paragraph{A semi-structural approach to matching observed inertia}
Can we test the consistency of consumption-savings models with observed consumption inertia 
without relying on any particular theory of adjustment frictions or expectation formation?
I address this question by building on the approaches in 
\cite{auclert2020micro} and \cite{bardoczy2023unemployment} 
to solve and estimate heterogeneous-agent models without the rational expectations 
assumption. 
I utilize an ``aggregate consumption function'' representation
implied by these models that remains agnostic to how expectations are formed.
This function takes in the full history of (cross-sectional) 
average subjective expectations formed for all future horizons and the realizations of aggregate 
income and interest rates. 
I then replace the average 
subjective expectation of agents within the model with the average expectation 
reported in the survey data. 
Given that these data are available\footnote{Far-horizon expectations that are not reported 
in the survey data need to be extrapolated. 
I discuss the details of this procedure briefly in Section \ref{subsec:irf_income_rates} 
and also in Appendix \ref{asubsec:extrap}.}, 
one can simply plug them into this model 
representation and evaluate model-implied aggregate consumption.

To assess model-implied consumption impulse responses against empirical 
ones, I build on the estimation approach from 
\cite{barnichon2020identifying} and \cite{lewis2022dynamic}.
In contrast to traditional impulse response matching (\citealt{christiano2005nominal}), 
which requires a fully specified equilibrium model, I directly use the 
empirical impulse responses of realized and expected income and interest rates 
to compute the model-implied consumption response to the same identified shock.
``Regressing'' the model-implied impulse response of consumption on its 
observed response to estimate model parameters can be interpreted 
as an impulse response matching method that relies on fewer structural assumptions. 
This approach narrowly focuses on the consistency of consumption models 
with consumption inertia without introducing a potentially misspecified 
model of expectation formation or confounding components of the equilibrium environment 
that may independently induce inertia.

Given the dynamics of survey expectations are sufficient to generate 
model-implied consumption inertia that matches the data, 
what features or biases explain the expectations data? 
The patterns of forecast errors in survey expectations are inconsistent with full-information 
rational expectations, which requires 
the ex-ante unpredictability of ex-post forecast errors. 
Models of bounded rationality that uniformly predict under- or overreaction relative 
to the full-information rational expectations cannot explain the 
varying degrees of under- and overreaction of survey expectations across variables and time.

The simplest explanation for this bias turns out to be persistent 
over-extrapolation of the current observation to expectations of future horizons. 
For example, upon observing the current period income realization conditional 
on the shock, average forecaster expectations of future income anchor on the 
current realization. 
If income is low today, they expect it to remain at the same level tomorrow and so on. 
If the path of income displays a hump-shaped profile, over-extrapolation entails 
initial underreaction and eventual overreaction.
This over-extrapolation can also be fairly long-lived, in certain cases 
persisting for many quarters after the initial shock impact period. 

\paragraph{Inertia as an equilibrium phenomenon} 
Why do expectations that over-extrapolate result in aggregate consumption inertia?
To jointly explain these features of the data and to consider policy counterfactuals, 
I now impose additional structure in the form of a model of expectation formation, 
disciplined to the expectations data, and a standard New Keynesian equilibrium environment.
Expectations are formed via Bayesian learning\footnote{
This form of learning has also been shown to explain systematic patterns in 
expectations bias in cross-sectional, experimental, and unconditional time-series evidence
(\citealt{afrouzi2023overreaction}, \citealt{crump2023term}, \citealt{farmer2024learning}, \citealt{nagel2024lean}). 
} in an information 
environment where fundamental shocks are comprised of transitory and persistent components 
that are imperfectly observed.
When agents mistakenly attribute a transitory change in the observed 
variable to a persistent shock, their expectations over-extrapolate 
from the current observation. 



Even in the rational learning baseline, agents cannot immediately disentangle 
the transitory from the persistent shock components. 
However, the impact on equilibrium output of their consumption decisions 
based on these imperfect beliefs does not distort their ability to 
learn about these shocks. 
They understand how equilibrium output responds endogenously to their 
actions and therefore can learn about the shocks by observing output over time 
just as effectively as if it were a pure tracking problem.

I consider an important deviation from this baseline by imposing agents' perceived 
equilibrium output law of motion is ``truncated'' relative to the actual equilibrium 
output law of motion\footnote{
One interpretation of this friction is that agent cognition faces a complexity limit 
as in \cite{molavi2022simple} or a memory constraint as in \cite{azeredo2024optimally}.
}. 
For example, if the actual equilibrium output law of motion is a function of the full history of  
shocks, the perceived law of motion may only account for a few recent periods.
The consequence is that agents cannot fully internalize the general equilibrium 
consequences of their misinformed actions on the evolution of their future beliefs.
Hence, overly responsive consumption to a transitory demand shock increases 
realized output (income) in equilibrium by more than expected, 
thereby reinforcing beliefs that the shock itself was persistent.
This positive feedback loop results in the endogenous unanchoring of expectations, which 
prolongs amplification, impedes learning, and results in inertia.

The two key factors that determine consumption inertia, 
the size of equilibrium amplification and the degree of unanchoring, 
are connected by a ``belief multiplier'' $\chi$. 
The multiplier $\chi$ increases in the income sensitivity of 
consumption demand but decreases in the interest rate sensitivity. 
Consequently, it tends to be large in heterogeneous-agent economies 
but small or even negative in representative-agent ones.
I demonstrate formally that inertia emerges when the belief multiplier 
$\chi$ is positive and sufficiently large but is absent otherwise.

\paragraph{How inertia influences policy transmission}
New trade-offs in stabilization policy arise because shocks are imperfectly observed 
and their equilibrium consequences are imperfectly understood. 
A more responsive Taylor rule lowers the belief multiplier $\chi$ and 
tends to reduce inertia.
However, an overly-responsive Taylor rule can destabilize the economy.
As the Taylor rule coefficient on output\footnote{
For the monetary policy examples, I focus on real output stabilization 
with respect to demand shocks, given that the divine coincidence of 
output and inflation stabilization holds in my setting (\citealt{blanchard2007real}). 
Therefore, to simplify analysis I adopt a ``real'' Taylor rule, which sets the 
ex-ante real interest rate, conditional on current 
inflation expectations, as a function of real output (\citealt{auclert2024intertemporal}).
}
crosses an upper threshold, the 
contribution of positive future output beliefs to current output 
are outweighed by expected future interest rate contractions. 
This produces a negative feedback loop between output and beliefs 
that increases output volatility.
I show that a Taylor rule that is not overly-responsive can 
balance the reduction of inertial propagation against the risk of 
destabilizing output and beliefs.
However, to achieve this balance the monetary authority must 
allow partial pass-through of the initial shock\footnote{
\cite{eusepi2024shortrun} and \cite{christiano2020anchoring} 
derive similar results showing overly-restrictive policy can be undesirable.
}.

I next consider the choice of a lagged or ``inertial'' term in 
the Taylor rule. 
With this specification, interest rate policy responds 
to current output fluctuations but also passes through 
changes in past policy rates.
In standard rational expectations settings, forward-looking agents can 
understand the dynamic equilibrium implications of interest rate changes 
stemming from both of these causes equally well. 
Because agents are able to accurately reason about the far-horizon 
equilibrium effects of interest rates, highly inertial policy rules 
can have powerful stabilizing effects on current output even if 
the current response of interest rates is muted.

I compare the counterfactual implications of demand shock transmission under 
two monetary policy regimes that vary by the degree of policy inertia. 
These regimes are chosen to obtain the same discounted sum of 
squared output deviations, a proxy measure for welfare loss, under rational 
expectations. 
However, under frictional learning, the ``gradual regime'' that has higher 
policy rule inertia results in larger welfare losses than the low policy inertia regime.
The reason is that learning frictions impair agents' ability to comprehend 
the dynamic equilibrium effects that make inertial Taylor rules effective 
under rational expectations.
When the policy response to a demand shock is delayed, agents instead perceive 
policy to be less responsive overall. 
This increases the degree to which expectations are unanchored and reduces the effective 
amount of stabilization.

Learning frictions also alter the transmission of deficit-financed fiscal policy 
through the same mechanism. 
\cite{angeletos2023can} show that the delayed financing of a one-time, unanticipated 
transfer can substantially amplify the output response to this policy shock 
under rational expectations. 
I consider this exercise in a model with learning frictions and show that the peak 
and cumulative responses of output to a transfer shock shift further out in time 
as financing is delayed.
This shift may be undesirable if policymakers aim to enact a timely and short-lived fiscal stimulus.

\paragraph{Related literature}

This paper relates to a large literature that seeks to understand and quantify the sources 
of macroeconomic inertia.
A major strand of this literature focuses on preference- and technology-based explanations 
for the slow adjustment of aggregate variables, such as consumption, inflation, and investment. 
The main preference-based explanation for consumption takes the form of habit formation in consumption 
spending (\citealt{fuhrer2000habit}, \citealt{dynan2000habit}, 
\citealt{chetty2016consumption}, \citealt{havranek2017habit}).
A separate strand relaxes the full-information, rational expectations (FIRE) assumption, 
dampening the responsiveness of forward-looking decisions to generate inertia. 
Theories that depart from FIRE and generate inertia include adaptive learning 
(\citealt{evans1999learning}, \citealt{williams2003adaptive}, \citealt{eusepi2011expectations}, 
\citealt{milani2011expectation}), 
incomplete information (\citealt{woodford2001imperfect}), 
sticky information and expectations (\citealt{mankiw2002sticky}, \citealt{carroll2020sticky}), 
and rational and behavioral inattention (\citealt{sims2003implications}, \citealt{luo2008consumption}, 
\citealt{mackowiak2015business}, \citealt{gabaix2019behavioral}). 

\cite{auclert2020micro} was the first paper to point out that models with 
consumption habits and FIRE cannot simultaneously produce sluggish aggregate consumption 
adjustment and high marginal propensities to consume (MPCs) in line with microeconomic evidence. 
They demonstrate that heterogeneous-agent New Keynesian (HANK) models capable of matching high MPCs 
must therefore relax the FIRE assumption to be able to match aggregate consumption inertia. 
\cite{bardoczy2023unemployment} extend the methodological approach introduced by 
\cite{auclert2020micro} and demonstrate that HANK models can be estimated 
by replacing a model of expectation formation directly with expectations data 
using the impulse response matching estimation framework 
of \cite{christiano2005nominal}.

My paper builds on the approaches of \cite{auclert2020micro} 
and \cite{bardoczy2023unemployment} by showing that 
the same methodology can be applied to estimate parameters of the 
heterogeneous-agent consumption-savings decision without imposing 
a model of expectation formation or the New Keynesian equilibrium 
assumptions. 
I do this by adopting the instrumental variables approach of 
\cite{barnichon2020identifying} and \cite{lewis2022dynamic}.
This approach allows me to retain the impulse response matching interpretation 
of my results without needing to compute model-implied impulse responses 
within a fully-specified equilibrium model or impose a model of 
expectation formation. 

My paper extends the literature that 
use survey expectations data for solving and estimating 
structural models following 
\cite{manski2004measuring}, \cite{manski2018survey}.
\cite{piazzesi2009momentum} use expectations of housing market and business 
conditions to characterize boom-bust cycles in a search model of housing 
transactions.
\cite{del2011fitting} incorporate survey inflation expectations into
a Bayesian estimation framework for representative-agent equilibrium models 
with rational expectations. 
\cite{kosar2023expectations} discuss wide-ranging applications of expectations data 
in estimating models of individual and household behavior. 

The theoretical results in my paper complement the work of 
\cite{angeletos2021myopia} and \cite{christiano2024slow}, which contain 
similar mechanisms. 
In these papers, when measures of equilibrium amplification or complementarity 
are large, learning is delayed and inertia can be prolonged. 
In contrast to \cite{angeletos2021myopia}, the model of expectation formation 
I adopt admits a tractable form for the belief law of motion, where the 
amplification parameter is a simple function of structural primitives. 
This allows me to analytically characterize the joint evolution of 
beliefs with equilibrium outcomes. 
\cite{christiano2024slow} focuses on characterizing 
the speed of convergence of the perceived equilibrium law of motion to 
the rational expectations equilibrium. 
The focus of my analysis is instead on the learning behavior of Bayesian 
agents with fixed, potentially-flawed perceived laws of motion 
trying to infer imperfectly observable shocks.

\cite{molavi2022simple} is a closely related paper that demonstrates that 
inertia can arise when agents are constrained to entertain low-dimensional 
state-space representations of the equilibrium law of motion. 
I show that this inertia can be exaggerated when distorted beliefs 
formed by a similar, low-dimensional state-space model are reinforced 
due to an equilibrium feedback loop that is particularly strong in 
heterogeneous-agent economies.

\cite{eusepi2024shortrun} demonstrates the equilibrium implications of 
the same model of expectation formation in a representative-agent 
New Keynesian model but with a different focus.
While these authors prioritize illustrating the limits of short-run stabilization policy 
when expectations over-extrapolate, a theme I revisit briefly in the policy 
section of this paper, I focus on the contribution of over-extrapolating 
expectations in generating inertia and the consequences of this inertia 
for policy conduct.

\section{Impulse responses of measured expectations}

I begin by documenting empirical patterns of impulse responses of 
measured expectations from survey data and their systematic biases. 
I demonstrate how existing theories that deviate from the full-information 
rational expectations (FIRE) benchmark struggle to rationalize these patterns. 
This observation highlights the usefulness of the semi-structural estimation method that 
I propose in the next two sections, where I directly condition on the path of measured 
expectations in place of a model of expectation formation 
to assess the fit of structural models of consumption-savings against consumption data.
I then offer supportive evidence for a form of extrapolation bias, which can rationalize 
the empirical patterns of measured expectations.

\begin{figure}[t]
    {
    \centering
    \begin{subfigure}{0.495\textwidth}
        \centering
        \includegraphics[width=\linewidth]{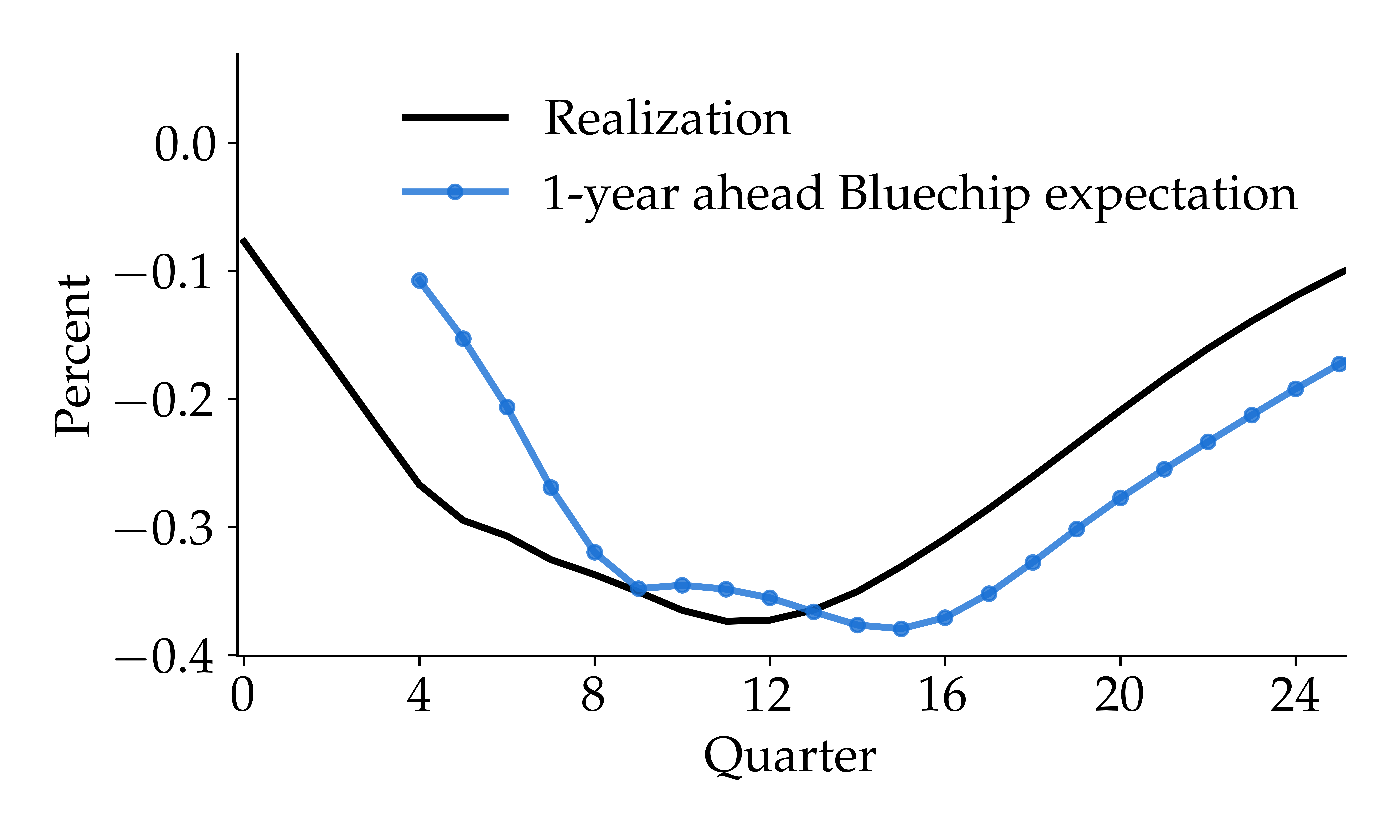}
    \end{subfigure}
    \hfill
    \begin{subfigure}{0.495\textwidth}
        \centering
        \includegraphics[width=\linewidth]{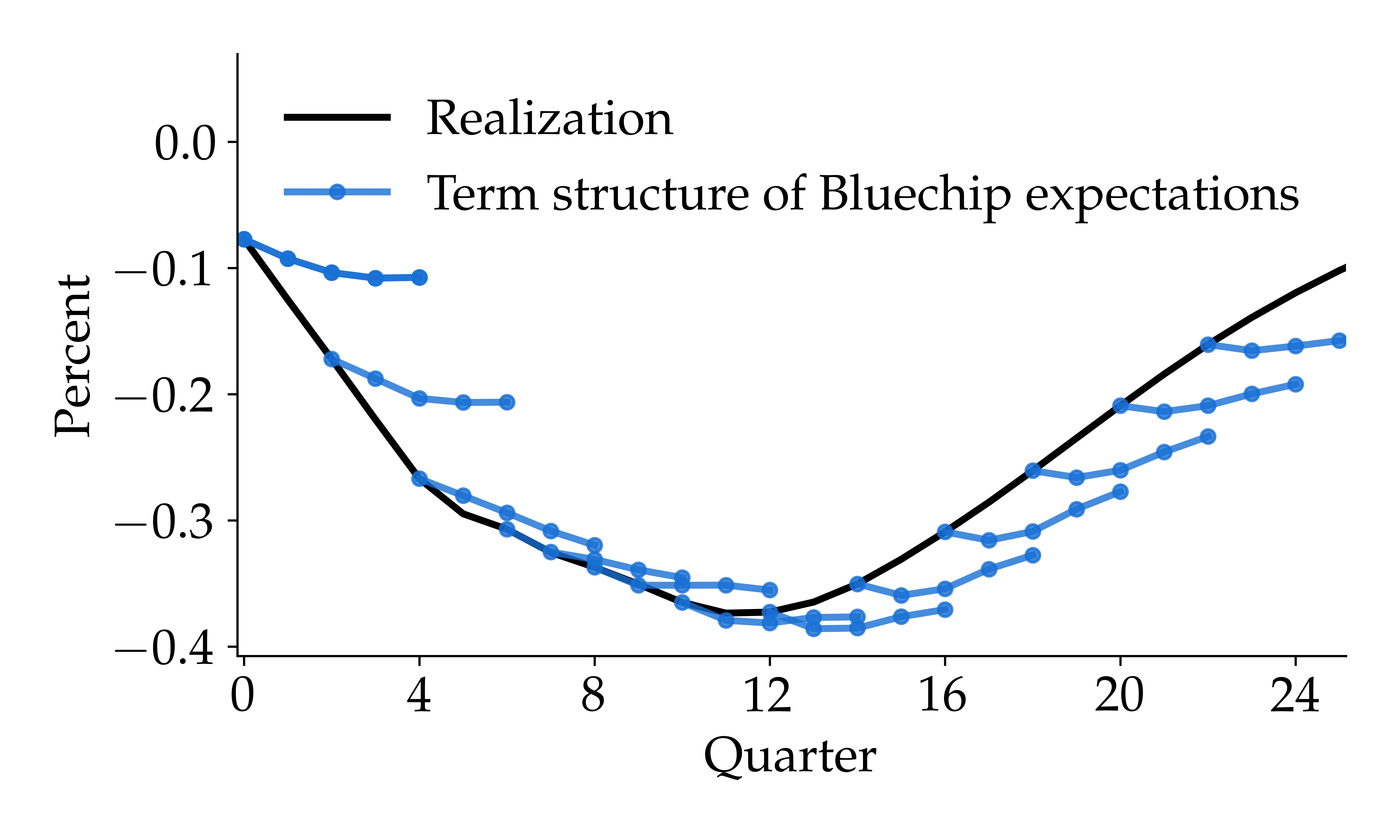}
    \end{subfigure}

    \caption{Realized and expected real output responses to a \cite{kanzig2021macroeconomic} oil shock}
    \label{fig:output_rlzd_bluechip_exp}
    }

    \caption*{\footnotesize 
    \emph{Note}: 
    Both panels plot the realized real output impulse response (black) to a positive 
    \cite{kanzig2021macroeconomic} oil supply news shock, which raises real oil prices by ten 
    percent on impact.
    The left panel plots the impulse response function of the consensus Bluechip one-year ahead 
    expectation of real output (blue).
    The right panel plots the impulse responses of the one-quarter through one-year ahead expectations (blue dots), 
    connected across horizons as opposed to time.
    }
\end{figure}

As discussed in \cite{kuvcinskas2022measuring} and \cite{angeletos2021imperfect}, impulse responses 
of measured expectations can be useful moments to diagnose biases in expectation formation.
In the left panel of Figure \ref{fig:output_rlzd_bluechip_exp}, I plot the impulse response of realized and one-year ahead expected real output 
to an oil supply news shock from \cite{kanzig2021macroeconomic}, using a similar structural 
VAR specification and realized data as in \cite{kanzig2021macroeconomic} and augmented with consensus 
survey expectations data from the Bluechip Economic Indicators and Financial Forecasts.
I postpone discussion of the details of the data and the relevance of this particular choice of 
shock until Section \ref{sec:estimation}.



Real output displays an inertial contraction reaching its peak response after nearly twelve quarters 
in response to an oil supply news shock that raises real oil prices on impact.
One-year ahead expectations initially underreact, with the impact period one-year ahead 
output expectation falling to -0.1 percent compared to the ex-post 
realization of -0.3 percent. 
However, similarly to the response of other macroeconomic variables to 
other shocks as in \cite{angeletos2021imperfect}, output expectations here 
eventually overreact after twelve quarters.



This pattern of delayed overreaction holds for the responses of expectations across various horizons, 
as shown in the right panel of Figure \ref{fig:output_rlzd_bluechip_exp}, 
where each blue dot extending from the black line represents the impulse response 
coefficient of the one-quarter through one-year ahead real output expectations.

\begin{figure}[t]
    {
    \centering
    \begin{subfigure}{0.495\textwidth}
        \centering
        \includegraphics[width=\linewidth]{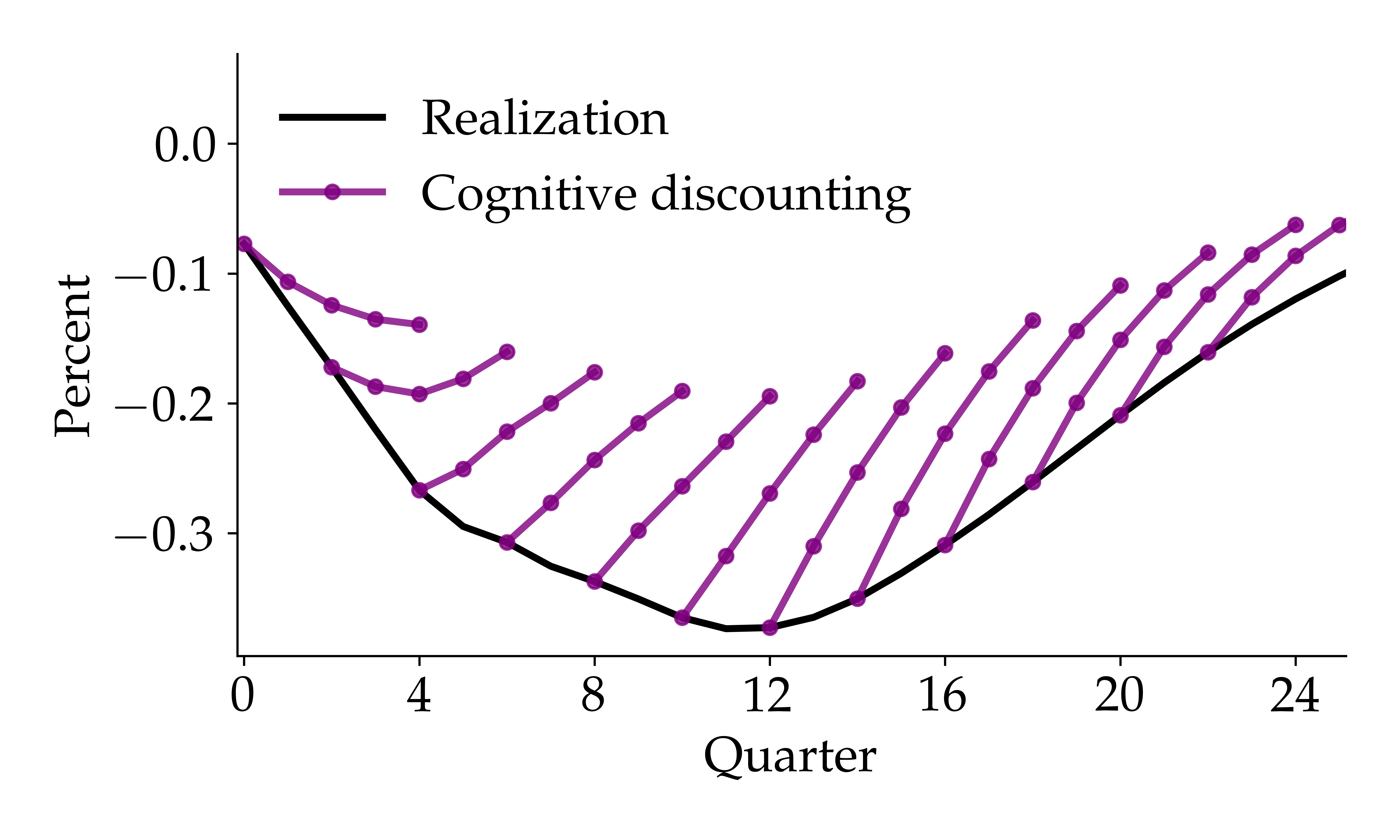}
    \end{subfigure}
    \hfill
    \begin{subfigure}{0.495\textwidth}
        \centering
        \includegraphics[width=\linewidth]{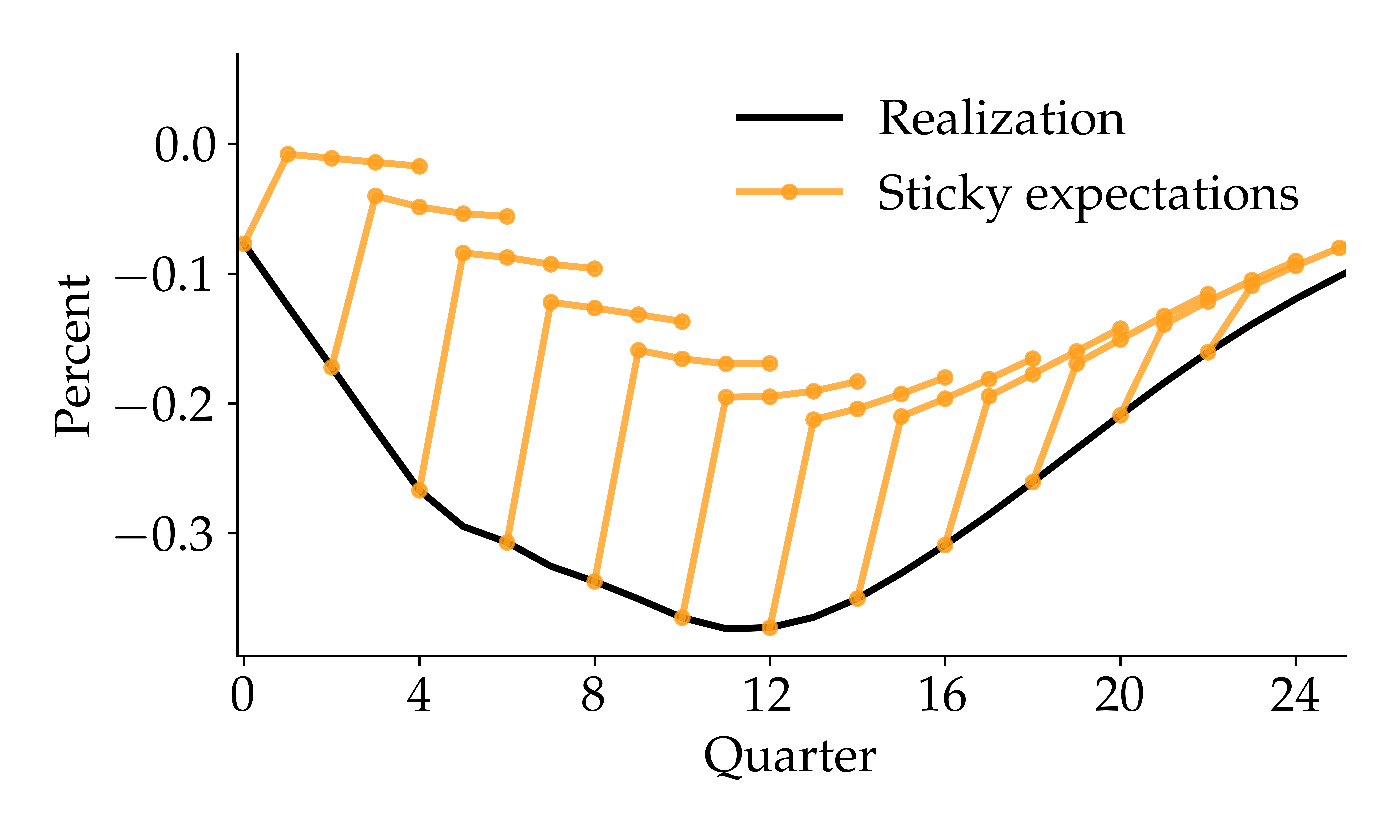}
    \end{subfigure}

    \caption{Systematic underreaction in expected real output responses to a \cite{kanzig2021macroeconomic} oil shock}
    \label{fig:output_cog_disc_and_sticky_exp}
    }

    \caption*{\footnotesize 
    \emph{Note}: 
    In addition to the realized real output impulse response (black), both figures plot the one-quarter 
    to one-year term structure of output expectations implied by two models of expectation underreaction, 
    cognitive discounting (purple) from \cite{gabaix2020behavioral} and sticky expectations (gold)
    from \cite{carroll2020sticky}.
    }
\end{figure}

Models of expectation underreaction have frequently been employed to rationalize inertia 
in realized outcomes. 
The economic rationale is that underreaction by definition dampens the initial responsiveness of 
economic agents, for example the consumption responses of households 
at the onset of a contractionary supply shock. 
However, over time as the contractionary output surprises continue this 
causes households to further spend down their wealth, causing output fall to its eventual trough.

Let us define the realized (linear) impulse response function of a variable $Y_t$ as
$$
\Psi(Y_t ; \varepsilon_{t-\ell}) = \mathbb{E}[Y_t | \varepsilon_{t-\ell} = 1, \mathbf{X}_{t-\ell}] - \mathbb{E}[Y_t | \varepsilon_{t-\ell} = 0, \mathbf{X}_{t-\ell}]
$$
Likewise we can define the response of a subjective expectation $E_t[Y_{t+h}]$ as
$$
\Psi(E_t[Y_{t+h}] ; \varepsilon_{t-\ell}) = \mathbb{E}[E_t[Y_{t+h}] | \varepsilon_{t-\ell} = 1, \mathbf{X}_{t-\ell}] - \mathbb{E}[E_t[Y_{t+h}] | \varepsilon_{t-\ell} = 0, \mathbf{X}_{t-\ell}]
$$
If $E_t[Y_{t+h}]$ is the rational expectation $\mathbb{E}_t[Y_{t+h}]$ then by the law of iterated expectations, 
we have
$$
\Psi(Y_{t+h}; \varepsilon_{t-\ell}) = \Psi(\mathbb{E}_t[Y_{t+h}]; \varepsilon_{t-\ell})
$$

Figure \ref{fig:output_cog_disc_and_sticky_exp} plots the behavior of the term structure of expectations 
implied by two common models of expectation underreaction: 
cognitive discounting as in \cite{gabaix2020behavioral} and 
sticky expectations as in \cite{carroll2020sticky}. 
These expectations were produced by applying each behavioral bias to the realized 
impulse response path of real output, which we treat as the rational expectations benchmark
\footnote{This relies on the assumption that the estimated impulse response from our empirical specification is the efficient 
linear conditional expectation of real output given the observable variables included in the 
specification and the realization of the exogenous shock.
If the data were generated from a structural model, which yielded a given structural moving average 
representation with respect to the exogenous shocks, then this assumption relies on our empirical 
specification being able to accurately recover this representation. 
}.

Because cognitive discounting and sticky information are constructed off of a rational expectations benchmark, 
where cognitive discounting anchors more to zero across horizons-$h$ and sticky expectations 
anchors less to zero as the time-$\ell$ following the shock passes, we can manipulate the 
time-$t$ rational expectation of $t+h$ impulse response, or equivalently the $t+h$ realization impulse 
response, to construct the response of expectations implied by these models of underreaction.
The impulse responses of subjective expectations from these models are then given by
\begin{align*}
    \Psi(E^{\text{CD}}_t[Y_{t+h}]; \varepsilon_{t-\ell}) &= \theta^h \Psi(Y_{t+h}; \varepsilon_{t-\ell}) \\
    \Psi(E^{\text{SE}}_t[Y_{t+h}]; \varepsilon_{t-\ell}) &= (1 - \vartheta^{\ell + 1}) \Psi(Y_{t+h}; \varepsilon_{t-\ell})
\end{align*}
where Figure \ref{fig:output_cog_disc_and_sticky_exp} uses a cognitive discount $\theta = 0.85$ 
as in \cite{gabaix2020behavioral} and the sticky expectation $\vartheta = 0.935$ 
as estimated in \cite{auclert2020micro}. 

\begin{figure}[t]
    {
    \centering
    \includegraphics[width=0.6\linewidth]{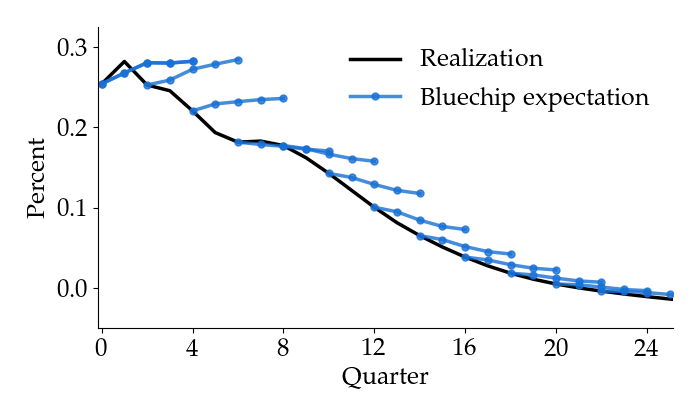}
    \caption{Realized and expected consumer price response to a \cite{kanzig2021macroeconomic} oil shock}
    \label{fig:cpi_rlzd_bluechip_exp}
    }

    \caption*{\footnotesize 
    \emph{Note}: This figure displays the consumer price response (black) to a positive 
    \cite{kanzig2021macroeconomic} oil supply news shock, which raises real oil prices by ten 
    percent on impact, and the impulse responses of the one-quarter through one-year ahead expectations (blue dots), 
    connected across horizons.
    }
\end{figure}

While both of these models of expectation underreaction are capable of generating inertia in realized outcomes 
when embedded in a standard DSGE model, they are unable to explain key patterns of bias in 
measured expectations shown in Figure \ref{fig:output_rlzd_bluechip_exp}, most notably 
the eventual overreaction in measured expectations. 
Existing models that have been proposed to explain delayed overreaction combine elements of 
models of pure overreaction, such as baseline diagnostic expectations (\citealt{bordalo2018diagnostic}), 
with models of underreaction, where the underreaction bias disappears over time. 
Examples include combinations of noisy information and diagnostic expectations 
(\citealt{bordalo2020overreaction}, \citealt{bardoczy2023unemployment}) or over-persistence bias 
(\citealt{angeletos2021imperfect}).
These combined models of underreaction and overreaction, however, are unable to replicate patterns of 
pure overreaction in the response of other variables, such as the consumer prices 
plotted in Figure \ref{fig:cpi_rlzd_bluechip_exp}, which has also independently been 
documented in experimental settings (\citealt{afrouzi2023overreaction}).

\begin{figure}[t]
    {
    \centering
    \begin{subfigure}{0.495\textwidth}
        \centering
        \includegraphics[width=\linewidth]{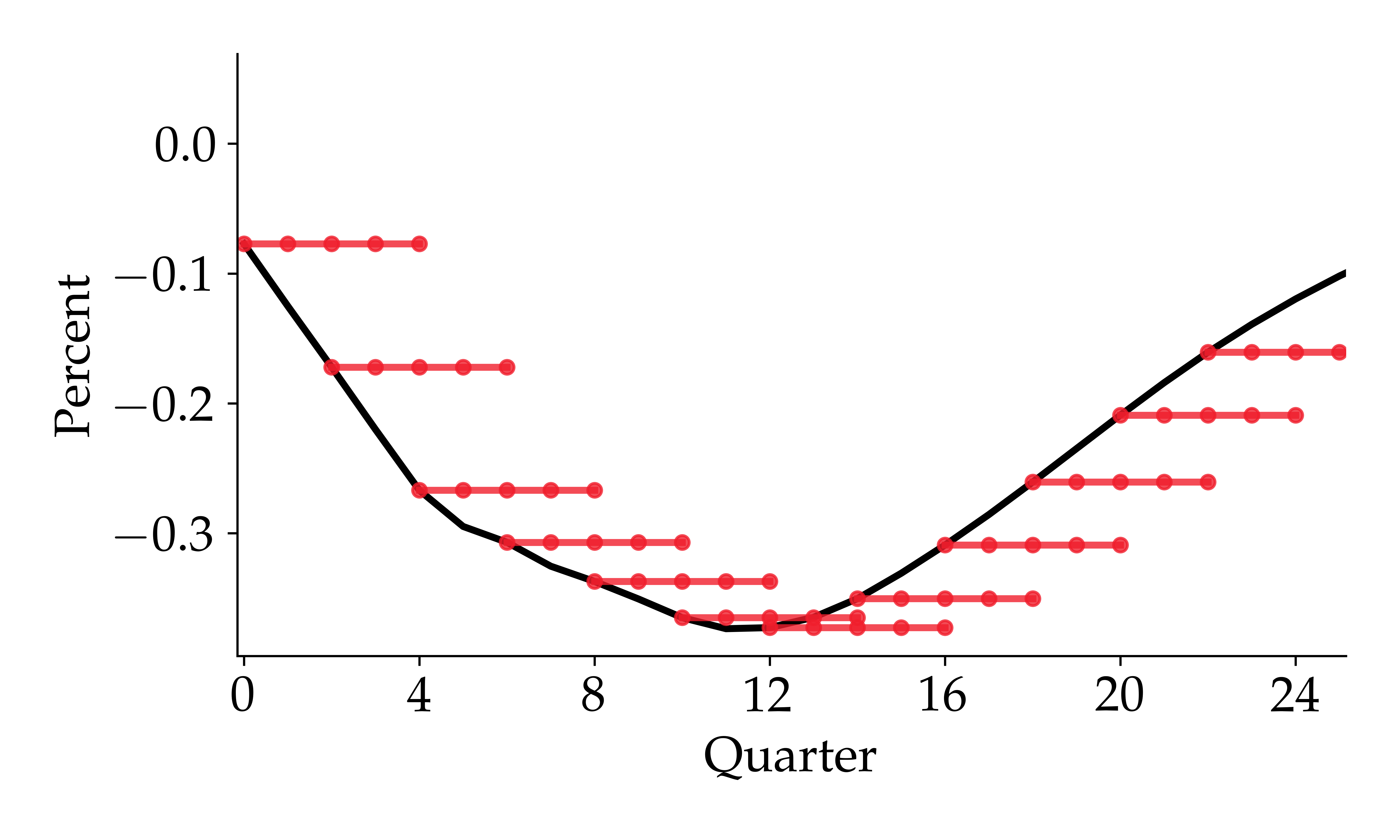}
        \caption{Real output}
    \end{subfigure}
    \hfill
    \begin{subfigure}{0.495\textwidth}
        \centering
        \includegraphics[width=\linewidth]{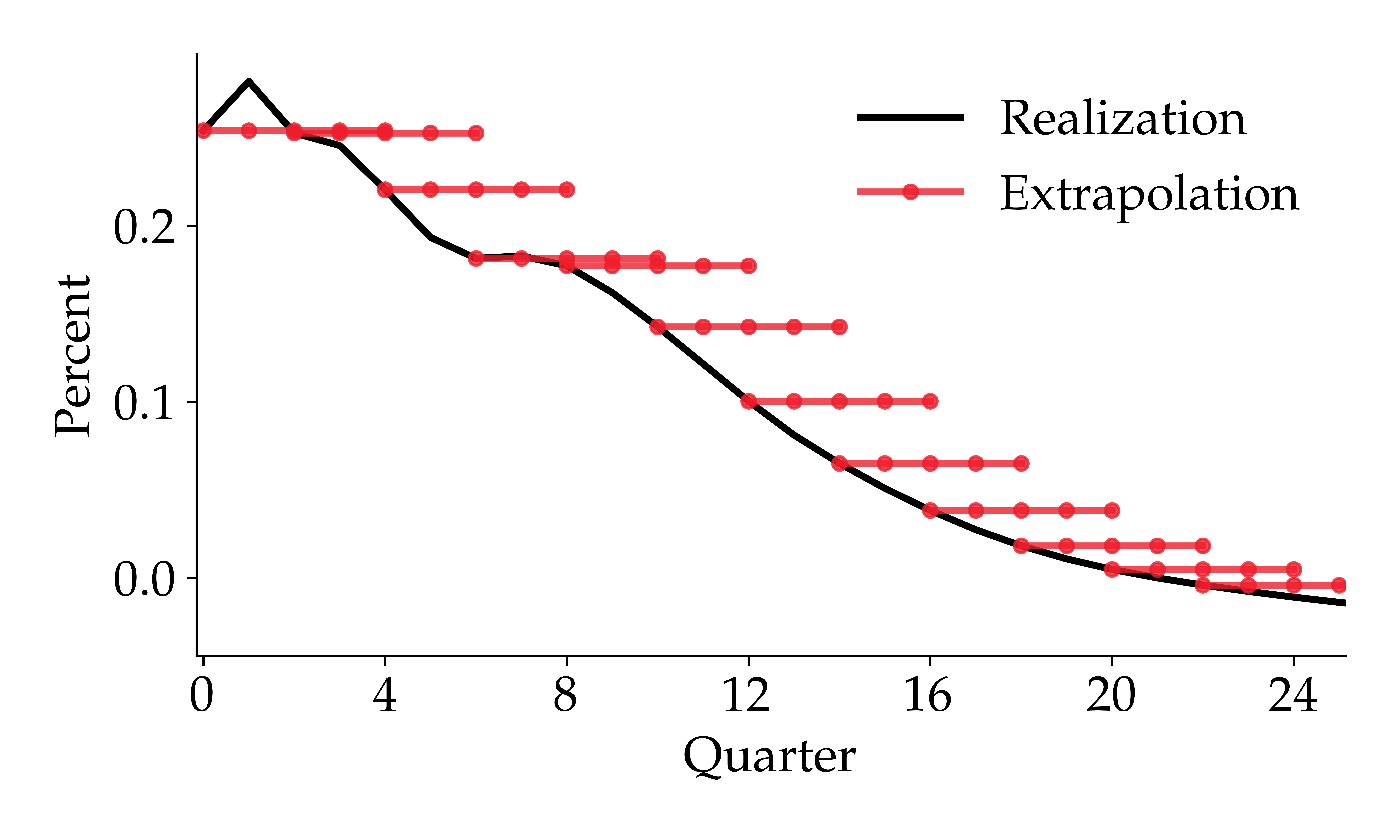}
        \caption{Consumer price index}
    \end{subfigure}

    \caption{Extrapolation in the response of expectations to a \cite{kanzig2021macroeconomic} oil shock}
    \label{fig:output_cpi_extrap}
    }

    \caption*{\footnotesize 
    \emph{Note}: The panels display the real output and consumer price response (black) to a positive 
    \cite{kanzig2021macroeconomic} oil supply news shock, which raises real oil prices by ten 
    percent on impact. 
    Each panel additionally plots the time-$t$ one-quarter through one-year ahead expectations 
    generated by extrapolating forward the time-$t$ realized impulse response (red).
    }
\end{figure}

While differing degrees of under- and overreaction across variables and over time are 
difficult to rationalize with many existing models of expectation formation, 
models that imply a form of extrapolation bias are able to explain observed 
patterns of forecast errors. 
Figure \ref{fig:output_cpi_extrap} illustrates the explanatory power of extrapolation bias, 
displaying the realized real output and consumer price impulse responses alongside 
a subjective expectation constructed by mechanically extrapolating forward the 
current realization response, that is
\begin{equation}\label{eq:simple_overextrap}
\Psi(E^{\text{Ex}}_t[Y_{t+h}]; \varepsilon_{t-\ell}) = \Psi(Y_t; \varepsilon_{t-\ell}) \text{ for all } h > 0
\end{equation}
The key insight to replicating both delayed overreaction in real output expectations and 
pure overreaction in consumer price expectations rests on the shape of the 
underlying realized impulse response. 
For hump-shaped realized responses, extrapolation results in delayed overreaction, 
but for realized responses that are peaked on impact and monotonically declining thereafter, 
extrapolation results in pure overreaction.

Hence, existing theories of expectation formation, which relied purely on underreaction to generate 
inertia in realized outcomes, may have done so at the cost of missing important features 
of measured expectations, namely extrapolation bias.
Instead of taking a stand on a particular theory of expectation formation, 
I next demonstrate how one can evaluate the aggregate consumption path implied 
by a given structural model of consumption-savings by conditioning directly on the 
paths of measured expectations and realizations of income and interest rates. 


\section{Model-implied impulse responses using expectations data}\label{sec:mod_irfs}

In this section I construct moment conditions, which 
represent the distance between unrestricted, empirical impulse responses 
and structural model-implied impulse responses to an externally identified exogenous shock. 
The main innovation relative to existing approaches, such as Euler equation estimation 
in \cite{hansen1982generalized}, \cite{barnichon2020identifying}, \cite{lewis2022dynamic}, is 
the use of a linearized, aggregate, sequential representation of the structural model, 
which removes the need to impose rational expectations or any other 
theory of expectation formation.
In lieu of a model of expectation formation, I directly use expectations data of 
the decision-relevant variables, such as income and interest rates for consumption, 
to evaluate the moment condition. 
I first illustrate this approach in the familiar representative-agent example 
and then proceed to the heterogeneous-agent case. 

\subsection{Representative-agent example}
Consider a representative household consumption-savings problem given 
arbitrary, subjective expectations $E_t[\cdot]$ of future income and interest rates $Y_s, r_s$ for $s > t$.
\begin{align*}
    \max_{C_t, A_t} \sum_{t=0}^\infty \beta^t \zeta_t E_0\left[\frac{C_t^{1-\sigma^{-1}}}{1-\sigma^{-1}}\right]\\
    \text{s. to }C_t + A_t = Y_t + (1 + r_{t-1}) A_{t-1}
\end{align*}
The household consumes $C_t$ and saves in a one-period, risk-free asset $A_t$, 
taking income and interest rates as given. 
Additionally, their choices are subject to an exogenous discount factor shock $\zeta_t$, 
which serves as an unobserved source of endogeneity from the perspective 
of the econometrician, who is aiming to estimate the structural elasticity of substitution parameter, $\sigma$.

Taking first-order conditions and linearizing around the steady-state $\beta(1 + r) = 1$, 
we obtain the subjective expectations consumption function, where $W_t := r_{t-1} A + (1 + r)A_{t-1}$ 
denotes financial wealth, $\gamma := \sigma \beta - (1 - \beta) A$ denotes 
the net interest rate elasticity, and $\varepsilon_t$ is the exogenous ``demand shock'', which is 
a function of the discount factor shock $\zeta_t$.
\begin{equation}\label{eq:ra_cons_func}
    C_t = (1 - \beta) \left(\sum_{h=0}^\infty \beta^h E_t[Y_{t+h}] + W_t\right) - \gamma \sum_{h=0}^\infty \beta^{h+1} E_t[r_{t+h}] + \varepsilon_t
\end{equation}
I assume current-period variables $Y_t, r_t$ are directly observed by households 
in that period and equivalently denoted as horizon $h=0$ expectations.
That is, $E_t[Y_t] \equiv Y_t, E_t[r_t] \equiv r_t$.
This serves to simplify the information structure and ease interpretation of the 
problem. I relax this assumption in the heterogeneous-agent case.

Suppose we want to estimate the structural model given in Equation (\ref{eq:ra_cons_func}) to match 
the unrestricted, empirical impulse response of consumption $C_t$ to 
an externally-identified, exogenous shock $z_t$, obtained from a local projection or 
vector auto-regression specification.

One classic approach in the literature, as used in \cite{christiano2005nominal},
is ``impulse response matching'', which imposes three key assumptions to estimate 
Equation (\ref{eq:ra_cons_func}). 
First, assume the structure of the rest of the surrounding general equilibrium environment: 
the firm problem, policy rules, market clearing, and the other exogenous shocks.
Second, assume that the identified shock from the empirical specification has a model 
counterpart in the equilibrium model.
Third, assume full-information rational expectations, or more generally some other 
model of expectation formation such as sticky expectations in \cite{auclert2020micro}. 
With these three assumptions, one can simulate the shock in the model to replicate the 
targeted responses produced by the empirical specification.

To avoid the specification bias that may arise due to the first two assumptions 
in impulse response matching, single-equation estimation approaches to estimating 
structural impulse responses, as in \cite{barnichon2020identifying}, 
\cite{lewis2022dynamic}, can be employed.
However, a recursive, aggregate representation of a structural model such as an Euler equation 
which is used in these previous papers cannot generically be obtained when relaxing the full-information, 
rational expectations assumption, such as in models of dispersed, private information 
and a lack of common knowledge (\citealt{angeletos2018forward}).

To remain agnostic to the correct model of expectation formation and move away 
from the rational expectations assumption, we must work directly with the sequential representation 
of the model. Denote the net present value of subjective expected income and rates as
\begin{align*}
    \mathcal{Y}_t &:= \sum_{h=0}^\infty \beta^h E_t[Y_{t+h}]\\
    \mathcal{R}_t &:= \sum_{h=0}^\infty \beta^{h+1} E_t[r_{t+h}]
\end{align*}

Suppose we have imperfectly measured expectations data for income and interest 
rates for all future horizons, denoted by the expectations operator $E_t^{\text{data}}[\cdot]$. 
Substituting them into the net present value expressions
denoting the new quantities as $\mathcal{Y}^{\text{data}}_t, \mathcal{R}^{\text{data}}_t$ 
and collecting the differences into a measurement error term 
$\mu_t := \mathcal{Y}_t - \mathcal{Y}_t^{\text{data}} + \mathcal{R}_t - \mathcal{R}_t^{\text{data}}$, 
we obtain
\begin{equation}\label{eq:ra_cons_func_data}
    C_t = (1 - \beta) \left(\mathcal{Y}^{\text{data}}_t + W_t\right) - \gamma \mathcal{R}^{\text{data}}_t + \varepsilon_t + \mu_t
\end{equation}

Given the potential endogeneity of $\varepsilon_t, \mu_t$ to 
$\mathcal{Y}^{\text{data}}_t, \mathcal{R}^{\text{data}}_t$, we need to find a 
valid instrumental variable $z_t$ that satisfies the standard assumptions\footnote{
$\mathcal{Y}^{\text{data}}_t$ does not show up in the relevance condition because 
the discount factor $\beta = (1 + r)^{-1}$ is a known quantity.
}
\begin{align*}
    \text{Exogeneity} \quad &\mathbb{E}[\varepsilon_t z_t] = \mathbb{E}[\mu_t z_t] = 0\\
    \text{Relevance} \quad &\mathbb{E}\left[\mathcal{R}^{\text{data}}_{t} z_t\right] \neq 0
\end{align*}

The net interest rate elasticity $\gamma$ contains the structural elasticity of intertemporal 
substitution, henceforth the EIS, $\sigma$, which is the only free parameter to estimate.

What instruments will satisfy these assumptions?
As in \cite{barnichon2020identifying} and \cite{lewis2022dynamic}, I will use contemporaneous values 
and lags of suitable, externally identified structural shocks as instruments, denoted $\{z_{t-\ell}\}$.
For example, to instrument for a demand shock $\varepsilon_t$ and satisfy the exogeneity condition, 
a natural candidate for $z_{t-\ell}$ is a supply shock. 
Additionally, to satisfy exogeneity with respect to the measurement error $\mu_t$, 
the response of the expectations data to the shock $z_{t-\ell}$ that I use to construct proxy measures 
$\mathcal{Y}^{\text{data}}_t, \mathcal{R}^{\text{data}}_t$ must resemble that of 
the response of true, subjective expectations of households $\mathcal{Y}_t, \mathcal{R}_t$.

A nice interpretation from using structural shocks as instruments is the instrumented version of 
Equation (\ref{eq:ra_cons_func_data}) can be interpreted as an ordinary least-squares 
``regression on impulse responses''. 
That is, post-multiplying and taking expectations we see that the instrumental variables 
moment condition can be interpreted as a comparison of impulse responses\footnote{The explanatory and response variables in the below expression are 
unnormalized estimands of a simple local projection onto the shock $z_{t-\ell}$, which is the approach 
used by \cite{barnichon2020identifying}. 
\cite{lewis2022dynamic} demonstrate how to obtain an analogous representation using standard LP/VAR specifications with lags 
and controls.}.
\begin{equation*}
    \mathbb{E}[C_t z_{t-\ell}] = (1 - \beta) \left(\mathbb{E}[(\mathcal{Y}^{\text{data}}_t + W_t)z_{t-\ell}]\right) - \gamma \mathbb{E}[\mathcal{R}^{\text{data}}_t z_{t-\ell}], \,\, \ell \geq 0
\end{equation*}
The left-hand side of the above expression can be interpreted as 
the unrestricted impulse response of consumption and 
the right-hand side as the representative-agent, model-implied impulse response 
of consumption to the shock $\{z_{t-\ell}\}_{\ell \geq 0}$. 

\paragraph{A consumption function of histories of realizations and expectations}
The aggregate consumption function written in Equation (\ref{eq:ra_cons_func}) had a single dynamic 
state variable, aggregate financial wealth $W_t$, which contains incoming period assets $A_{t-1}$.
To write a more general representation of the consumption function that will provide intuition for 
the heterogeneous-agent case, let us re-consider Equation (\ref{eq:ra_cons_func_data}). 
Model-implied consumption is the sum of responses due to endogenous and exogenous variables, denoted
$\tilde{\mathcal{C}}_t, \tilde{\mathcal{E}}_t$ respectively
\begin{equation}\label{eq:cons_func_gen_ra_w_Atm1}
    C_t = \tilde{\mathcal{C}_t}(\mathcal{Y}^{\text{data}}_t, \mathcal{R}^{\text{data}}_t, W_t; \sigma) + \tilde{\mathcal{E}_t}(\varepsilon_t, \mu_t)
\end{equation}
Using the household budget constraint $C_t + A_t = Y_t + (1 + r_{t-1}) A_{t-1}$,
we can recursively substitute the asset state $A_{t-1}$ out of Equation (\ref{eq:cons_func_gen_ra_w_Atm1}) to 
obtain an alternate representation
\begin{equation}\label{eq:cons_func_gen_ra}
    C_t = \mathcal{C}_t(\{ \mathcal{Y}^{\text{data}}_\tau, \mathcal{R}^{\text{data}}_\tau, r_{\tau-1} A \}_{\tau \leq t}; \sigma) + \mathcal{E}_t(\{\varepsilon_\tau, \mu_\tau\}_{\tau \leq t})
\end{equation}

Equipped with a set of externally identified structural shocks as instrumental variables as before, 
we can derive the ``regression on impulse responses'' moment condition for the general 
consumption function representation in Equation (\ref{eq:cons_func_gen_ra})
\begin{equation}\label{eq:mom_cond_cons_func_gen_ra}
    \mathbb{E}[(C_t - \mathcal{C}_t(\{ \mathcal{Y}^{\text{data}}_\tau, \mathcal{R}^{\text{data}}_\tau, r_{\tau-1} A \}_{\tau \leq t}; \sigma))z_{t-\ell}] = 0, \,\, \ell \geq 0
\end{equation}
when $\sigma = \sigma_0$, the true parameter value.

While representations (\ref{eq:cons_func_gen_ra_w_Atm1}) 
and (\ref{eq:cons_func_gen_ra}) resulted from a simple algebraic manipulation, 
they are connected by a useful intuition.
Current financial wealth represents the \emph{total} accumulated effect on current consumption 
of past consumption-savings decisions, which were based on past expectations and realizations 
of income and interest rates.
By integrating out wealth as the dynamic state variable, we can explicitly consider 
the \emph{individual} contribution of each past expectation and realization in isolation, 
say $E^{\text{data}}_\tau[Y_{\tau+h}]$ at some time period $\tau < t$, on past 
consumption and savings $C_\tau, A_\tau$, and in turn how that change in outgoing savings 
$A_\tau$ propagated forward to eventually impact current consumption $C_t$. 
This chain of events is summarized by the coefficient of $\mathcal{C}_t$ on $E^{\text{data}}_\tau[Y_{\tau+h}]$
in Equation (\ref{eq:cons_func_gen_ra}). 

Dynamic state variables like $W_t$ are often low-dimensional in representative-agent models, 
which allows for this backward algebraic substitution to derive Equation (\ref{eq:cons_func_gen_ra}) 
in explicit form.
However, when dynamic state variables are function-valued distributions as in heterogeneous-agent models, 
this procedure becomes analytically intractable.
I will show in the following section, with a few additional assumptions,
that one can obtain an analogous aggregate consumption function representation 
to Equation (\ref{eq:cons_func_gen_ra}) in heterogeneous-agent models with distributional state variables, 
like incomplete markets models with idiosyncratic risk and borrowing constraints.

\subsection{Heterogeneous-agent consumption functions and moment conditions}\label{sec:mom_conds_gen_models}



I now demonstrate how a class of heterogeneous-agent models that include 
idiosyncratic risk, incomplete markets, and borrowing constraints 
(\citealt{bewley1986stationary}, \citealt{imrohorouglu1989cost}, 
\citealt{huggett1993risk}, \citealt{aiyagari1994uninsured}) admit 
analogous moment conditions to Equation (\ref{eq:mom_cond_cons_func_gen_ra})\footnote{The dynamic programming setup that leads to the following 
consumption function derivation is 
also outlined in \cite{auclert2020micro}, albeit applied to sticky expectations and other parametric 
models of expectation formation, and in \cite{bardoczy2023unemployment}, who derive a 
similar representation but in terms of forecast errors and revisions.}.

\paragraph{Dynamic programming problem setup} 
Suppose individuals-$i$ have time-$t$ subjective expectations $E_{i, t}$.
Let an individual-$i$'s optimization problem be defined by the following 
value function with common structural parameters $\boldsymbol{\theta}$ that 
we seek to estimate
\begin{equation}\label{eq:gen_backward_eqn}
    v_{i, t} = v(E_{i, t}[v_{i, t+1}], \mathbf{S}_{i, t}; \boldsymbol{\theta})
\end{equation}

An individual-$i$'s state $\mathbf{S}_{i, t} = (\mathbf{s}_{i, t}, \mathbf{X}_{t})$ 
has an idiosyncratic component $\mathbf{s}_{i, t}$ and an aggregate one $\mathbf{X}_t$.
Individuals-$i$ take aggregates $\mathbf{X}_t$ as given\footnote{That is, they act as if they are atomistic with respect to aggregate outcomes $\mathbf{X}_t$, even though their actions may collectively determine $\mathbf{X}_t$ variables in equilibrium.}, 
which represent either aggregate 
exogenous shocks or aggregate endogenous variables determined in equilibrium. 
I deviate from the usual definition of the state $\mathbf{S}_{i, t}$, which includes 
the full information set of expectation $E_{i, t}$, to separate variables which directly 
influence the decision problem in Equation (\ref{eq:gen_backward_eqn}) via 
current constraints and flow payoffs, denoted $\mathbf{S}_{i, t}$, from 
those that may also influence expectation formation, which I leave implicit 
in the operator $E_{i, t}$. 
This distinction allows me to state assumptions more clearly.

{
\assumption{
\normalfont{(\textbf{Idiosyncratic rationality})} \label{assump:idio_rationality}
Exogenous idiosyncratic states $\mathbf{x}_{i, t} \subset \mathbf{s}_{i, t}$ are 
stochastic processes whose functional form is common and commonly known across individuals-$i$.
}
}
\vspace{0.25cm}

\noindent Because I will focus on the effects of measured biases in average survey expectations of aggregate 
realizations, I reduce the generality of this problem setup by only considering deviations from full-information, 
rational expectations about aggregate variables. 
By Assumption \ref{assump:idio_rationality}, subjective expectations of functions of only exogenous 
idiosyncratic states are therefore taken with respect to their true probability density.
I place the following additional structure on subjective expectations.

{
\assumption{Individual-$i$, time-$t$ subjective conditional expectations $E_{i, t}[\cdot]$ satisfy
\begin{enumerate}
    \item[\normalfont{a)}] \normalfont{\textbf{Dynamic consistency}}\, $E_{i, t}[v_{i, t+h}] = E_{i, t}[v(E_{i, t+h}[v_{i, t+h+1}], \mathbf{S}_{i, t+h}; \boldsymbol{\theta})]$ for $h > 1$
    \item[\normalfont{b)}] \normalfont{\textbf{Law of iterated expectations}}\, $E_{i,t}[E_{i, t^\prime}[X_{i, T}]] = E_{i, t}[X_{i, T}]$ for all $t < t^\prime < T$ and stochastic variables $X_{i, t}$ 
    \item[\normalfont{c)}] \normalfont{\textbf{Independence}}\, variables inducing heterogeneity in $E_{i, t}$ across $i$ are independent of $\mathbf{s}_{i, t}$
\end{enumerate}
\label{assump:indiv_exp}
}
}

Assumptions \ref{assump:indiv_exp}a) and \ref{assump:indiv_exp}b) build on 
the idea of individual rationality from Assumption \ref{assump:idio_rationality}. 
Assumption \ref{assump:indiv_exp}a) requires agents to be aware of their dynamic programming problem 
as written in Equation (\ref{eq:gen_backward_eqn}) and to form expectations in a manner that is consistent 
with its recursive structure.
Assumption \ref{assump:indiv_exp}b) is a typical assumption requiring agents' own perceptions of 
their subjective expectations to be efficient and hence not systematically subject to future revision.
Notably, this is not the same thing as assuming law of iterated expectations of the cross-sectional 
average expectation $\int_{i \in \mathcal{I}} E_{i, t}$, which will not hold generically, for example
 when relaxing the common knowledge assumption as in \cite{angeletos2018forward}.

Assumption \ref{assump:indiv_exp}c) is the main substantive assumption, restricting all forms of 
correlated heterogeneity in expectations with individual characteristics or states. 
An example of a model of expectation formation where this assumption holds is sticky 
information as in \cite{mankiw2002sticky}, where the prior information update period is 
independent of all individual characteristics and states even though it is heterogeneous across the population.
Hence, the individual state variable encoded in the subjective expectation $E_{i, t}$ would be the prior 
information update period $\tau \leq t$, which is assumed to be independent of individual 
states $\mathbf{s}_{i, t}$ such as income.
An example of a violation of this assumption is if individual attention correlated 
with the incidence of aggregate income on individual income, as in \cite{guerreiro2023belief}. 

While this assumption is not innocuous, I would need additional data on the covariances of individual-$i$ 
subjective expectations and characteristics, which are not be available in the survey expectations 
data that I use.
Nonetheless, given that a large class of models of imperfect information and/or non-rational 
expectations abides by Assumptions \ref{assump:idio_rationality} and \ref{assump:indiv_exp}\footnote{Some examples 
include diagnostic expectations \cite{bordalo2018diagnostic} applied to aggregate variables, sticky 
information \cite{mankiw2002sticky} and sticky expectations \cite{carroll2020sticky}, 
exogenous noisy and dispersed information \cite{angeletos2021imperfect}.} and that my focus 
is to explain deviations of cross-sectional average expectations about aggregate variables, I 
believe it is useful to operate under these commonly maintained assumptions.



Let $\mathcal{I}$ denote the index set of all individuals-$i$. 
The law of motion of the full distribution of individual-$i$ state variables $\mathbf{D}_t$
is then defined by the transition equation
\begin{equation}\label{eq:gen_forward_eqn}
    \mathbf{D}_{t+1} = \Lambda(\{ v_{i, t} \}_{i \in \mathcal{I}}, \mathbf{X}_t, \mathbf{D}_t)
\end{equation}
The aggregation of individual decision rules $y_{i, t}(E_{i, t}[v_{i, t+1}], \mathbf{X}_t; \boldsymbol{\theta}_i)$ is given by
\begin{equation}\label{eq:gen_aggr_eqn}
    Y_t = \mathcal{Y}(\{ y_{i, t}(v_{i, t})\}_{i \in \mathcal{I}}, \mathbf{D}_t)
\end{equation}
for a scalar, aggregated variable $Y_t$.

\definition{A \textbf{steady state} ($\{v_i, \mathbf{s}_i \}_{i \in \mathcal{I}}, \mathbf{D}, \mathbf{X}, Y$) is a constant-valued fixed point  
consistent with (\ref{eq:gen_backward_eqn}, \ref{eq:gen_forward_eqn}, \ref{eq:gen_aggr_eqn}).
}
\vspace{0.25cm}

\vspace{0.25cm}

\noindent Given Assumptions \ref{assump:idio_rationality}, \ref{assump:indiv_exp} and the system 
of equations (\ref{eq:gen_backward_eqn}, \ref{eq:gen_forward_eqn}, \ref{eq:gen_aggr_eqn}), the 
first-order response (locally around a steady state) of aggregated decisions $Y_t$ to 
an aggregate shock $\mathbf{X}_t$ is given by 
\begin{equation}\label{eq:gen_fakenews_func}
    Y_t = \sum_{\tau \, \leq \, t} \sum_{h \geq 0} \sum_{X \in \mathbf{X}} \mathcal{F}^X_{t - \tau, h}(\boldsymbol{\theta}) E_\tau[X_{\tau+h}]
\end{equation}
where $\mathcal{F}^X_{t - \tau, h}(\boldsymbol{\theta})$ is the $(t - \tau, h)$ entry of the ``fake news'' matrix $\mathcal{F}^X$,
defined in \cite{auclert2021using}, for an aggregate output variable $Y$ with respect to an aggregate input variable $X$. 

\vspace{0.25cm}


Our earlier representative-agent consumption function 
fits into the general representation expressed by Equation (\ref{eq:gen_fakenews_func}).
Given the standard incomplete markets model of household consumption-savings 
(\citealt{bewley1986stationary}, \citealt{imrohorouglu1989cost}, 
\citealt{huggett1993risk}, \citealt{aiyagari1994uninsured}) can be 
defined by the above system, we can derive an aggregate consumption function 
for this class of models under the same limited set of assumptions 
on expectation formation.

Consider the heterogeneous-agent aggregate consumption function implied by Equation (\ref{eq:gen_fakenews_func}) 
and the coefficients on its typical inputs: income, interest rates, and 
demand shocks\footnote{The fake news matrix for $\varepsilon_t$ depends on the particular micro-foundation 
one uses for the primitive demand shock that comprises this exogenous intercept term. 
For example, if $\varepsilon_t$ is a linear combination of discount factor shocks, then 
$\mathcal{F}^\varepsilon$ will be functions of the interest rate matrix $\mathcal{F}^r$, since 
discount factor shocks alter consumption similarly to perturbations to the ex-ante 
interest rate.}.
\begin{equation}\label{eq:ha_fakenews}
    C_t = \sum_{\tau \leq t} \sum_{h \geq 0} \mathcal{F}^Y_{t-\tau, h} E_\tau[Y_{\tau+h}] + \mathcal{F}^r_{t - \tau, h} E_\tau[r_{\tau+h}] + \mathcal{F}^\varepsilon_{t-\tau, h}E_\tau[\varepsilon_{\tau+h}]
\end{equation}
As in the representative-agent case in Equation (\ref{eq:cons_func_gen_ra}), aggregate consumption 
here is also a function of both current and past (cross-sectional average) expectations of income, 
interest rates, and demand shocks.
The sum across horizons-$h$ of the terms $\sum_{h \geq 0}\mathcal{F}^Y_{t-\tau, h} E_\tau[Y_{\tau, h}]$ is equivalent 
to the net present value of subjective expected income in the representative-agent model, previously 
denoted $\mathcal{Y}_\tau$.
A single coefficient $\mathcal{F}^Y_{t-\tau, h}$ can be interpreted as the contribution 
of the distribution of \emph{individuals}-$i$
consumption-savings responses at time-$\tau$, based on a change in their average, aggregate income expectation 
(or realization if $h=0$), to time-$t$ aggregate consumption, holding beliefs and realizations in other periods fixed.
These time-$\tau$ consumption-savings responses propagate forward and 
affect time-$t$ aggregate consumption because of their effects on
the evolution of wealth in the intervening periods.


The reason why coefficients in the matrices $\mathcal{F}^Y, \mathcal{F}^r$ represent the responses of 
aggregate consumption to a change in arbitrary, average subjective expectations 
relies principally on certainty equivalence.
As long as agents have certainty equivalence, then the response of their decisions to changes in 
their expectations at a given time-$\tau$ are the same, regardless of whether those expectations are rational 
or not.

If households collectively expect future income will be higher, the 
effect of that average belief\footnote{The average subjective belief 
is the only moment from the subjective probability distribution over aggregate variables 
that determines household consumption decisions because of certainty equivalence 
from the linearization.} on aggregate consumption
is the same irrespective of whether the belief is accurate or distorted. 
Likewise, when the next period arrives and realized income is lower or 
higher than expected, the change in aggregate consumption can be decomposed 
into two components: the response of aggregate consumption to an 
unanticipated aggregate income shock (without any change in future average expectations), 
and the response of aggregate consumption to a potentially revised set of average expectations.
As before, the effect of the latter component on the aggregate consumption response will be the 
same regardless of whether the average beliefs are accurate or distorted.

\paragraph{Moment conditions} 
As in the previous section, we can proceed to construct moment conditions from the 
aggregate consumption function in Equation (\ref{eq:ha_fakenews}).
For simplicity, I proceed by following the notation and distributed lag specification in 
\cite{barnichon2020identifying}, but the same procedure applies for more general 
VAR/LP specifications as in \cite{lewis2022dynamic}, which I will use later in the 
actual estimation\footnote{The moment conditions in this case would be written with all variables written 
as differences from their projection onto the set of controls, i.e. $Y^{\perp} := Y - \mathbb{E}[Y | X]$ 
for controls $X$.}.

Suppose we have a vector of current and lagged structural shocks 
$\mathbf{z}_t = \{z_{t-\ell}\}_{\ell = 0, ..., N_\ell}$ to use as instruments.
{
\assumption{\label{assump:serially_uncorr}
{\normalfont{(\textbf{Serially uncorrelated})}} \, $z_{t-\ell}$ are serially uncorrelated across $\ell$
}
}
\vspace{0.25cm}

Partition the aggregate state vector $\mathbf{X}_t$ into variables unobserved by the econometrician, 
$\boldsymbol{\varepsilon}_t$, and those observed, $\mathbf{W}_t$. 
{
\assumption{\label{assump:exog}
{\normalfont{(\textbf{Exogeneity})}} \, $\mathbb{E}[E_\tau[\varepsilon_{\tau+h}] \, z_{t-\ell}] = 0$, \quad $\forall \, \tau \leq t, h \geq 0, \, \varepsilon_{\tau+h} \in \boldsymbol{\varepsilon}_{\tau+h}, \, z_{t-\ell} \in \mathbf{z}_t$
}
}
\vspace{0.25cm}

This exogeneity condition is slightly more general than the one in the representative-agent 
example, encompassing the case where shocks may be imperfectly observed by economic agents. 
For $\tau < t - \ell$, if $z_{t-\ell}$ instruments are not systematically predictable by information 
available prior to time-$t-\ell$, it is natural to assume orthogonality 
to measurable functions of earlier information sets, 
i.e. $E_\tau[\varepsilon_{\tau+h}]$ for $\tau < t-\ell$. 
For $\tau \geq t-\ell$, this assumption requires agents to be aware that the instrument 
$z_{t-\ell}$ is orthogonal to time-$t$ information relevant for determining $\varepsilon_{\tau+h}$.
Alternatively, one could directly assume the shocks $\varepsilon_{\tau+h}$ are observable by agents 
but not the econometrician, and that the shocks $\varepsilon_{\tau+h}$ are known to be 
orthogonal to $z_{t-\ell}$. 
An example of this would be if $\varepsilon_{t+h}$ were household preference shocks known 
to be orthogonal to a supply or policy shock $z_{t-\ell}$.

\vspace{0.25cm}


\noindent Assume assumptions \ref{assump:idio_rationality}, \ref{assump:indiv_exp}, \ref{assump:serially_uncorr}, \ref{assump:exog}. 
Given cross-sectional average, subjective expectations $E_t[\cdot]$ and a vector of instruments $\mathbf{z}_t$, 
apply Equation (\ref{eq:gen_fakenews_func}), where $\mathbf{W} \subseteq \mathbf{X}$ denotes the 
subset of aggregate inputs $\mathbf{X}$ that are observable to the econometrician to obtain the following $N_\ell$ moment conditions
\begin{equation*}
    \mathbb{E}\left[\left(Y_t - \sum_{\tau=t - \ell}^t \sum_{h \geq 0} \sum_{W \in \mathbf{W}} \mathcal{F}^W_{t-\tau, h}(\boldsymbol{\theta}_0) E_{\tau}[W_{\tau+h}]\right) z_{t-\ell} \right] = 0, \,\, \text{ for } z_{t-\ell} \in \mathbf{z}_t
\end{equation*}
Where the moment condition equals zero uniquely at $\boldsymbol{\theta} = \boldsymbol{\theta}_0$.
\vspace{0.25cm}


\paragraph{Moment conditions with missing data for distant horizons}
Suppose we have data on realizations and cross-sectional average expectations of income 
and interest rates up to a finite horizon $H$, denoted by $\{E^{\text{data}}_t[\mathbf{W}_{t+h}]\}_{h \leq H}$. 
To be able to evaluate the moment conditions involving the full term structure of expectations, 
we need to extrapolate the missing horizons in the expectations data.

Let $F_t[\mathbf{W}_{t+h}; \boldsymbol{\vartheta}]$ denote an auxiliary model of the subjective expectation $E_t[\mathbf{W}_{t+h}]$, 
with parameter vector $\boldsymbol{\vartheta}$ to be used for extrapolation. 
We must then make an assumption that this auxiliary model produces an unbiased fit to the impulse response of 
expectations data for farther horizons. 
{
\assumption{{\normalfont \textbf{(Shape restriction)}} \label{assump:shape_restr} \, 
\begin{equation}\label{eq:shape_restr}
    \mathbb{E}[(E^{{\text{\normalfont data}}}_t[W_{t+h}] - F_t[W_{t+h}; \boldsymbol{\vartheta}_0]) z_{t-\ell}] = 0, \,\, {\normalfont \text{for }} h \leq H, W_{t+h} \in \mathbf{W}_{t+h}, \, z_{t-\ell} \in \mathbf{z}_t
\end{equation}
}
}
We can then estimate $\boldsymbol{\vartheta}$ from the auxiliary model using the $H N_\ell$ moment conditions 
implied by assumption \ref{assump:shape_restr}. 
For example, $\boldsymbol{\vartheta}$ could be coefficients of an AR(p) process fit to the impulse response 
of each subjective expectation $E^{\text{data}}_t[W_{t+h}]$ across horizons-$h$ and impulse response 
periods-$\ell$ to shocks $z_{t-\ell}$. 

Defined expectations data with missing horizons extrapolated from the auxiliary model as
$$
E^{\text{data}}_t[\mathbf{W}_{t+h}; \boldsymbol{\vartheta}] := 
\begin{cases}
    E_t^{\text{data}}[\mathbf{W}_{t+h}] & \text{ if } h \leq H \\
    F_t[\mathbf{W}_{t+h}; \boldsymbol{\vartheta}] & \text{ if } h > H
\end{cases}
$$
\vspace{0.10cm}

Then as before, assuming that the expectations data is an unbiased fit to the true model's average, subjective expectation
{
\assumption{{\normalfont \textbf{(Measurement error exogeneity)}} \label{assump:meas_error_exog} \, 
$$
\mathbb{E}[(E_t[W_{t+h}] - E^{{\text{\normalfont data}}}_t[W_{t+h}; \boldsymbol{\vartheta}_0]) z_{t-\ell}] = 0, \,\, {\normalfont \text{for }} W_{t+h} \in \mathbf{W}_{t+h}, \, z_{t-\ell} \in \mathbf{z}_t
$$}
}

\noindent With assumptions \ref{assump:idio_rationality}, \ref{assump:indiv_exp}, \ref{assump:serially_uncorr}, \ref{assump:exog}, \ref{assump:shape_restr}, \ref{assump:meas_error_exog}
and given data $\{\mathbf{z}_t, Y_t, \{E^{\text{data}}_t[\mathbf{W}_{t+h}]\}_{h \leq H}\}$, we obtain our desired moment conditions

\begin{equation}\label{eq:gen_fake_news_moment_cond_edata}
    \mathbb{E}\left[\left(Y_t - \sum_{\tau = t-\ell}^t \sum_{h \geq 0} \sum_{W \in \mathbf{W}} \mathcal{F}^W_{t-\tau, h}(\boldsymbol{\theta}_0) E^{\normalfont \text{data}}_\tau[W_{\tau+h}; \boldsymbol{\vartheta}_0]\right) z_{t-\ell} \right] = 0, \,\, \text{ for } z_{t-\ell} \in \mathbf{z}_t
\end{equation}
\vspace{0.25cm}

Equation (\ref{eq:gen_fake_news_moment_cond_edata}) is a collection of unconditional moment 
conditions, which can be estimated with two-step generalized method of moments (\citealt{newey1994large}). 
The first step is the estimation of the nuisance parameter $\boldsymbol{\vartheta}$ used to 
extrapolate missing data using condition (\ref{eq:shape_restr}). 
Following that one can evaluate the moment condition to estimate the structural parameters of 
interest $\boldsymbol{\theta}$.
Applying Equation (\ref{eq:gen_fake_news_moment_cond_edata}) to the case of aggregate consumption, 
I construct the following set of moment conditions, which encompass both representative- and 
heterogeneous-agent models, that I estimate in the following section. 
\begin{equation}\label{eq:gen_fake_news_moment_cond_edata_cons}
    \mathbb{E}\left[\left( C_t - \sum_{\tau = t - \ell}^t \sum_{h \geq 0} \mathcal{F}^Y_{t-\tau, h}(\boldsymbol{\theta}_0) E_\tau[Y_{\tau+h}; \boldsymbol{\vartheta}_0] + \mathcal{F}^r_{t - \tau, h}(\boldsymbol{\theta}_0) E_\tau[r_{\tau+h}; \boldsymbol{\vartheta}_0] \right) z_{t-\ell} \right] = 0, \,\, \text{ for } z_{t-\ell} \in \mathbf{z}_t
\end{equation}

There is typically sparse availability of expectations data with a large set of horizons $H$. 
Therefore, choosing an auxiliary model with a large number of parameters $\boldsymbol{\vartheta}$ 
may overfit the noise in expectations data.
I err on the side of caution by testing robustness of $\boldsymbol{\theta}$ estimates 
against multiple auxiliary models that are sparsely parameterized.
The choice of these models is informed by the impulse response 
interpretation of moment conditions (\ref{eq:shape_restr}), (\ref{eq:gen_fake_news_moment_cond_edata}). 
Relying on the typical assumption that impulse response functions are smooth (\citealt{barnichon2019impulse})
provides some additional justification for focusing on low-dimensional, smooth auxiliary models for extrapolation.

\section{Semi-structural estimation of model-implied impulse responses}\label{sec:estimation}

The aim of this section is to estimate and evaluate the fit of a standard set of 
consumption-savings models against consumption data 
using moment conditions described in Equation (\ref{eq:gen_fake_news_moment_cond_edata_cons}), 
substituting assumptions on a model of expectation formation with expectations data.


\subsection{Consumption functions from structural consumption-savings models}\label{subsec:cons_funcs}
The structural models of consumption-savings I consider are the standard representative-agent 
model and two heterogeneous-agent models.

The first heterogeneous-agent model I consider is the perpetual youth, overlapping 
generations model because of its analytical tractability. 
I use the variation of the model developed in \cite{angeletos2023can}, 
which builds on the original \cite{yaari1965uncertain}, \cite{blanchard1985olg}.
The consumption function from the model solution linearized around steady state $\beta (1 + r) = 1$ is
\begin{equation}\label{eq:olg_cons_func}
    C_t = (1 - \beta \omega) \left(\sum_{h=0}^\infty (\beta \omega)^h E_t[Y_{t+h}] + W_t \right) - \gamma \sum_{h=0}^\infty (\beta \omega)^h E_t[r_{t+h}]
\end{equation}
where aggregate wealth is given by $W_t$ and the net 
interest elasticity $\gamma := \sigma \beta \omega - (1 - \beta \omega) \beta A$.

This model is attractive because it closely mirrors its representative-agent counterpart, 
albeit with an additional degree of freedom, $\omega \in [0, 1]$, the perpetual youth hazard rate. 
When $\omega < 1$ the overlapping generations model exhibits greater 
income sensitivity of consumption as measured by the current 
marginal propensity to consume (MPC) out of unearned income. 
As in the representative-agent model, the discount factor $\beta$ is pinned down 
by the steady state real interest rate $r$ and is therefore not a degree of freedom for estimation. 

The second heterogeneous-agent model I consider is the standard incomplete markets 
model of \cite{bewley1986stationary}, \cite{imrohorouglu1989cost}, \cite{huggett1993risk}, 
\cite{aiyagari1994uninsured}. 
In this model, a unit mass of households face idiosyncratic income risk, borrowing constraints, and 
incomplete markets in the form of a one-period risk-free asset.
The individual-$i$ household problem is
\begin{align*}
    V(e_{i, t}, a_{i, t-1}) &= \max_{c_{i, t}, a_{i, t}} \frac{c_{i, t}^{1 - \sigma^{-1}}}{1 - \sigma^{-1}} + \beta E_{i, t}[V(e_{i, t+1}, a_{i, t}) | e_{i, t}] \\
    c_{i, t} + a_{i, t} &= e_{i, t} Y_t + (1 + r_{t-1}) a_{i, t-1} \\
    a_{i, t} &\geq 0
\end{align*}
where $E_{i,t}$ denotes the time-$t$ subjective expectation of household $i$. 
Idiosyncratic productivity $e_{i, t}$ is a stationary, commonly-known Markov process 
with persistence $\rho_e$ and variance $\sigma^2_e$. 
The process has a fixed transition matrix $\Pi(e, e^\prime)$ with an associated 
stationary distribution $\pi(e)$ and a stationary mean normalized to one, 
i.e. $\sum_{e} \pi(e) e = 1$.

As mentioned in the previous section, all models can be represented by a linearized aggregate 
consumption function of the functional form
\begin{equation}\label{eq:ha_cons_func}
    C_t = \sum_{\tau \leq t} \sum_{h \geq 0} \mathcal{F}^Y_{t-\tau, h} E_\tau[Y_{\tau+h}] + \mathcal{F}^r_{t - \tau, h} E_\tau[r_{\tau+h}]
\end{equation}
where the coefficients $\mathcal{F}^Y, \mathcal{F}^r$ vary across each model.


\begin{table}[t]
    \centering
    \begin{tabular}{ccc} 
        \toprule
        Calibrated parameters & Description & Value \\ 
        \midrule
        \midrule
        $r$ & Real interest rate & 0.005 \\ 
        $\beta$ & Discount factor & See note below \\
        $\rho_e$ & Idiosyncratic productivity persistence  & 0.966 \\
        $\sigma_e^2$ & Idiosyncratic productivity variance & 0.504 \\
        \midrule
        \midrule
        Estimated parameters & & \\ 
        $\sigma$ & Elasticity of intertemporal substitution (EIS) & \\ 
        $\omega$ & Perpetual youth hazard rate & \\
        \bottomrule

    \end{tabular}
    \caption{Model parameters}
    \label{tab:parameters}
    \vspace{0.25cm}
    {
    \footnotesize \emph{Note}: The real interest rate and disposable income values 
    are calibrated to a quarterly frequency.\\ 
    The discount factor $\beta$ is calibrated to $(1 + r)^{-1}$ in the representative-agent 
    and perpetual youth, overlapping generations model and to hit the asset-to-disposable 
    income ratio of 1.4 in the standard incomplete markets model.
    The hazard rate $\omega$ is only available for estimation in the perpetual youth overlapping 
    generations model.
    }
\end{table}

All linearized consumption functions are local approximations 
about a steady state, which I calibrate following \cite{mckay2016power}.
The discount factor $\beta$ in the standard incomplete markets model must be calibrated 
to hit the asset-to-disposable income calibration target of 1.4.
This leaves at most two parameters available for estimation, which are listed in the 
lower panel of Table \ref{tab:parameters}: the elasticity of intertemporal 
substitution and the hazard rate in the perpetual youth model.

\subsection{Data}
To evaluate the consumption functions of all models requires data on 
realized and expected real disposable income and interest rates.

\paragraph{Realizations data}
The income measure I use is real disposable personal income (DSPIC96), 
sourced from the Bureau of Economic Analysis. 
The interest rate measure I use is the nominal federal funds rate (DFF), 
deflated by one-period ahead realized consumer price index inflation (CPIAUCSL), 
sourced from the Bureau of Labor Statistics.

In typical linearized macroeconomic models without financial frictions, there is a single 
interest rate due to no-arbitrage conditions on asset choice.
In reality, households face different interest rates for saving or borrowing products, 
which makes the choice of a single interest rate data series non-obvious.
Given our consumption functions are local approximations around a steady state 
where all households hold weakly positive wealth, a savings rate is the 
best analog to the model interest rate.
Due to this, the federal funds rate is a reasonable proxy for the model-relevant 
savings rate, given savings rates move closely with the federal funds rate.

\paragraph{Expectations data} I use consensus expectations reported by the Bluechip Economic Indicators 
and Financial Forecasts for real disposable personal income, the nominal federal funds rate, 
and CPI inflation. 
As mentioned earlier, because I am estimating household consumption functions, 
a household-level survey like the Michigan Survey of Consumers or 
the Survey of Consumer Expectations conducted by the Federal Reserve Bank of New York 
would be most ideal. 
However, given neither of these sources nor other commonly-used household surveys 
elicit point forecasts (expectations) of interest rates, I would have to make 
auxiliary assumptions to map qualitative responses about interest rates in these 
surveys to point forecasts. 
In addition, household surveys tend to report one short-horizon forecast,
typically one-year ahead, and one longer five-to-ten year horizon forecast. 
Because of the need to extrapolate horizons it is also useful to have a more complete 
term structure of near-term expectations from the Bluechip.
For consistency across available data periods, I use Bluechip expectations 
for one through four-quarters ahead for each forecasted variable.

\paragraph{Instrumental variables}
There are other substantive reasons that may alleviate some concerns of 
using forecaster as opposed to household survey expectations.
\cite{rozsypal2023overpersistence} analyze household income expectations from the Michigan 
survey and find evidence of over-persistence bias, 
where households extrapolate expectations of future income from realized current income. 
This is precisely the form of bias I document empirically in upcoming results on 
forecaster expectations of real disposable income. 
In recent work, \cite{desilva2024selective} document that household interest rate expectations 
tend to be close to forecaster expectations during periods where they make durables purchases.

Importantly, household and forecaster expectations have been documented to exhibit 
some systematic differences.
\cite{candia2020communication} and \cite{kamdar2023supplyside} find that household 
expectations overweight ``supply-side'' narratives, which emphasize the 
negative co-movement of real variables like real output and inflation, 
and underweight ``demand-side'' narratives. 
\cite{andre2022subjective} document the mental models households 
use to understand and form expectations of the economic effects of supply shocks, 
such as sudden changes in oil prices, are similar to those of forecasters 
but differ materially for monetary and fiscal policy shocks.

Using shocks which are interpreted in a systematically different way by households and 
forecasters would amount to a violation of measurement error exogeneity, as stated in 
Assumption \ref{assump:meas_error_exog}.
Using a supply shock to instrument forecaster expectations 
is the best way to address this concern given forecaster 
and household expectations exhibit qualitatively similar co-movements in response 
to these shocks.
Therefore, I estimate model-implied impulse responses with respect to an 
oil supply news shock from \cite{kanzig2021macroeconomic}.
Identified using a high-frequency identification approach, this shock captures variation in 
oil futures prices around a narrow time window of OPEC production announcements. 

\subsection{Empirical impulse response estimation}
To estimate impulse responses of macroeconomic variables and their forecasts, I adopt the proxy structural vector autoregression (VAR) approach
(\citealt{stock2012disentangling}, \citealt{mertens2013dynamic})
and follow the empirical setup and notation from \cite{kanzig2021macroeconomic} closely. 
I first estimate a reduced-form VAR, with a constant and a deterministic linear trend
$$
    \mathbf{Y}_t = \boldsymbol{\alpha} + \boldsymbol{\delta} t + \sum_{l=1}^p \boldsymbol{\beta}_l \mathbf{Y}_{t-l} + \mathbf{u}_t
$$
where $\boldsymbol{\alpha}, \boldsymbol{\delta} t, \{\boldsymbol{Y}_{t-l}\}_{l=0}^p, \boldsymbol{u}_t$ are vectors of length $n$ and $\boldsymbol{\beta}_l$ is a matrix of dimension $n \times n$.

I assume invertibility, in that the reduced-form residuals $\mathbf{u}_t$ are a linear combination of i.i.d structural shocks $\boldsymbol{\varepsilon}_t$
$$
\boldsymbol{u}_t = \mathbf{S} \boldsymbol{\varepsilon}_t
$$
where $\mathbb{E}[\boldsymbol{\varepsilon}_t] = \mathbf{0}$ and $\mathbb{E}[\boldsymbol{\varepsilon}_t \boldsymbol{\varepsilon}_t^\prime] = \boldsymbol{\Omega}$, a positive, diagonal matrix.

Assuming an instrument $z_t$ satisfies the standard identifying assumptions
\begin{align*}
    \mathbb{E}[\varepsilon_{1, t} z_t] &= \alpha \neq 0 \\
    \mathbb{E}[\boldsymbol{\varepsilon}_{2:n, t} z_t] &= \mathbf{0}
\end{align*}
where the structural shock we are identifying is ordered first in the VAR, without loss of generality.
I can identify the first column of $\mathbf{S}$ up to sign and scale, which I denote $\mathbf{s}_1$, given by
$$
\mathbf{s}_1 = \mathbb{E}[\mathbf{u}_t z_t]
$$
Finally, to pin down the sign and scale factor $s_{1, 1} := \frac{\mathbb{E}[u_{1, t} z_t]}{x}$ for the econometrician's desired value $x$, 
I can normalize the impact effect of the identified shock on variable $y_{1, t} = x$, using the re-scaled structural impact vector $\tilde{\boldsymbol{s}}_1 = \boldsymbol{s}_1 / s_{1, 1}$ provided $\mathbb{E}[u_{1, t} z_t] \neq 0$. 

\paragraph{Specification} 
The variables included in the baseline specification are real gross domestic product, 
real disposable income, the consumer price index (CPI), 
the nominal federal funds rate, real oil price and world oil production measures. 
The real oil price is the WTI crude oil price deflated by CPI inflation 
and will be the proxy SVAR's first-stage variable, scaled such that the on-impact 
effect of a positive oil supply news shock increases real oil prices by 10 percentage points.
I then augment the baseline specification with real personal consumption expenditures 
and Bluechip expectations at each horizon-$h$ period ahead 
one variable at a time. 
The data are measured at a quarterly frequency and in log-levels, aside from 
the federal funds rate. 
The time period spans 1985-Q1 through 2017-Q3, due to availability of 
the Bluechip data.

\subsection{Impulse responses of income and interest rates to an oil shock}\label{subsec:irf_income_rates}
\begin{figure}[t]
    {
    \centering

    \begin{subfigure}{0.495\textwidth}
        \centering
        \includegraphics[width=\linewidth]{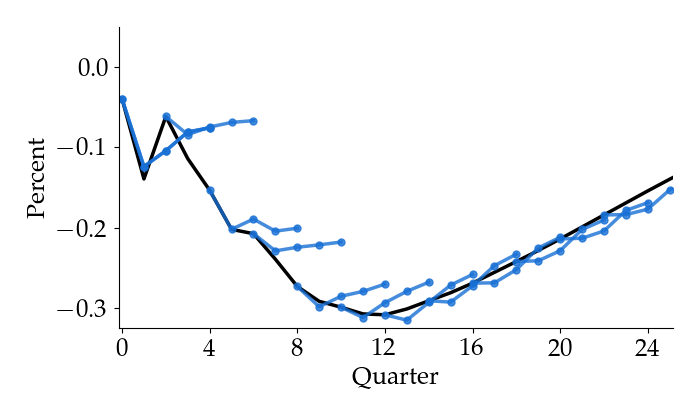}
        \caption{Real Disposable Income}
    \end{subfigure}
    \hfill
    \begin{subfigure}{0.495\textwidth}
        \centering
        \includegraphics[width=\linewidth]{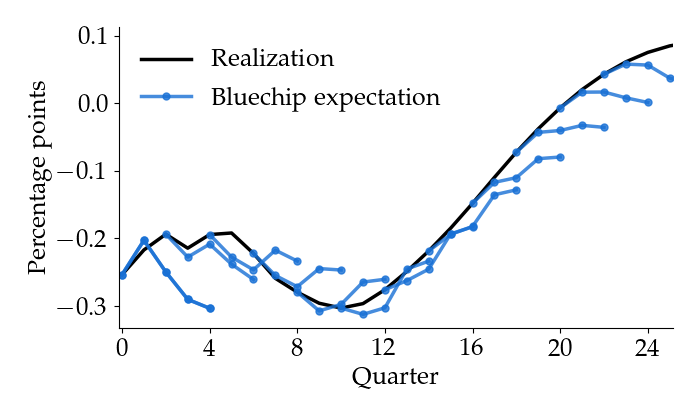}
        \caption{Real Federal Funds Rate}
    \end{subfigure}



    \caption{Income and interest rate responses to a \cite{kanzig2021macroeconomic} oil shock}
    \label{fig:exp_irfs_income_rates_cpi}
    }

    \caption*{
    \footnotesize \emph{Note}: each panel contains an impulse response function of realizations (black) 
    to a positive \cite{kanzig2021macroeconomic} oil price news shock, which raises real oil prices 
    by ten percent on impact, and the impulse responses of the one-quarter through one-year ahead 
    expectations (blue dots), connected across horizons.
    The real federal funds rate is the nominal federal funds rate deflated by consumer price inflation.
    }
\end{figure}

Figure \ref{fig:exp_irfs_income_rates_cpi} displays impulse responses of 
realizations and Bluechip expectations of real income and real interest rates, the two 
main input variables into consumption-savings decisions in standard models. 
In response to an inflationary oil shock, realized 
real income and real interest rates (black lines) exhibit a prolonged decline.
Counter to results using a time sample with an earlier start date 
(\citealt{bernanke1997systematic}, \citealt{gagliardone2023oil}), 
real interest rates decline as a result of the oil supply news shock due to a 
largely accommodative nominal federal funds rate response.


I consider a battery of simple parametric models fit to the observed term structure of expectations 
to extrapolate missing horizons. 
As a baseline, I use an estimated AR(2) process, constrained to be stationary, to 
extrapolate missing horizons.
The results in the following section on consumption function estimation are robust to alternate choices. 
Details for the choice of auxiliary models for extrapolation, their estimation, and resulting 
structural parameter estimates are in Appendix \ref{asubsec:extrap}.

    


\subsection{Empirical vs. model-implied impulse responses of consumption}
\begin{figure}[htbp]
    {
    \centering
    \includegraphics[width=0.75\linewidth]{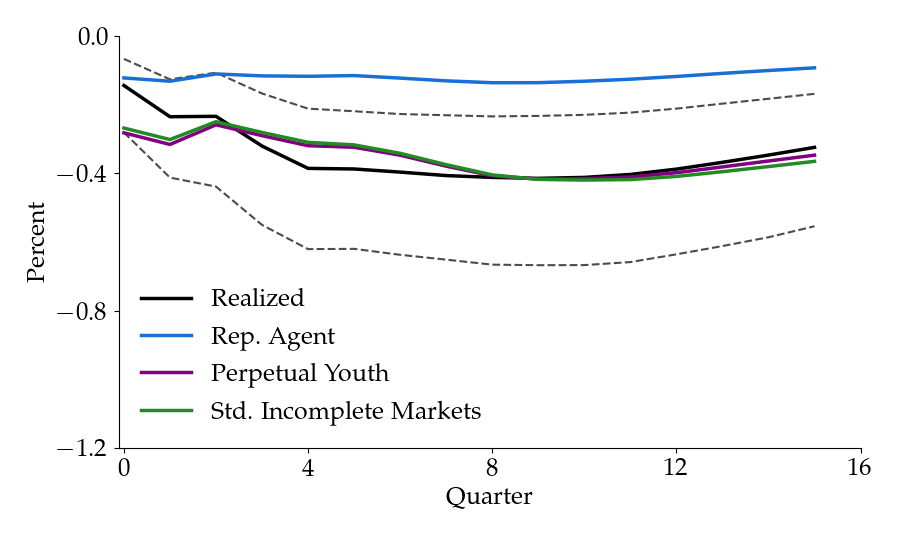}
    \caption{Estimated model-implied consumption impulse responses to a \cite{kanzig2021macroeconomic} oil shock}
    \label{fig:cons_irfs}
    }

    \caption*{\footnotesize 
    \emph{Note}: the realized (black) consumption impulse response 
    and model-implied consumption impulse responses to a positive 
    \cite{kanzig2021macroeconomic} oil price news shock. 
    Model-implied responses are produced by evaluating each models 
    consumption function using empirical impulse responses of realized and 
    expected income and interest rates. 
    The dashed lines are 68\% confidence bands produced using moving block bootstrap by
    \cite{jentsch2019dynamic}.
    }
\end{figure}

The expectation impulse responses plotted in Figure \ref{fig:exp_irfs_income_rates_cpi} 
correspond to the impulse response estimands reported in the model-implied consumption moment 
condition (\ref{eq:gen_fake_news_moment_cond_edata}). 
Figure \ref{fig:cons_irfs} displays the model-implied impulse responses of consumption from 
the estimated representative-agent, perpetual youth overlapping generations, and 
standard incomplete markets models. 

My baseline estimates for the elasticity of intertemporal substitution (EIS) across models 
are low. For the representative-agent benchmark the estimated EIS approaches zero, while for 
the heterogeneous-agent models the estimated EIS is around 0.1. 
The main intuition stems from observing in Figure \ref{fig:exp_irfs_income_rates_cpi}b that 
realized and expected real interest rates decline in response to the oil price shock. 
The estimated EIS is pushed downward to mitigate the positive intertemporal substitution response 
of model-implied consumption and to amplify the negative income effects from lower rates.

While an EIS estimate of 0.1 is low it is not unprecedented. 
In a quasi-experimental setting, \cite{best2020estimating} exploit 
borrower bunching behavior around loan-to-value thresholds used to price mortgages 
and also find an estimated EIS of 0.1. 
Likewise, \cite{ring2024wealth} finds evidence for a similarly low EIS 
using Norwegian administrative data and geographic variation to investigate 
the relative size of substitution and income effects of wealth taxation on savings behavior.

One concern might be that forecasters' inflation expectations respond differently than 
household expectations due to this shock. 
Figure 7 of \cite{kanzig2021macroeconomic} provides suggestive evidence that the 
magnitude of households' inflation expectation responses to oil shocks may be 
larger and more positive than that of forecasters'. 
However, this would imply an even lower expected real rate taking account of household 
expectations reinforcing the need for a low estimated EIS.





The income sensitivity of consumption, as measured by the current marginal 
propensity to consume (MPC) out of unearned income, is small in the 
representative-agent model by construction. 
In contrast, heterogeneous-agent models can have substantially higher MPCs, 
and indeed I find this to be the case in this estimation.
The estimated hazard rate for the perpetual youth model implies an MPC 
of approximately five percent at a quarterly frequency. 
While the standard incomplete markets model did not have an independent degree of freedom 
from the EIS to estimate, due to the discount factor being used to target the steady-state level of assets, 
its MPC nonetheless matches that of the perpetual youth model at five percent.
As Figure \ref{fig:cons_irfs} demonstrates, the higher MPC in the heterogeneous-agent 
models proves crucial to match the pronounced consumption contraction due to the 
oil shock. 
Due to the much longer effective horizons for income smoothing, the representative-agent 
models' response to the shock is less severe.

\begin{table}[t]
    \centering
    \begin{tabular}{cccc} 
        & \multicolumn{3}{c}{\textbf{Consumption-Savings Models}} \\
        \toprule
        Parameter & Perpetual Youth & Standard Incomplete Markets & Rep. agent \\ 
        \midrule
        \midrule
        EIS &  0.08  &   0.09  & 0.00 \\ 
        MPC &  0.04  &  0.05  & 0.005 \\ 
        $\frac{\text{Assets}}{\text{Income}}$ & 1.4 & $1.4^*$ & 1.4 \\
        \midrule
        \midrule
        EIS & 0.80 & 0.05 & - \\ 
        MPC & $0.2^*$ & $0.2^*$ & - \\ 
        $\frac{\text{Assets}}{\text{Income}}$ & 1.4 & 0.425 & - \\
        \bottomrule
    \end{tabular}
    \caption{Estimated/targeted parameters from consumption-savings models}
    \label{tab:mod_moments}

    \caption*{\footnotesize 
    \emph{Note}: 
    The top panel contains estimated parameters enforcing that the 
    steady state assets-to-income ratio is equal to the initial calibration target. 
    The bottom panel contains estimated parameters when models instead target 
    a higher marginal propensity to consume (MPC).
    The EIS $\sigma$ is the elasticity of intertemporal substitution. 
    MPC and income are reported at a quarterly frequency.
    }
\end{table}


\paragraph{MPCs in micro-calibration versus macro-estimation}

It is well-known that the standard incomplete markets model is unable to simultaneously match 
typical microeconomic estimates of the current MPC and the 
steady state level of household assets (\citealt{kaplan2022marginal}). 
By restricting the estimated model to match the latter, I attain an 
implied MPC of around 0.05 at a quarterly frequency, which is lower than 
typical microeconomic estimates which range from 0.15 to 0.25. 
However, this MPC that is consistent with our targeted macroeconomic 
impulse responses and has been shown to be consistent with a broader range of macroeconomic moments 
in full-information HANK estimation on macroeconomic time series \cite{bayer2024shocks}.

\begin{figure}[htbp]
    {
    \centering
    \includegraphics[width=0.75\linewidth]{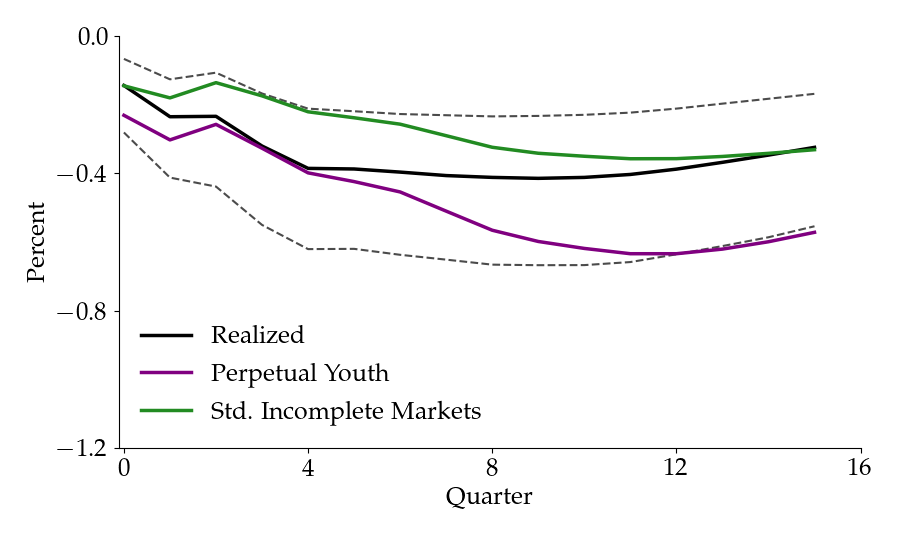}
    \caption{Calibrated model-implied consumption impulse responses to a \cite{kanzig2021macroeconomic} oil shock}
    \label{fig:cons_irfs_match_mpcs}
    }

    \caption*{\footnotesize 
    \emph{Note}: the realized (black) consumption impulse response 
    and model-implied consumption impulse responses to a positive 
    \cite{kanzig2021macroeconomic} oil price news shock. 
    Model-implied responses are produced by evaluating each models 
    consumption function using empirical impulse responses of realized and 
    expected income and interest rates. 
    The dashed lines are 68\% confidence bands produced using moving block bootstrap by
    \cite{jentsch2019dynamic}.
    }
\end{figure}

I consider how calibrating the MPC in both the perpetual youth and standard incomplete markets models 
to 0.2, in line with microeconomic evidence, affects their implied impulse response fit. 
In the standard incomplete markets model I calibrate the discount factor $\beta$ 
to now match the MPC target.
While the fit deteriorates, as shown in Figure \ref{fig:cons_irfs_match_mpcs}, 
they both still remain within a one standard deviation bound of the 
empirical impulse response of consumption. 
However, the estimated parameters for the EIS now diverge between these models. 
The estimated EIS is an order of magnitude larger in the perpetual youth model, 
while in the standard incomplete markets model it is slightly lower. 
In addition, the perpetual youth model now overshoots the empirical response, whereas 
the standard incomplete markets model undershoots.

To explain the reason for this change, consider an important difference 
between these two models: given the linearization in aggregates, 
the perpetual youth model lacks a precautionary savings motive. 
Recall in the perpetual youth model that the steady state level of assets 
is independent of the MPC.
Therefore, changing the EIS only scales the relative size of the substitution 
versus income effects in response to the discounted value of expected interest rate changes, 
as shown in Equation (\ref{eq:olg_cons_func}). 
Given the large, prolonged decline in realized and expected real interest rates in 
response to the shock shown in Figure \ref{fig:exp_irfs_income_rates_cpi}, 
higher MPCs in the perpetual youth model at the original, 
lower EIS estimate would have excessively amplified the negative income effect from lower rates.

The standard incomplete markets model requires a lower discount factor to attain a high 
MPC, which in turn reduces steady state asset demand because agents are less patient. 
Whether the magnitude of interest rate income effects increases due to the higher MPC 
or decreases due to the lower stock of steady state assets is a quantitative question. 
In this case, the lower discount factor reduces the magnitude of the 
negative interest rate income effect, requiring an even lower EIS.
Because I directly use the canonical standard incomplete markets model, I 
cannot resolve the fundamental tension between these parameter calibrations.
Nonetheless, I show Figures \ref{fig:cons_irfs} and \ref{fig:cons_irfs_match_mpcs} 
that conditioning directly on expectations data, 
the models are similarly able to rationalize the observed inertia in aggregate consumption.


\paragraph{Full-information, rational expectations comparison}

It is natural to consider how model-implied consumption responses may differ in
comparing those using expectations data with those formed 
via full-information, rational expectations (FIRE). 
However, without the complete specification of an equilibrium model we are not 
able to consider this counterfactual because of the Lucas critique. 
If the true data generating process for observed real income and interest rates 
is an economy where agents' expectations are biased, as demonstrated by the Bluechip expectations, 
then one cannot answer this counterfactual by simply setting the impulse response 
of expectations equal to the realized response of income and interest rates, i.e. the 
``rational expectation''.

\section{An equilibrium model with extrapolation in expectations and inertia in realizations}\label{sec:macro_model_inertia}

To better understand the determinants of the documented biases in expectation formation 
and how they result in inertia in aggregate consumption, I proceed to analyze a specific 
model of expectation formation embedded in a general equilibrium environment.

\subsection{Temporary equilibrium definition}\label{sec:temp_eqm}
I begin by defining a temporary equilibrium, an intermediate step toward 
a fully-specified general equilibrium that does not yet place restrictions 
on how forward-looking agents form expectations. 
One can consider the earlier mentioned consumption functions from 
Section \ref{subsec:cons_funcs} as the aggregate demand block of 
the temporary equilibrium environment, given we will not consider investment for 
the sake of simplicity.
As in \cite{woodford2013macroeconomic}, I immediately resort to using 
the linearized equilibrium\footnote{The consumption functions written earlier are in level as opposed to log deviations. To maintain this notation, I normalize steady state output $Y = 1$ such that level and log deviations for the below-defined variables can be interpreted identically.}, 
whose deviations are given by time-indexed variables, e.g. $C_t$, 
around a non-stochastic steady state, whose notation is given by 
non-time-indexed variables, e.g. $C$.
The rest of the equilibrium environment closely follows \cite{angeletos2023can}, 
although I simplify along a few dimensions that are not central to my analysis. 
I only briefly discuss the shared equilibrium ingredients, such as the firm problem, policy rules, and 
market clearing conditions and elaborate only on my points of departure. 

\paragraph{Households and firms} 
The household sector is exactly the same as the \cite{angeletos2023can} specification 
of the perpetual youth overlapping generations model.
Labor unions intermediate labor markets, ensuring households supply an identical quantity 
of labor and equalizing the real wage and the average marginal rate of substitution 
between aggregate consumption and labor supply. 
Households therefore receive the identical labor income. 
Firm production follows the textbook 
New Keynesian model (\citealt{gali2015monetary}), where identical monopolistically-competitive 
firms operate a linear-in-labor production technology and face Calvo price-setting frictions. 
This gives rise to an aggregate price inflation New Keynesian Phillips curve 
linearized around a zero-inflation steady state.

\begin{equation}\label{eq:price_nkpc}
    \pi_t = \kappa Y_t + \beta E_t[\pi_{t+1}]
\end{equation}

Firms distribute dividends evenly, ensuring all households receive the same profit income.

\paragraph{Monetary policy and market clearing} 
The real interest rate is determined by the real Taylor rule
\begin{equation}\label{eq:real_taylor_rule}
    i_t - E_t[\pi_{t+1}] \equiv r_t = \phi Y_t
\end{equation}
The monetary authority sets nominal interest rates accounting for the equilibrium consequences 
on subjective inflation expectations $E_t[\pi_{t+1}]$ to achieve a real interest rate target of $\phi Y_t$. 
A rule of this form allows the monetary authority to conduct policy as if it 
maintained direct control of the ex-ante real interest rate. 
Adopting this rule therefore allows us to focus on the 
equilibrium determination of household consumption as a function of real interest rates without 
needing to also analyze the dynamics of subjective inflation expectations. 

Market clearing in the goods market is given by
\begin{align}
    C_t &= Y_t \label{eq:goods_mkt}
\end{align}

{
\definition{A (linearized) \textbf{temporary equilibrium} consists of sequences of prices $\{i_t, \pi_t\}$ 
and quantities $C_t, Y_t$ that satisfy (\ref{eq:olg_cons_func}), (\ref{eq:price_nkpc}), (\ref{eq:real_taylor_rule}), (\ref{eq:goods_mkt}) 
for all periods $t$, given subjective expectations 
of $\{E_t[Y_{t+h}], E_t[\pi_{t+h}], E_t[i_{t+h}]\}_{h > 0}$.
}
}
\vspace{0.25cm}

Assets are in zero net supply, requiring households to hold zero net wealth in equilibrium.
Given output, or equivalently aggregate income, is equal to consumption in equilibrium, 
moving forward I will simply refer to output as the real quantity of interest.

% ...

\subsection{General equilibrium with learning about unobserved components}
Closing the temporary equilibrium defined in Section \ref{sec:temp_eqm}, 
I assume households form expectations with a particular class of learning models.
These models have been shown to be consistent with multiple dimensions of 
empirical evidence on expectation formation, including in the 
cross-section of households and forecasters (\citealt{nagel2024lean}, \citealt{chen2025expectation}), 
experiments (\citealt{afrouzi2023overreaction}),
and unconditional time-series (\citealt{crump2023term}, \citealt{farmer2024learning}). 

In these models, agents observe aggregate variables such as output, 
perceiving their dynamics to be driven by the sum of 
a persistent and a transitory component\footnote{Some of the cited literature adopt 
the convention that the persistent component is a non-stochastic, long-run mean parameter, 
which agents are nonetheless uncertain about. The inference problem is similar 
to the case I study here and results in similar forms of extrapolation 
that lie at the core of my analysis.}, also referred to as ``trend'' 
and ''cycle'' components.
Because they cannot immediately distinguish these components, they gradually 
update their beliefs about each component by solving a filtering problem given 
the history of observations.

For example, consider an unobserved components model, where households 
form beliefs about realized output $Y_t$ via their perceived law of motion for output $\tilde{Y}_t$, 
which is given by
\begin{equation}\label{eq:unobs_comp_meas}
    \tilde{Y}_t = \tilde{\lambda}_t + \tilde{\eta}_t
\end{equation}
where $\tilde{\lambda}_t, \tilde{\eta}_t$ are two independent AR(1) processes, 
parameterized by persistence parameters $\tilde{\rho}_\lambda, \tilde{\rho}_\eta$ and 
standard deviations $\tilde{\sigma}_\lambda, \tilde{\sigma}_\eta$ with Gaussian innovations.
Given observed output $Y_t$, households update their mean estimates of each component 
by the standard Kalman update equation
\begin{equation}\label{eq:unobs_kal_update}
    \begin{bmatrix}
        E_t[\lambda_{t+1}] \\
        E_t[\eta_{t+1}]
    \end{bmatrix}
    = 
    \mathbf{F}
    \begin{bmatrix}
        E_{t-1}[\lambda_t] \\
        E_{t-1}[\eta_t]
    \end{bmatrix}
    + 
    \mathbf{g} (Y_t - E_{t-1}[Y_t])
\end{equation}
where $\mathbf{F}$ collects $\tilde{\rho}_\lambda, \tilde{\rho}_\eta$ into a diagonal 
matrix, $\mathbf{g}$ is the stationary Kalman gain, and
$E_{t-1}[Y_t] = E_{t-1}[\lambda_t] + E_{t-1}[\eta_t]$ is the expectation induced by 
the perceived law of motion in Equation (\ref{eq:unobs_comp_meas}).

However, instead of treating $Y_t$ as exogenous and 
given by a simple functional form like an AR(1), as is commonly done 
in the forecasting literature on unobserved components models, I close 
the previously defined temporary equilibrium with this unobserved components model of 
expectation formation. 
Therefore, the dynamics of $Y_t$ are endogenous to households' component beliefs 
$E_t[\lambda_{t+1}], E_t[\eta_{t+1}]$ because of the equilibrium feedback of these beliefs 
into output, inflation, and interest rates via households' decisions.

To simplify the learning representation and clarify the key model mechanism that 
results in both extrapolation in expectations and inertia in realizations 
I make two further assumptions.

I assume that the policy rule in 
Equation (\ref{eq:real_taylor_rule}) is common knowledge and therefore that the 
perceived laws of motion for inflation and rates, $\tilde{\pi}_t, \tilde{i}_t$, 
are consistent with the rule. 
This allows us to isolate the determination of equilibrium output 
from inflation and nominal interest rates. 
Additionally, I assume that the only exogenous shock that directly alters consumption 
and consequently output $Y_t$ is a demand shock $\varepsilon_t$, 
which is itself the sum of two AR(1) components
$$
\varepsilon_t = \lambda_t + \eta_t
$$
which is itself the sum of two AR(1) components with persistence parameters 
$\rho_\lambda > \rho_\eta$ and mean-zero Gaussian i.i.d innovations 
with variances $\sigma_\lambda^2, \sigma_\eta^2$. 

Consequently, because the perceived and realized real rate are the only ways that 
inflation and the nominal rate affect output and because these real rates are 
determined only by perceived and realized output, inflation and the nominal rate 
provide no additional information about the components $\lambda_t, \eta_t$. 
Hence, we can treat realized output $Y_t$ as the only observable agents learn from 
to infer the unobserved components, $\tilde{\lambda}_t, \tilde{\eta}_t$ in the example above, 
via the perceived law of motion $\tilde{Y}_t$.

I also make the typical learning assumption that the information set for 
time-$t$ decisions determining equilibrium $Y_t$ is the history of past 
realizations $\{Y_{t-\ell}\}_{\ell \geq 1}$. 
To reflect this staggered timing, subjective expectations that inform time-$t$ decisions 
are labeled $E_{t-1}$. 
I now consider a generic equilibrium definition, which nests certain special cases that I will study further.
{
\definition{A \textbf{learning equilibrium} is a temporary equilibrium and 
a collection of subjective expectations $\{E_{t-1}[Y_{t+h}]\}_{h > 0}$, 
which are induced by a perceived law of motion $\tilde{Y}_t$, itself a function 
of a vector of latent states, and the history $\{Y_{t-\ell}\}_{\ell > 1}$.
}
}
\vspace{0.25cm}

This equilibrium definition is generic in that I have not imposed any particular consistency 
criterion between the perceived law of motion $\tilde{Y}_t$ that generates subjective expectations 
and the resulting equilibrium process for realized $Y_t$. 
In the following sections I will consider different perceived laws of motion, 
which range in the consistency criterion, and study their implications for whether 
inertia in $Y_t$ results as an equilibrium outcome.





Consolidating equilibrium conditions into a single equation determining real output
\begin{equation*}
    Y_t \propto (1 - \beta \omega - \beta \omega \sigma \phi) \sum_{h=1}^\infty (\beta \omega)^h E_{t-1}[Y_{t+h}] + \varepsilon_t
\end{equation*}

Note that $Y_t$ is not exactly equal to the right-hand side because of 
the within-period general equilibrium feedback of $Y_t, r_t$. 
The constant of proportionality that I have omitted, $\beta \omega (1 + \sigma \phi)$, affects the 
overall level of $Y_t$ but not the shape of its impulse response to $\varepsilon_t$ across periods. 
In the following sub-sections I focus on characterizing the shape and not the overall level 
of impulse responses.
Therefore, without loss of generality I normalize the variance of $\varepsilon_t$ 
and proceed denoting Equation (\ref{eq:olg_agg_demand_prop}) with equality for convenience.
When considering policy counterfactuals in Section \ref{sec:policy} I undo this normalization to ensure 
the level contribution of counterfactuals is properly accounted for.

Let $\chi := (1 - \beta \omega - \beta \omega \sigma \phi)$ 
and the subjective expectation of permanent income $\mathcal{Y}_t := \sum_{h=1}^\infty (\beta\omega)^h E_{t-1}[Y_{t+h}]$. 
Re-writing the aggregate demand equation we obtain the following compact expression for realized output.
\begin{equation}\label{eq:olg_agg_demand_prop}
    Y_t = \chi \mathcal{Y}_t + \varepsilon_t
\end{equation}

In the simple unobserved components model described above, the full set of equilibrium 
conditions determining realized $Y_t$ can be summarized by Equations 
(\ref{eq:unobs_comp_meas}), (\ref{eq:unobs_kal_update}), (\ref{eq:olg_agg_demand_prop}).
We can see that if $\varepsilon_t$ does not exhibit inertia in the form of a hump-shaped 
impulse response, then for $Y_t$ to exhibit inertia $\mathcal{Y}_t$ must exhibit inertia and 
$\chi$ must be sufficiently large such that $Y_t$ inherits its shape. 
Given $\mathcal{Y}_t$ is the variable that summarizes the effect of future output beliefs 
on current realized output, I will refer to $\chi$ as the \emph{belief multiplier}. 
To formalize this intuition, I consider different perceived 
laws of motion to demonstrate how $Y_t$ inertia arises endogenously due to 
the evolution of beliefs.

\subsection{Simple learning} 
As a tractable starting point, suppose households' perceived law of motion of output 
is given by the simple form
\begin{equation}\label{eq:Y_simple_plm}
    \tilde{Y}_t = \lambda_t + \eta_t
\end{equation}
where the parameters of persistent component $\lambda_t$ and transitory component $\eta_t$, 
the additive components of the exogenous demand shock $\varepsilon_t$, are known.
This perceived law of motion implies that households fully understand the partial 
equilibrium or direct effect of the shocks on output but disregard the 
general equilibrium\footnote{\cite{bastianello2025expectations} study this idea of ``partial equilibrium thinking'' 
in an application to learning from endogenously determined asset prices.} or indirect effect of the shocks on output through the shocks' 
feedback into future beliefs, future spending, and back into current spending\footnote{This feedback loop is also known as the intertemporal Keynesian cross, 
as in \citealt{auclert2024intertemporal}, whereas here I am also incorporating the counterveiling 
effects of interest rates into }.
While households perceive the output process to be exogenous, 
realized output is in fact endogenous to households' beliefs through this 
general equilibrium feedback.


Given the perceived law of motion in Equation (\ref{eq:Y_simple_plm}), we can describe the 
evolution of component beliefs given by Equation (\ref{eq:unobs_kal_update}) by
\begin{equation}\label{eq:kal_update_simple_plm}
\begin{bmatrix}
    E_t[\lambda_{t+1}] \\
    E_t[\eta_{t+1}]
\end{bmatrix}
= 
\mathbf{F}
\begin{bmatrix}
    E_{t-1}[\lambda_t] \\
    E_{t-1}[\eta_t]
\end{bmatrix}
+ 
\mathbf{g} 
\left(
\overbrace{
\underbrace{\chi \mathcal{Y}_t}_{\text{Misspecified }\tilde{Y}_t\text{ wedge}} + 
\underbrace{
\mathbf{1}^\prime\left(
\begin{bmatrix}
    \lambda_t \\
    \eta_t
\end{bmatrix}
-
\begin{bmatrix}
    E_{t-1}[\lambda_t] \\
    E_{t-1}[\eta_t]
\end{bmatrix}
\right)
}_{\text{Standard (rational) Kalman update}}
}^{\text{Forecast error }Y_t - E_{t-1}[Y_t]}
\right)
\end{equation}
where $\mathbf{1}$ is a (2 $\times$ 1) vector of ones.

Permanent income implied by Equation (\ref{eq:Y_simple_plm}) takes the form
$$
\mathcal{Y}_t = \frac{\beta \omega \rho_\lambda}{1 - \beta \omega \rho_\lambda} E_{t-1}[\lambda_t] +  \frac{\beta \omega \rho_\eta}{1 - \beta \omega \rho_\eta} E_{t-1}[\eta_t]
\equiv \mathbf{h}^\prime
\begin{bmatrix}
    E_{t-1}[\lambda_t] \\
    E_{t-1}[\eta_t]
\end{bmatrix}
$$
With $\rho_\lambda > \rho_\eta$ the same-sized belief update of $E_{t-1}[\lambda_t]$ 
raises expected future income $\mathcal{Y}_t$ by more than a comparable change in $E_{t-1}[\eta_t]$ 
because it corresponds to the belief that future income $\{E_t[Y_{t+h}]\}_{h>1}$ will be persistently higher.
Hence, I will denote the coefficients preceding component beliefs $E_{t-1}[\lambda_t], E_{t-1}[\eta_t]$ 
as the ``effective horizons'' $\mathbf{h}^\prime:= [h_\lambda, h_\eta]$ of the current beliefs.

The key difference in the evolution of component beliefs 
relative to rational learning, i.e. $Y_t = \tilde{Y}_t$, 
is the wedge $\chi \mathcal{Y}_t$ that stems from the misspecified 
perceived law of motion $\tilde{Y}_t$.
This wedge can be interpreted as households misattributing the portion of a given 
realized forecast error $Y_t - E_{t-1}[Y_t]$ that was actually due to the endogenous feedback of 
component beliefs into output instead to perceived, larger exogenous shock realizations $\lambda_t, \eta_t$. 
This causes a larger component belief revision, if $\chi \mathcal{Y}_t > 0$, 
than would be warranted under rational learning.

\begin{figure}[t]
    {
    \centering
    \includegraphics[width=0.625\linewidth]{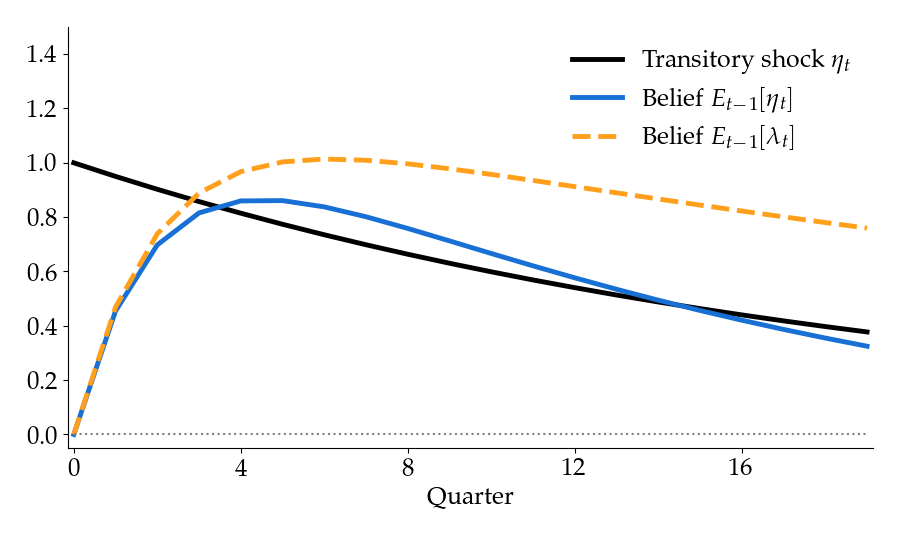}
    \caption{Component belief impulse responses to a transitory demand shock $\eta_t$}
    \label{fig:shock_exp_decomp}
    }
\end{figure}

One can obtain an equivalent but informative interpretation by collecting terms
$\tilde{\mathbf{F}} := \mathbf{F} + \chi \mathbf{g} \mathbf{h}^\prime$, 
noting that all entries of $\tilde{\mathbf{F}} > \mathbf{F}$ as long as the belief multiplier $\chi$ 
is positive. 
The rearranged expression in Equation (\ref{eq:kal_update_simple_plm}) is
$$
\begin{bmatrix}
    E_t[\lambda_{t+1}] \\
    E_t[\eta_{t+1}]
\end{bmatrix}
= 
\tilde{\mathbf{F}}
\begin{bmatrix}
    E_{t-1}[\lambda_t] \\
    E_{t-1}[\eta_t]
\end{bmatrix}
+ 
\mathbf{g}\, 
\mathbf{1}^\prime\left(
\begin{bmatrix}
    \lambda_t \\
    \eta_t
\end{bmatrix}
-
\begin{bmatrix}
    E_{t-1}[\lambda_t] \\
    E_{t-1}[\eta_t]
\end{bmatrix}
\right)
$$
where the effective persistence of beliefs $\tilde{\mathbf{F}}$ is higher (and correlated across 
components, since $\tilde{\mathbf{F}}$ is no longer diagonal) because of the endogenous 
feedback from the misspecification wedge $\chi \mathcal{Y}_t$.
Figure \ref{fig:shock_exp_decomp} displays the evolution of component beliefs given a shock to the 
transitory component $\eta_t$.
Note that in either interpretation the only reason why there are larger belief revisions or 
effectively more belief persistence is because of the general equilibrium feedback of 
beliefs into realizations governed by $\chi$. 

The size of general equilibrium belief feedback $\chi \mathcal{Y}_t$ influences not only the 
evolution of component beliefs by increasing their effective persistence but also the 
extent to which changes in component beliefs alter realized outcomes $Y_t$. 
In simple learning, the exact same term $\chi \mathcal{Y}_t$ amplifies both effective persistence 
as described above and realized output in Equation (\ref{eq:olg_agg_demand_prop}).

\begin{figure}[t]
    {
    \centering
    \begin{subfigure}{0.495\textwidth}
        \centering
        \includegraphics[width=\linewidth]{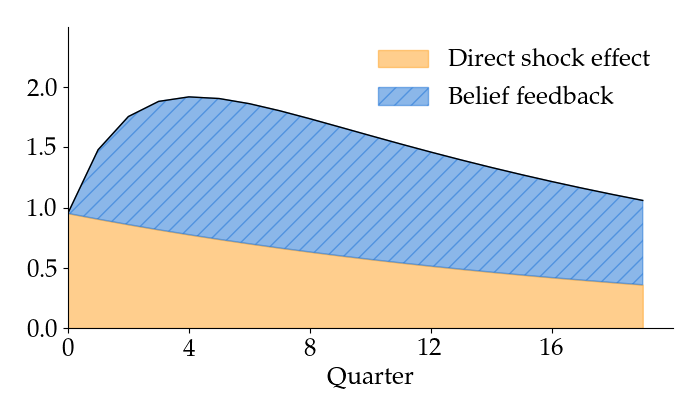}
        \caption{Output Decomposition}
    \end{subfigure}
    \hfill
    \begin{subfigure}{0.495\textwidth}
        \centering
        \includegraphics[width=\linewidth]{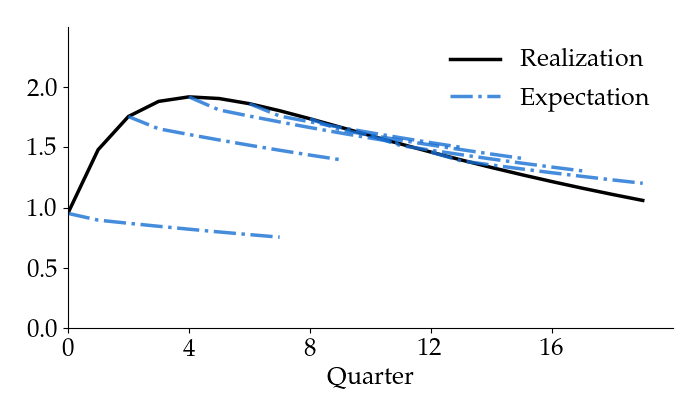}
        \caption{Output Expectation Hair Plot}
    \end{subfigure}

    \caption{Output impulse response to a transitory demand shock $\eta_t$}
    \label{fig:output_decomp_and_hair}
    }

    \caption*{\footnotesize 
    \emph{Note}: 
    Both panels plot the realized output $Y_t$ impulse response to a transitory $\eta_t$ demand shock in black.
    The left panel decomposes the output response 
    into the direct effect of the shock (gold solid) and the belief feedback effect (blue hatched) 
    in Equation (\ref{eq:olg_agg_demand_prop}). 
    The right panel expectation ``hairs'' (blue dot-dash) collect impulse response coefficients 
    of output expectations across horizons.
    }
\end{figure}

The left panel of Figure \ref{fig:output_decomp_and_hair} visualizes the intuition stated earlier 
that realized output $Y_t$ displays inertia only if belief feedback is sufficiently large.
Notice also that in the effective persistence of past beliefs, $\tilde{\mathbf{F}}$, 
the upper row is larger than the bottom row, because the persistent component 
belief $E_{t-1}[\lambda_t]$ has a larger effective horizon $h_\lambda > h_\eta$. 
As belief feedback increases the size of belief revisions due to 
the misattribution of equilibrium effects to further shocks, it also shifts the composition of beliefs 
toward the persistent component belief, $E_{t-1}[\lambda_t]$, over time.
Extrapolation bias arises as a consequence of this compositional shift.
Hence, delayed overreaction arises endogenously when the response of 
output is inertial and as persistent component beliefs become 
further reinforced as shown in the right panel of Figure \ref{fig:output_decomp_and_hair}.
I will refer to the two-way feedback loop of beliefs into equilibrium output as ``\textbf{unanchoring}'', 
given the propensity for an initial shock to trigger an exaggerated and protracted response of 
beliefs and outcomes relative to the rational learning benchmark, which I will analyze further 
in the next section.

\paragraph{Inertia and the belief multiplier $\chi$}

Figure \ref{fig:output_decomp_and_hair} suggests that inertia in realized output 
and extrapolation bias in expectations arises when the belief multiplier $\chi$ is 
sufficiently large. 
Here, I formalize this argument and expound further on the intuition.

{
\definition{\label{defn:inertia}
Consider the moving average representation of $Y_t = \sum_{\ell=0}^\infty (a_{\ell} \lambda_{t-\ell} + b_{\ell} \eta_{t-\ell})$.
$Y_t$ exhibits \textbf{inertia} with respect to a component shock $\lambda_{t-\ell}$
if its corresponding coefficients $\{a_\ell\}$ are weakly increasing (decreasing) for $\ell \leq \bar{\ell} > 0$ 
and weakly decreasing (increasing) for $\ell > \bar{\ell}$\footnote{
Given the system (\ref{eq:olg_agg_demand_prop}), (\ref{eq:kal_update_simple_plm}), 
I impose regularity conditions on model parameters in Appendix \ref{asubsec:main_prop_proof} 
that prevent the impulse responses of output and component beliefs 
from exhibiting oscillation, such that this definition applies.
}.
Denote the impulse response period $\ell = \bar{\ell}$ as the \textbf{inertial peak}.
These definitions hold symmetrically for component shock $\eta_{t-\ell}$ and its coefficients $\{b_\ell\}$.
}
}
\vspace{0.25cm}



The above definition essentially states that a variable is inertial with 
respect to a shock if its maximal impulse response period, which I call the inertial peak, 
is not the initial shock period. 

{
\proposition{
\label{prop:main_prop}
For each component shock $e_t \in \{\lambda_t, \eta_t\}$ if $\chi > \underline{X}_e$,
then $Y_t$ will exhibit inertia and its inertial peak period $\bar{\ell}$ will be weakly increasing in $\chi$.\\
\noindent {\normalfont Proof in Appendix \ref{asubsec:main_prop_proof}}.
}
}
\vspace{0.25cm}

Proposition \ref{prop:main_prop} demonstrates the tight connection between $Y_t$ inertia and the belief 
multiplier $\chi$.
For $Y_t$ to exhibit inertia at all, the belief multiplier $\chi$ must be sufficiently large, 
such that the endogenous amplification and persistence contributed by equilibrium belief feedback 
exceeds the exogenous decay of the direct shock effect. 
In the initial period the criteria for inertia to arise simply requires the response of 
belief feedback to exceed initial exogenous decay of the direct shock effect. 
However, in later periods the relevant comparison is whether the endogenous persistence 
of the persistent belief component $E_{t-1}[\lambda_t]$ exceeds the combined decay 
from the transitory belief component $E_{t-1}[\eta_t]$, when 
this belief starts reverting to zero, and the transitory shock $\eta_t$ itself. 

\begin{figure}[t]
    {
    \centering
    \includegraphics[width=0.625\linewidth]{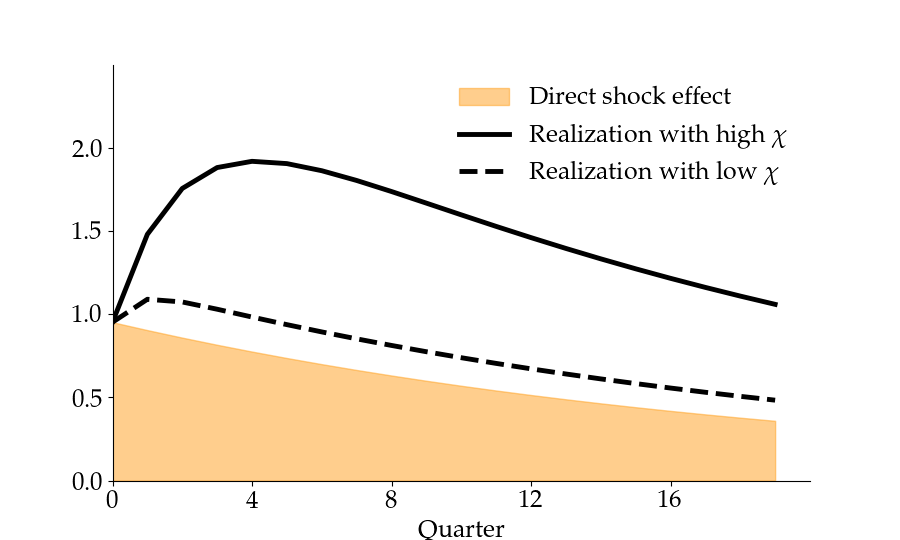}
    \caption{Output impulse responses to an $\eta_t$ shock under different belief multipliers $\chi$}
    \label{fig:output_high_low_chi}
    }

    \caption*{\footnotesize 
    \emph{Note}: 
    The black solid and dashed lines represent the output impulse response under high and low 
    belief multiplier calibrations. The direct shock effect (gold solid) is the same under 
    both calibrations. The gap between the direct shock effect and the black solid and 
    dashed lines represent the size of expectation feedback under each calibration.
    }
\end{figure}

\begin{figure}[t]
    {
    \centering
    \begin{subfigure}{0.495\textwidth}
        \centering
        \includegraphics[width=\linewidth]{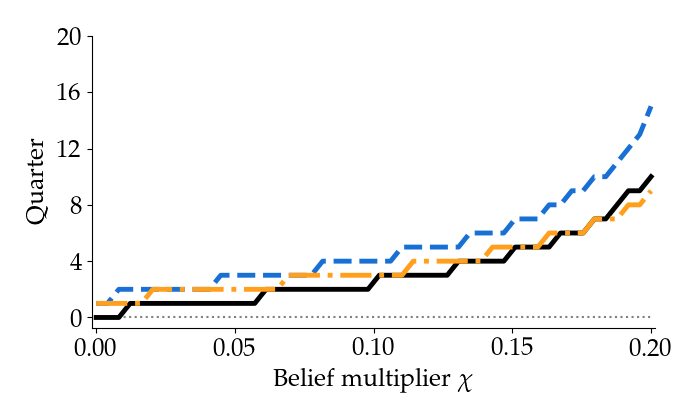}
    \end{subfigure}
    \hfill
    \begin{subfigure}{0.495\textwidth}
        \centering
        \includegraphics[width=\linewidth]{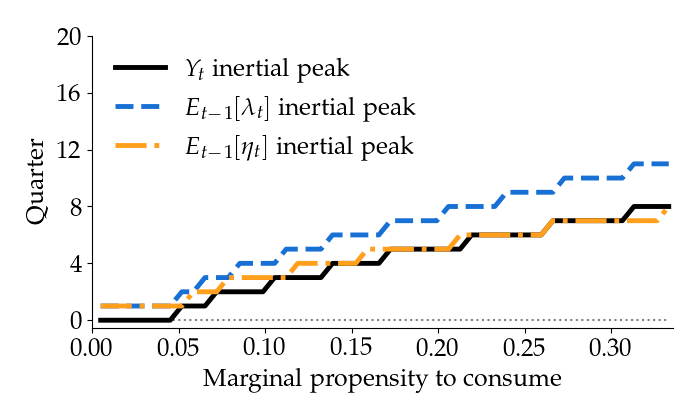}
    \end{subfigure}
    \caption{Inertial peaks of output and component belief impulse responses to a transitory $\eta_t$ shock}
    \label{fig:inertial_peaks}
    }

    \caption*{\footnotesize 
    \emph{Note}: Each line represents the inertial peak period, defined in Definition \ref{defn:inertia}, 
    for output (black), the persistent component belief (dashed blue), and the transitory component belief (dot-dashed gold) 
    to a transitory shock. 
    The left panel plots the inertial peaks across values of the belief multiplier, holding the 
    effective horizons fixed. 
    The right panel plots the inertial peaks across values of the current marginal propensity to consume, 
    which also alters the effective horizons.
    }
\end{figure}

\begin{figure}[t]
    {
    \centering
    \includegraphics[width=0.625\linewidth]{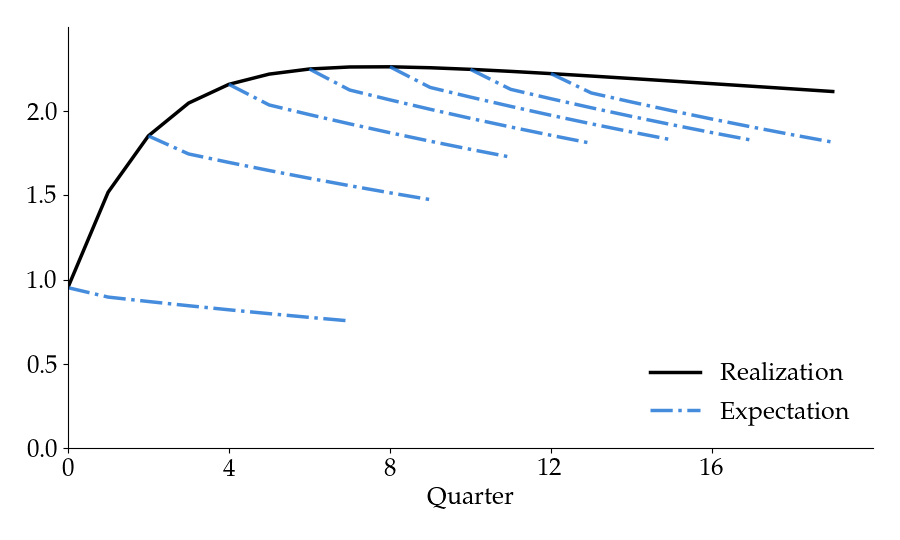}
    \caption{Realized and expected output response to a persistent demand shock $\lambda_t$}
    \label{fig:output_hair_persistent}
    }

    \caption*{\footnotesize 
    \emph{Note}: 
    The black line represents the realized output $Y_t$ impulse response to a persistent $\lambda_t$ demand shock.
    The expectation ``hairs'' (blue dot-dash) collect impulse response coefficients 
    of output expectations across horizons.
    }
\end{figure}

Figure \ref{fig:output_high_low_chi} displays the shape of the baseline output response under 
a high belief multiplier $\chi$ in the solid black line and the response 
under a counterfactual economy with a lower multiplier in the dashed line.
We can see that not only is the overall level of the output response lower, but 
the time profile of the response is also shifted earlier (to the left), since the 
amplification from belief feedback only occurs gradually through the dynamic unanchoring of 
beliefs.

What role does the ``heterogeneous-agent'' side of the economy play in generating inertia?
Recall the form of the belief multiplier $\chi$ from the perpetual youth model
$$
\chi := (1 - \beta \omega - \beta \omega \sigma \phi) \equiv (\text{MPC} - (1 - \text{MPC}) \text{EIS} \phi)
$$
It is useful to rewrite the parameter values in terms of more interpretable quantities, namely 
the current marginal propensity to consume (MPC) and the elasticity of intertemporal substitution 
(EIS). 
We see then that the belief multiplier $\chi$ is large when MPC is high, which is the key 
difference between heterogeneous and representative-agent models of consumption, 
and when the EIS is low. 

Figure \ref{fig:inertial_peaks} shows how the inertial peaks of output and the component beliefs 
change as the belief multiplier and the MPC change. 
The only place the EIS enters the equilibrium output law of motion is through the 
belief multiplier. 
Thus, the left panel of Figure \ref{fig:inertial_peaks} can be interpreted equivalently as 
how inertia changes with the EIS, where a lower EIS implies a higher multiplier and consequently 
more inertia. 
The reason is that when the EIS is low, equilibrium output is more insensitive to 
the counterveiling response of real interest rates via the monetary policy rule. 
Hence, a boom caused by an initial transitory shock is more likely to cause beliefs to 
unanchor and result in inertia in realized output.
A higher MPC results in a higher belief multiplier but a lower effective horizon for 
component beliefs, which makes its contribution toward inertia ambiguous in principle.
However in practice, as the right panel of Figure \ref{fig:inertial_peaks} demonstrates, 
the net effect of a higher MPC still results in more inertia.

This same intuition for how $Y_t$ inertia arises also applies to a persistent component 
shock $\lambda_t$.
However, the main difference in that case, as documented in Figure \ref{fig:output_hair_persistent}, 
is that expectations underreact at all horizons and time periods following a persistent 
$\lambda_t$ shock.
Even though the composition of beliefs still shift toward the persistent component belief, 
the portion of beliefs attributed to the transitory component generate underreaction 
that diminishes over time. 
This effect resembles the standard mechanisms that arise in models, 
such as sticky or noisy information, which also exhibit diminishing underreaction over time.


\subsection{Rational and constrained-rational learning}

This sub-section contrasts the simple perceived law of motion with more 
sophisticated beliefs and demonstrates that inertia may be absent in certain 
cases of learning while present in others.
The first case I consider is rational learning, where the perceived and 
actual laws of motion coincide. 
With rational learning households are able to account for the equilibrium impacts of 
their decisions on their own expectations and hence optimally incorporate past 
observations of output into their component forecasts. 
The rational learning equilibrium is given by
\begin{equation}\label{eq:rational_learning_eqm}
    Y_t \equiv \tilde{Y}_t = \left(\frac{\chi h_\lambda}{1 - \chi h_\lambda} E_{t-1}[\lambda_t] + \frac{\chi h_\eta}{1 - \chi h_\eta} E_{t-1}[\eta_t]\right) + \lambda_t + \eta_t
\end{equation}

The second case, which I call ``constrained-rational'' learning, still restricts 
households' beliefs to be functions of the contemporaneous components $\lambda_t, \eta_t$ 
but permits their coefficients to be optimally estimated given their economic environment. 
This allows me to later consider policy counterfactuals that are robust to the Lucas 
critique within this class of models with imperfect learning, while still retaining 
similar limitations and implications as simple learning. 
Constrained-rational learning takes the below functional form
$$
\tilde{Y}_t = \tilde{a} \lambda_t + \tilde{b} \eta_t
$$
Given the moving average representation of realized output 
$Y_t = \sum_{\ell=0}^\infty (a_\ell u_{\lambda, t-\ell} + b_\ell u_{\eta, t-\ell})$, 
we can solve for the optimal coefficients $\tilde{a}, \tilde{b}$ as the linear projection 
coefficients of the constrained-rational perceived law of motion $\tilde{Y}_t$ 
onto the actual one $Y_t$. 
The projection yields the following pair of implicit equations, which define $\tilde{a}, \tilde{b}$ as 
the solution to their fixed point. 
\begin{align*}
    \tilde{a} &= (1 - \rho_\lambda^2) \sum_{\ell=0}^\infty \rho_\lambda^\ell a_\ell(\tilde{a}, \tilde{b}) \\
    \tilde{b} &= (1 - \rho_\eta^2) \sum_{\ell=0}^\infty \rho_\eta^\ell b_\ell(\tilde{a}, \tilde{b})
\end{align*}

Let us now consider the differences between these simple, constrained-rational, and rational 
learning by analyzing their implications on agents' ``forward-looking reasoning''.
I have started by writing down ``backward-looking'' or ``statistical'' learning rules 
that represent households' perceived laws of motion, 
as opposed to starting with a perceived ``forward-looking'' or ``structural'' model 
of the economy, whose reduced form solution is given by these learning rules. 
A standard notion of ``reasoning'' within a structural model is an agents' ability to 
understand the dynamic equilibrium implications of a particular exogenous shock occurring 
today on current and future outcomes.
This typically consists of the shock's forward propagation, e.g. through the exogenous persistence 
of the shock itself, and then backward propagation, e.g. through the endogenous response of 
output (consumption) today understanding future output (income) will rise due to 
the persistence of the shock into the future---that is, the intertemporal Keynesian cross.

These equilibrium feedbacks, both forward and backward, are encoded in the reduced form, 
or statistical, representation of the perceived structural model as the coefficient on the 
particular shock, e.g. the moving average coefficient (or loading) $a_\ell$ of the shock 
innovation $u_{\lambda, t-\ell}$ on realized output $Y_t$. 
Hence, the assumption of a particular ``backward-looking'' learning rule, such as simple learning 
where $\tilde{Y}_t = \lambda_t + \eta_t$ with the loading
$\tilde{a}_\ell = \rho_\lambda^\ell$, implies that households beliefs behave \emph{as-if} 
they are not ``reasoning'' at all about the general equilibrium feedback of a $\lambda_t$ 
shock into future output (income) and its effects on current output (consumption) when 
conducting inference on $\lambda_t, \eta_t$ after observing realized output $Y_t$.
In other words, assuming learning rules are restricted, in the sense that 
they are too low-dimensional to capture the true stochastic process for output $Y_t$, 
can be treated analogously as an assumption that agents' general equilibrium reasoning 
is distorted.

\begin{figure}[t]
    {
    \centering
    \includegraphics[width=0.625\linewidth]{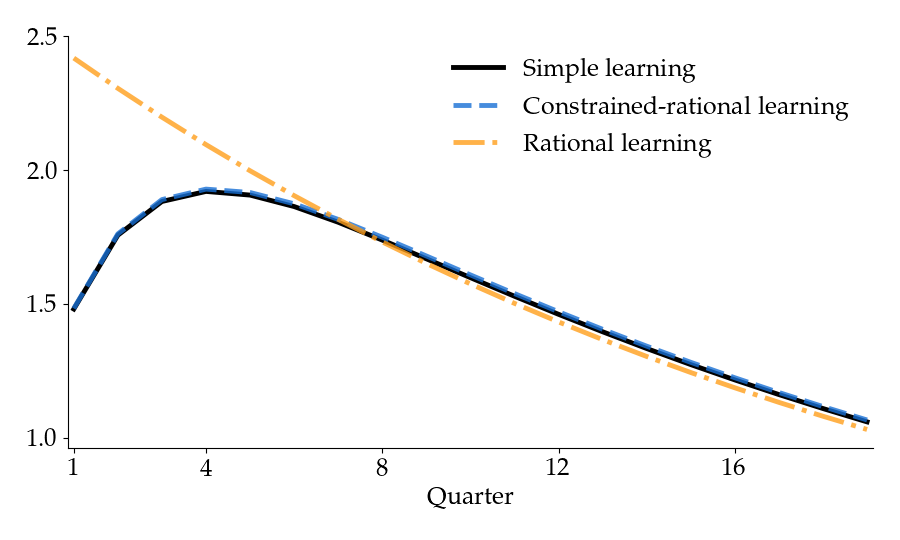}
    \caption{Output impulse responses to an $\eta_t$ shock under different forms of learning}
    \label{fig:output_all_plms}
    }

    \caption*{\footnotesize 
    \emph{Note}: 
    Each line represents the impulse response of output to a transitory demand shock 
    under different learning assumptions. 
    I omit period zero because it does not contain an expectational response due to the 
    staggered timing assumption.
    }
\end{figure}

Consider realized output at a future horizon-$h > 0$, $Y_{t+h} = \chi \mathcal{Y}_{t+h} + \varepsilon_{t+h}$.
The rational learning equilibrium is the solution to the fixed point 
of $Y_t = \tilde{Y}_t$, where subjective expectations of future output $E_{t-1}[Y_{t+h}]$ 
fully account for the amplification from anticipated future belief feedback $\chi \mathcal{Y}_{t+h}$ 
and in turn from further future belief feedback $\chi \mathcal{Y}_{t+h^\prime}$ for $h^\prime > h$ 
encoded in $\chi \mathcal{Y}_{t+h}$ and so on.
The simple perceived law of motion is on the opposite extreme, where it contemplates 
that future output is only affected by the direct impact of the shock itself. 

Constrained-rational learning constitutes a middle ground between these two polar cases.
While the persistence of the components in the constrained-rational learning rule 
are still fixed to be the actual, exogenous persistences of the shocks $\rho_\lambda, \rho_\eta$, 
this learning rule allows for the level impact of the shock to differ from the direct effect 
via the coefficients $\tilde{a}, \tilde{b}$.
Hence, while constrained-rational learning does not account for the additional endogenous 
persistence in output due to equilibrium belief feedback, it can partially account for the 
equilibrium amplification of shocks due to this feedback.

Figure \ref{fig:output_all_plms} displays the impulse responses of realized output $Y_t$ 
to a transitory $\eta_t$ shock under counterfactual economies with the three different 
learning rules just described.
Notably, the impulse response under rational learning does not display inertia. 
This is not a generic feature of rational learning, which in principle alone could generate 
inertia, but it will depend ultimately on the persistences and variances of 
the underlying shock components.
However, it is worth emphasizing that inertia is more likely to arise under simple 
and constrained-rational learning for a given pre-defined set of shock components 
because of the additional amplification and endogenous 
persistence in beliefs due to the feedback wedge $\chi \mathcal{Y}_t$ as 
in Equation (\ref{eq:kal_update_simple_plm}).

The output response under rational learning is initially much larger than 
under simple or constrained-rational learning, due to the belief feedback amplification, 
but it eventually decays below their responses.
In other words, the time profile of amplification of rational learning over simple learning is 
not uniform, where constrained rational learning seems more closely resemble simple 
learning rather than rational learning.

\begin{figure}[t]
    \centering
    \begin{subfigure}{0.495\textwidth}
        \centering
        \includegraphics[width=\linewidth]{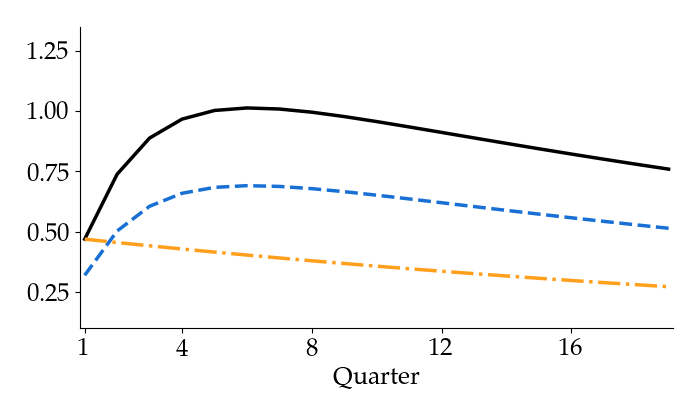}
        \caption{Persistent $E_{t-1}[\lambda_t]$}
    \end{subfigure}
    \hfill
    \begin{subfigure}{0.495\textwidth}
        \centering
        \includegraphics[width=\linewidth]{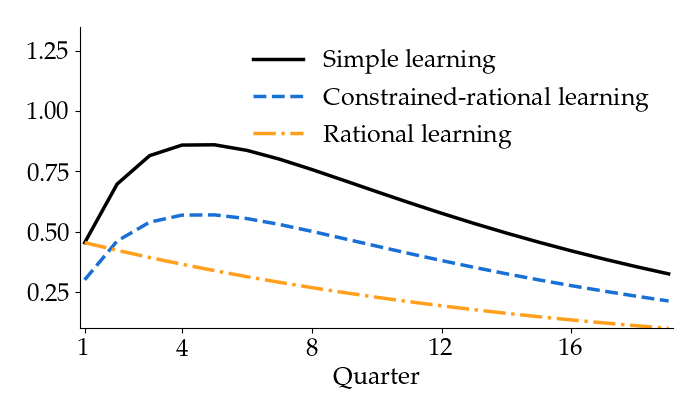}
        \caption{Transitory $E_{t-1}[\eta_t]$}
    \end{subfigure}
    \caption{Component belief impulse responses to an $\eta_t$ shock under different forms of learning}
    \label{fig:exp_irfs_all_plms}

    \caption*{\footnotesize 
    \emph{Note}: 
    Each line represents the impulse response of component beliefs to a transitory demand shock 
    under different learning assumptions. 
    I omit period zero because it does not contain an expectational response due to the 
    staggered timing assumption.
    }
\end{figure}

What accounts for the absence of amplification in the realized output response under constrained-rational learning 
relative to the simple learning benchmark?
Figure \ref{fig:exp_irfs_all_plms} displays the impulse responses of component beliefs under the 
three different forms of learning and provides some insight to address this question.
First, recall that under all three forms of learning the direct shock effect is the same. 
Therefore, if the realized output responses are similar this must mean that the 
size and shape of the belief feedback effects are similar.

Figure \ref{fig:exp_irfs_all_plms} shows that the responses of component beliefs 
under constrained-rational learning are smaller than under simple learning.
This means that the degree of equilibrium amplification in realized output $Y_t$ 
for a given level of belief $E_{t-1}[\lambda_t], E_{t-1}[\eta_t]$ must then be 
larger under constrained-rational learning for the overall belief feedback effect to 
be equal the feedback effect under simple learning.
In turn, if we treat the response of beliefs under rational learning as the 
rational benchmark, then we see that the degree of unanchoring of beliefs under 
constrained-rational learning is substantially smaller than under simple learning 
because of the ability of beliefs to partially account for equilibrium feedback.
Nonetheless, there is still some unanchoring of beliefs, and it is the resultant 
amplification from this unanchoring that eventually causes the output 
response under constrained-rational learning to exceed that of rational learning.
I consider this idea more formally in Proposition \ref{prop:rational_comp} 
in a special case with an i.i.d transitory shock $\eta_t$.

{
\begin{proposition}\label{prop:rational_comp}
    Suppose $\chi > 0$, $\eta_t$ is i.i.d, and $\sigma_\eta$ are normalized to 
    make constrained-rational and rational Kalman gains proportional.
    Let $Y_t^{R}, Y_t^{CR}$ denote output under rational learning and 
    constrained-rational learning respectively.
    \vspace{0.1cm}
    \begin{itemize}
        \item In response to a transitory $u_{\eta, 0}$ innovation to $\eta_0$, 
              $Y_t^{R} > Y_t^{CR}$ for time-$t \leq \bar{t}$, where $\bar{t} > 1$.
        \item In response to a persistent $u_{\lambda, 0}$ innovation to $\lambda_0$,
              $Y_t^{R} > Y_t^{CR}$ for time-$t > 1$.
    \end{itemize}

    \noindent {\normalfont Proof in Appendix \ref{asubsec:rational_comp_proof}.}
\end{proposition}
}

Beliefs under constrained-rational learning exhibit partial unanchoring, which makes 
the intuition from the simple learning example I developed earlier a helpful analogy. 
Unlike simple learning, constrained-rational beliefs still maintain a degree of 
consistency to realized output dynamics in being the best approximation in the class 
of learning rules that are linear in the shock components. 
Given beliefs are partly responsive to their surrounding economic environment, 
constrained-rational learning is a useful illustrative model of expectation formation 
that delivers inertia in realizations and extrapolation bias in expectations, 
while still satisfying a limited notion of the Lucas critique.
Therefore, we proceed to the next section to consider some simple policy 
counterfactuals and how alternative policies can alter the degree of inertia 
in realizations via their ability to anchor expectations.

\section{Policy implications of macroeconomic inertia}\label{sec:policy}

This section discusses novel policy considerations that arise under 
constrained-rational learning and contrast them with typical policy 
transmission outcomes in models with full-information rational expectations. 
Under constrained-rational learning, it is no longer desirable to be 
infinitely-responsive to demand-driven fluctuations because 
of the risk of destabilizing expectations. 
Gradual policy approaches to monetary policy, as in a highly inertial 
Taylor rule, fail to stabilize output as effectively as they would 
under the full-information rational expectations benchmark. 
In addition, the stimulus effects of a deficit-financed fiscal transfer in a heterogeneous-agent 
New Keynesian economy may no longer be front-loaded.

\subsection{Simple Taylor rules}
The typical monetary policy prescription in response to demand shocks is to 
completely close the output gaps they induce, which aligns with the 
welfare aims of inflation stabilization in standard New Keynesian economies 
(\citealt{blanchard2007real}). 
This divine coincidence also holds in my setting, assuming firms expectations 
are the same as households.
Therefore I use the discounted path of squared deviations of output from steady state 
as a simple welfare measure to contrast counterfactual policy rules in an economy 
with solely demand shocks.
\begin{equation}\label{eq:output_loss}
    \mathcal{L} = \sum_{t=0}^\infty \beta^t Y_t^2
\end{equation}
I consider first the full-information rational expectations equilibrium response to 
a transitory demand shock $\eta_t$.
Given the unnormalized aggregate demand equation
$$
Y_t = \frac{1}{\beta \omega (1 + \sigma \phi)} \left( \underbrace{(1 - \beta \omega - \beta \omega \sigma \phi)}_{\text{Belief multiplier }\chi} \sum_{h=1}^\infty (\beta \omega)^h \mathbb{E}_t[Y_{t+h}] + \eta_t \right)
$$
The equilibrium solution is given by $Y_t = b \eta_t$ where the coefficient $b$ is
$$
b = \underbrace{\frac{1}{\beta \omega (1 + \sigma \phi)}}_{\to \, 0 \text{ as } \phi \to \infty} \underbrace{\left( 1 - \frac{1 - \beta \omega - \beta \omega \sigma \phi}{1 + \sigma \phi} \right)^{-1}}_{\to (1 - \beta \omega)^{-1} \text{ as } \phi \to \infty}
$$
Because welfare is given by the discounted squared loss of output, the optimal choice of 
the Taylor coefficient that completely closes the output gap is for the 
monetary authority to be infinitely responsive $\phi \to \infty \implies b \to 0$. 
Further, welfare loss strictly decreases as $\phi$ increases for any finite $\phi$. 
This shows that in the standard full-information rational expectations setting, 
a counterfactual policy that is more responsive to demand shocks is always 
welfare-improving.

However, when agents form expectations with simple and constrained-rational learning 
the optimal policy prescription differs. 
Simple learning yields the aggregate demand equation
\begin{align*}
    Y_t &= \underbrace{\frac{1 - \beta \omega - \beta \omega \sigma \phi}{1 + \sigma \phi}}_{\to -\beta \omega \text{ as } \phi \to \infty} \left( \frac{\rho_\lambda}{1 - \beta \omega \rho_\lambda} E_{t-1}[\lambda_t] + \frac{\rho_\eta}{1 - \beta \omega \rho_\eta} E_{t-1}[\eta_t] \right) + \underbrace{\frac{1}{\beta \omega (1 + \sigma \phi)}}_{\to \, 0 \text{ as } \phi \to \infty} (\lambda_t + \eta_t)
\end{align*}

In the infinitely-responsive $\phi$ limit, output is perfectly stabilized at the steady state if 
component beliefs are fully anchored $E_{-1}[\lambda_0] = E_{-1}[\eta_0] = 0$. 
Unlike in the full-information rational expectations case, if $\phi$ is not taken 
fully to the infinite limit but instead is finite and sufficiently large then 
the component beliefs $E_t[\lambda_{t+1}], E_t[\eta_{t+1}]$ and consequently output 
$Y_t$ itself can become destabilized. 
Higher monetary responsiveness is therefore effective only up to a point, a 
limitation that is also demonstrated in \cite{eusepi2024shortrun}. 

\begin{figure}[h]
    {
    \centering
    \includegraphics[width=0.625\linewidth]{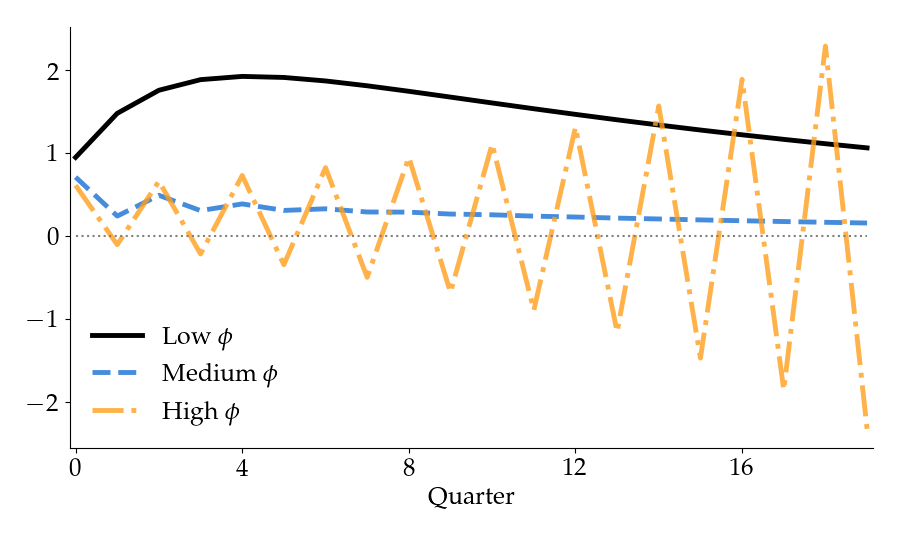}
    \caption{Output (de-)stabilization under different Taylor rule $\phi$}
    \label{fig:optimal_plm_diff_phis}
    }

    \caption*{\footnotesize 
    \emph{Note}: 
    Each line represents the impulse response of output to a transitory demand shock 
    in an economy with different Taylor rule coefficients $\phi$. 
    }
\end{figure}

Figure \ref{fig:optimal_plm_diff_phis} demonstrates that this behavior also holds 
in the constrained-rational learning case. 
The shared reason in both simple and constrained-rational learning is that the expectation feedback 
wedge that appears in the belief component law of motion is increasingly negative as 
$\phi$ increases. 
When households spend more due to optimistic beliefs about demand shocks, the monetary 
authority raises interest rates so significantly that it triggers a contraction in output. 
This causes households to mistakenly infer that the realized shocks were actually negative 
and larger in magnitude than they had previously anticipated. 
The result is an increasingly unstable negative feedback loop, resulting in the explosive 
oscillation of the high $\phi$ case shown in Figure \ref{fig:optimal_plm_diff_phis}. 

However, with mildly elevated responsiveness in the medium $\phi$ we see 
not only a reduced level response of output to the shock but also the absence 
of output inertia. 
Even though the expectation wedge still appears in the component belief law of motion, 
which in principle still contributes to inertia, a monetary authority that chooses 
a Taylor coefficient $\phi$ that sets the belief multiplier $\chi$ sufficiently 
close to zero effectively sets the wedge $\chi \mathcal{Y}_t$ equal to zero. 
This stops the expectation feedback loop that would induce inertia from 
starting in the first place.


In choosing an optimal level of responsiveness to demand shocks $\phi$, 
a monetary authority facing households with constrained-rational learning should not 
respond as forcefully as in the rational benchmark because of the risk of 
destabilizing expectations and therefore may not be able to fully shut down 
demand-driven fluctuations.

\subsection{Inertial Taylor rules and monetary policy gradualism}
A popular Taylor rule specification includes a lagged or ``inertial'' term
$$
r_t = \rho r_{t-1} + \phi Y_t
$$
Early justifications for this approach were based on observed inertia in interest rate policy 
(\citealt{clarida1998monetary}).
However, whether the policy rules themselves are inertial or are simply responding 
to inertial economic conditions was subject to debate (\citealt{rudebusch2005monetary}). 
Other justifications for inertial policy rules include uncertainty about the 
effects of policy (\citealt{sack1998uncertainty}) and their ability to implement 
optimal allocations when forward-looking agents understand 
the dynamic implications of policy commitments (\citealt{woodford1999optimal}).

I expand briefly on this latter reason by demonstrating the inability of 
constrained-rational and simple learning to map the effects of dynamic policy 
commitments to perceived output. 
Just as constrained households are unable to fully internalize the equilibrium feedbacks of 
future expected output changes on current output, which we discussed in the previous section, 
so too are they unable to internalize the effects of current policy commitments 
on future expected output.

The aggregate demand equation for output $Y_t$, 
where households understand the inertial form of the policy rule and its parameters yields
$$
Y_t = -\underbrace{\frac{\sigma \bar{\phi}}{1 + \sigma \bar{\phi}} \, \left( \sum_{\ell=1}^\infty \rho^\ell Y_{t-\ell} \right)}_{\text{Policy commitments in }r_{t-1}} + \frac{1 - \beta \omega - \beta \omega \sigma \bar{\phi}}{1 + \sigma \bar{\phi}} \sum_{h=1}^\infty (\beta \omega)^{h-1} E_{t-1}[Y_{t+h}] + \frac{1}{\beta \omega (1 + \sigma \bar{\phi})} \varepsilon_t
$$

The ``effective'' Taylor coefficient $\bar{\phi} = \frac{\phi}{1 - \beta \omega \rho}$ 
demonstrates that whether policy responds contemporaneously via $\phi$ or with a delay 
via $\rho$, one can equate their contribution to dampening the level of 
equilibrium output $Y_t$ due to their feedback from 
future expectations $\{E_t[Y_{t+h}]\}_{h > 0}$. 
Hence, in response to an unanticipated shock at time-0, absent pre-existing policy 
commitments $r_{-1} = 0$ and fixing a given path of future expectations 
$\{E_t[Y_{t+h}]\}_{h > 0}$, the response of time-0 output $Y_0$ should be the same 
for a continuum of regimes $(\rho, \phi)$ that induce the same effective $\bar{\phi}$. 

The crucial step in the above explanation was that the path of 
future expectations was held fixed. 
Consider if households correctly perceived time-$t+h$ output used to inform 
time-$t$ consumption which determines time-$t$ output in equilibrium. 
$$
\tilde{Y}_{t+h} = -\underbrace{\frac{\sigma \bar{\phi}}{1 + \sigma \bar{\phi}} \, \left( \sum_{\ell=1}^\infty \rho^\ell \tilde{Y}_{t+h-\ell} \right)}_{\text{Function of } \{\varepsilon_{t+h-\ell}\}_{\ell > 0}} + \frac{1 - \beta \omega - \beta \omega \sigma \bar{\phi}}{1 + \sigma \bar{\phi}} \underbrace{\sum_{j=1}^\infty (\beta \omega)^{j-1} E_{t+h-1}[Y_{t+h+j}]}_{\text{Function of }\{ \varepsilon_{t+h-\ell}\}_{\ell > 0}} + \frac{1}{\beta \omega (1 + \sigma \bar{\phi})} \varepsilon_{t+h}
$$
By correctly perceiving future output at time-$t+h$, households understand that current policy 
decisions which respond to current shocks will persist into time-$t+h$ with persistence $\rho$. 
Two regimes with the same effective $\bar{\phi}$ would exhibit different 
equilibrium responses of output at time-0 if households correctly perceived 
that the regime with higher policy rule persistence $\rho$ would continue to respond more forcefully 
in future periods and that the future equilibrium output change would propagate backward 
to time-0 output via the intertemporal Keynesian cross.

However, when learning rules are restricted to load on contemporaneous shocks 
as in the constrained-rational and simple learning cases where $\tilde{Y}_{t+h}$ can 
only be a function of $\varepsilon_{t+h}$, we see that both belief 
feedback in $\{E_{t+h}[Y_{t+h+j}]\}_{j > 0}$ and effects of continued policy 
commitments on future output $\{\tilde{Y}_{t+h-\ell}\}_{\ell > 0}$ are imperfectly 
incorporated into beliefs.
This is again because simple linear functions of contemporaneous shocks 
$\varepsilon_{t+h}$ cannot capture the equilibrium dynamics of 
the contributions of past shocks $\{\varepsilon_{t+h-\ell}\}_{\ell > 0}$ through 
belief feedback and policy commitments. 

To demonstrate the consequences for welfare, I utilize the discounted squared 
output loss $\mathcal{L}$ in Equation (\ref{eq:output_loss}) from before and 
consider two policy regimes. 
I call the first policy regime the ``swift'' policy regime, 
$(\rho^{\text{S}}, \phi^{\text{S}})$, and the second one the ``gradual'' policy regime, 
$(\rho^{\text{G}}, \phi^{\text{G}})$, where the swift regime exhibits less inertia 
$\rho^{\text{S}} < \rho^{\text{G}}$ and greater contemporaneous responsiveness 
$\phi^{\text{S}} > \phi^{\text{G}}$. 
I choose these regimes to equate their welfare loss from a transitory shock under 
the full-information rational expectations benchmark, since there is a continuum 
of regimes $(\rho, \phi)$ that obtain the same welfare loss.

\begin{figure}[t]  
    \centering
    \begin{subfigure}[b]{0.495\textwidth}
        \centering
        \includegraphics[width=\linewidth]{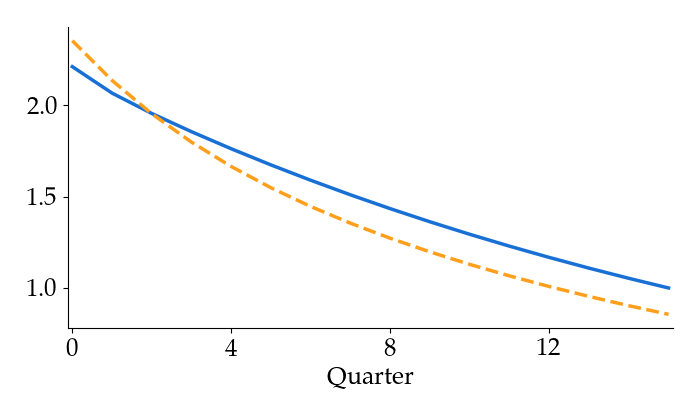}  
        \caption{Output response under FIRE}
    \end{subfigure}
    \hfill
    \begin{subfigure}[b]{0.495\textwidth}
        \centering
        \includegraphics[width=\linewidth]{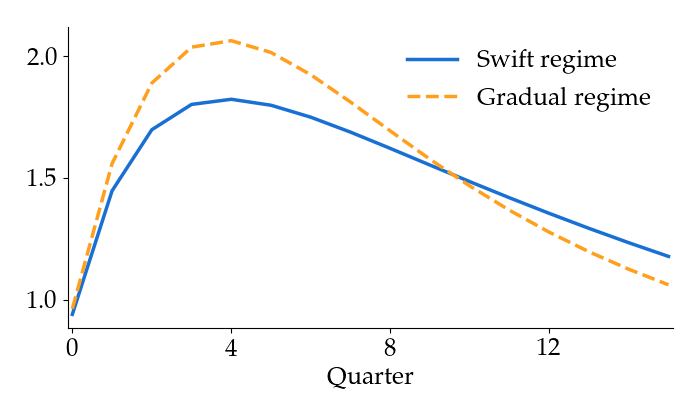}  
        \caption{Output response under learning}
    \end{subfigure}
    
    \vskip\baselineskip  

    \begin{subfigure}[b]{0.495\textwidth}
        \centering
        \includegraphics[width=\linewidth]{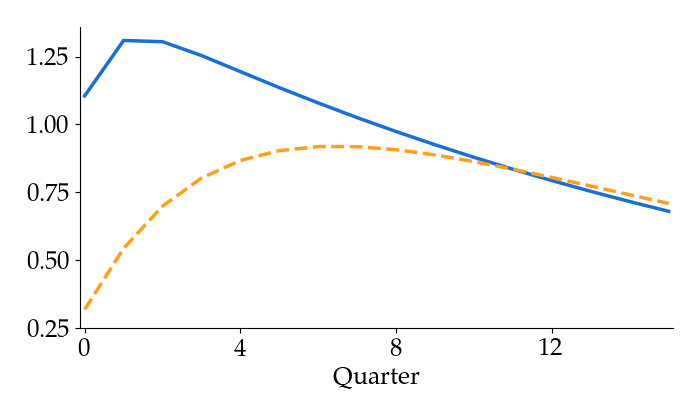}  
        \caption{Interest rate response under FIRE}
    \end{subfigure}
    \hfill
    \begin{subfigure}[b]{0.495\textwidth}
        \centering
        \includegraphics[width=\linewidth]{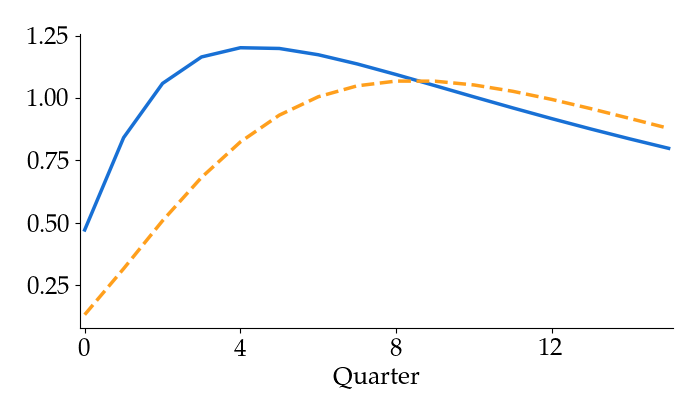}  
        \caption{Interest rate response under learning}
    \end{subfigure}

    \caption{Output and interest rate impulse responses to an $\eta_t$ shock across monetary regimes}
    \label{fig:rho_regime_responses}

    \caption*{\footnotesize 
    \emph{Note}: 
    The top panels represent the impulse responses of output to a transitory demand shock 
    and the bottom panels the analogous responses for the real interest rate. 
    The left column plots the impulse responses under full-information rational expectations (FIRE), 
    and the right column under constrained learning. 
    }
\end{figure}

Figure \ref{fig:rho_regime_responses} displays the impulse responses of output and interest 
rates to a transitory demand shock $\eta_t$ under swift and gradual regimes. 
The left column are the responses under the full-information, rational expectations (FIRE) benchmark and 
the right column are the responses under constrained-rational learning. 
Focusing first on the FIRE case, we see that output responds immediately under both regimes, 
even though the peak response of interest rates is not immediate under either regime. 
This is because time-0 beliefs instantly incorporate the equilibrium effects of 
future implied interest rate changes on future output, which feed back into time-0 output.

Because the gradual regime has a higher degree of policy inertia, we see the path of interest 
rates rising more slowly than in the swift regime.
Nonetheless, the path of output under the gradual regime falls by more than under the swift 
regime only a couple of quarters after the initial shock even though it takes interest 
rates nearly twelve quarters to finally exceed the rate path under the swift regime.
In contrast, under learning we see that output gap between the gradual and swift regimes is 
still positive until the interest rate gap between regimes also switches signs.

\begin{figure}[h]
    {
    \centering
    \includegraphics[width=0.625\linewidth]{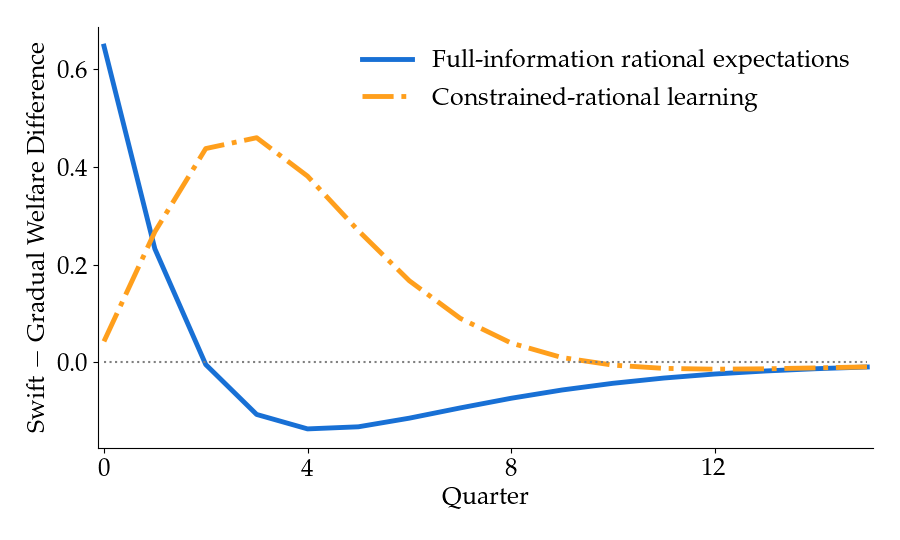}
    \caption{Monetary regime welfare differences under different models of expectation formation}
    \label{fig:rho_regime_welfare}
    }

    \caption*{\footnotesize 
    \emph{Note}: 
    The measure of welfare loss is the discounted squared output deviation each quarter. 
    Each line corresponds to the difference in welfare loss incurred between the swift 
    and gradual policy regime under a different model of expectation formation. 
    When the lines exceed zero, the swift regime incurred a lower discounted welfare loss in 
    that quarter and vice versa. 
    The regimes were chosen such that the area between the blue curve and zero 
    integrates to zero.
    }
\end{figure}

Figure \ref{fig:rho_regime_welfare} demonstrates the difference in current-period discounted 
welfare $-\beta^t Y_t^2$ between the two regimes. 
Because these regimes were chosen to equate total welfare loss under full-information 
rational expectations, the area under the blue curve in Figure \ref{fig:rho_regime_welfare} 
integrates to 1. 
We see that the swift regime achieves higher initial welfare for the first two quarters by 
responding more forcefully but the gradual regime slowly makes up for the welfare 
difference in the long-run. 
However, because the gradual regime is less effective at containing output under learning 
due to households limits in reasoning, the initial differences 
in welfare loss are too large to be offset by persistently higher rates and a smaller output 
response in later periods.

\subsection{Deferred financing and the delayed impacts of fiscal stimulus}

A key difference in the policy implications of 
heterogeneous-agent versus representative-agent macroeconomic models are the 
equilibrium responses of prices and quantities to deficit-financed fiscal transfers 
(\citealt{auclert2024intertemporal}). 
Representative-agent models with full-information, rational expectations 
have Ricardian equivalence, which implies that 
non-distortionary government spending and tax and transfer policies 
have identical effects on equilibrium outcomes regardless of the timing of financing. 
Conversely, deficit-financed fiscal policy can induce immediate and large 
equilibrium responses of quantities, such as consumption spending, 
in heterogeneous-agent models, where the magnitudes of these responses 
increase as financing is further delayed due to the compounding effects of 
future general equilibrium feedback into current demand (\citealt{angeletos2023can}).

When households form expectations with imperfect learning, there are 
additional implications of financing delays for the time profile of output 
in response to fiscal stimulus. 
In particular, deferred financing can delay the peak response to fiscal stimulus, potentially 
reducing its effectiveness if the policymaker desires immediacy. 
Prolonging deficits additionally stretches out the cumulative response of output across 
a much longer horizon, running the risk that stimulus will last 
beyond the initially desired period of fiscal support.
The propagation channels that induce these effects are similar to those in governing 
expectation feedback and prior monetary policy commitments.

\begin{figure}[h]  
    {
    \centering
    \begin{subfigure}[b]{0.495\textwidth}
        \centering
        \includegraphics[width=\linewidth]{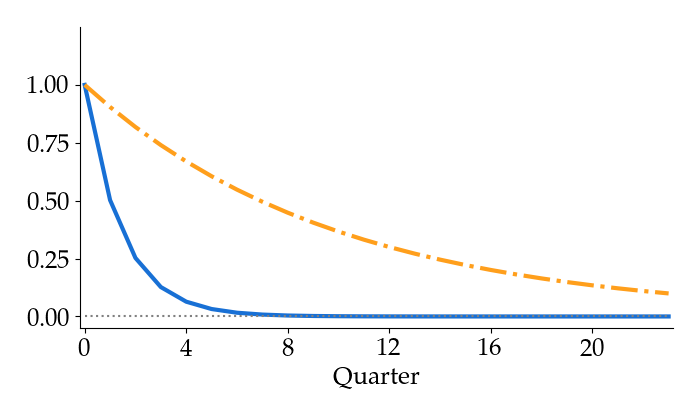}  
        \caption{Normalized response under FIRE}
    \end{subfigure}
    \hfill
    \begin{subfigure}[b]{0.495\textwidth}
        \centering
        \includegraphics[width=\linewidth]{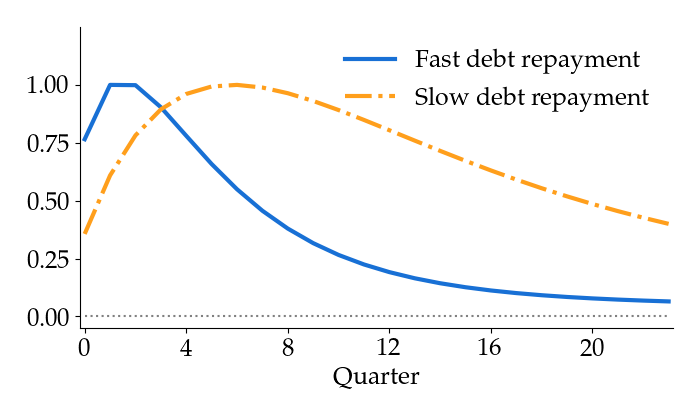}  
        \caption{Normalized response under learning}
    \end{subfigure}
    
    \vskip\baselineskip  

    \begin{subfigure}[b]{0.495\textwidth}
        \centering
        \includegraphics[width=\linewidth]{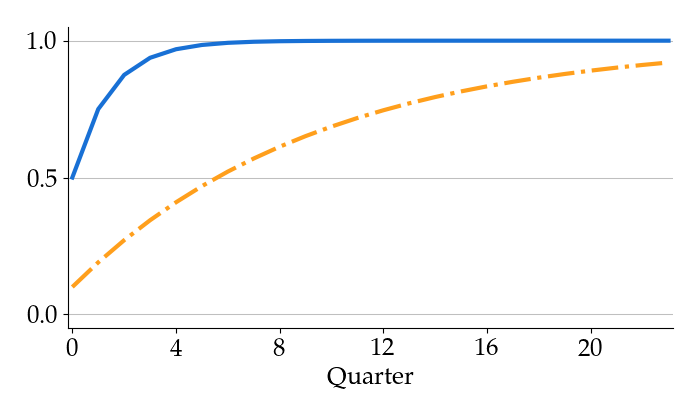}  
        \caption{Discounted cumulative response under FIRE}
    \end{subfigure}
    \hfill
    \begin{subfigure}[b]{0.495\textwidth}
        \centering
        \includegraphics[width=\linewidth]{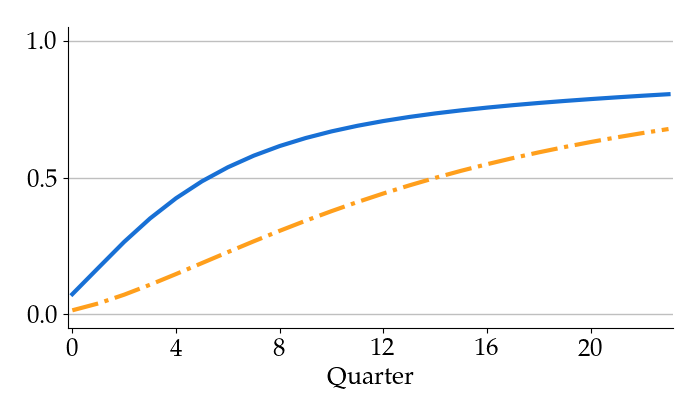}  
        \caption{Discounted cumulative response under learning}
    \end{subfigure}

    \caption{Output impulse responses to a transfer shock $\zeta_t$ across debt repayment speed regimes}
    \label{fig:delta_regime_responses}
    }

    \caption*{\footnotesize 
    \emph{Note}: The top panel displays the output impulse response under each 
    debt repayment regime, where the peak response is normalized to one. 
    The bottom panel displays the share of the cumulative impulse response of output, 
    discounted by the inverse gross interest rate. 
    }
\end{figure}

I consider a simple form of fiscal policy that resembles the setting in \cite{angeletos2023can}. 
The government issues real-valued debt $B_t$ that is financed by a lump-sum tax $T_t$. 
Lump-sum taxes adjust to repay debt gradually, where the speed of repayment is given by $\delta \in (0, 1)$. 
The linearized government budget constraint and tax rule are given by
\begin{align*}
    B_t &= \frac{1}{\beta}(B_{t-1} - T_t)\\
    T_t &= \delta B_{t-1} - (1 - \delta) \zeta_t
\end{align*}
I assume there is no outstanding government debt in steady state $B = 0$, and $\zeta_t$ denotes 
an i.i.d deficit shock which I will use in the following policy exercise.
By assuming debt is real-valued, I omit the possibility that surprise inflation 
erodes the real value of debt through nominal revaluation so we can still maintain our 
focus on real output alone. 
In addition, we now need to enforce asset market clearing between household wealth and government debt.
$$
A_t = B_t
$$
Given this setting we can again derive an aggregate demand equation that determines equilibrium output 
that is analogous to Equation (13) in \cite{angeletos2023can} but without the real interest rate peg.
$$
Y_t = \frac{1}{1 - \chi}\left( \frac{(1 - \beta \omega)(1 - \omega)(1 - \delta)}{1 - \omega (1 - \delta)} (B_{t-1} + \zeta_t) + \chi \sum_{h=1}^\infty (\beta \omega)^h E_t[Y_{t+h}] + \varepsilon_t \right)
$$
where the belief multiplier $\chi := (1 - \beta \omega - \beta \omega \sigma \phi)$ as before.

Figure \ref{fig:delta_regime_responses} displays the response of output to a time-0 deficit 
shock $\zeta_0$ under fast (large $\delta$) and slow (small $\delta$) debt repayment regimes. 
The top panel displays the response of output with its peak period normalized to one, 
and the bottom panel displays the cumulative response of output, discounted by the 
inverse steady state gross interest rate $(1 + r)^{-1} \equiv \beta$ with a total 
response normalized to one. 
The left column displays the responses under a full-information rational expectations (FIRE)
benchmark and the right column under constrained-rational learning.

In the FIRE case, we see that the initial response of output 
to a one-time fiscal transfer is peaked on impact and monotonically decreasing.
This is because the full dynamic effects of higher debt holdings and slower debt 
repayment are internalized on impact by household consumption decisions.
A large share of the discounted cumulative response of output $\sum_{t=0}^\infty (1 + r)^{-t} Y_t$, 
which is a commonly-used measure of the size of fiscal stimulus (\citealt{mountford2009effects}), 
also occurs at relatively short horizons. 
In the FIRE case, half of the discounted cumulative output response occurs immediately 
under the fast debt repayment regime and after five quarters in the slower repayment regime. 
In contrast, with constrained-learning there is a difference of one year in the peak output 
response between regimes and a difference of ten quarters for the discounted cumulative 
response, more than double the gap under FIRE. 


\section{Conclusion}

This paper argues that canonical heterogeneous-agent models are not just 
consistent with aggregate consumption inertia but fundamentally 
contribute to its emergence.
I first show that the minimal structure imposed by these models when supplemented 
with measured expectations data yield model-implied impulse responses 
of consumption that closely resemble observed consumption inertia. 

Guided by new empirical evidence on the extrapolative tendencies of measured 
expectations, I embed an unobserved components model of expectation formation 
into a tractable heterogeneous-agent, general equilibrium environment.
Learning is imperfect when the model of expectation formation is too simple 
to perfectly perceive the dynamic equilibrium effects of shocks.
I show that this model can endogenously generate inertia in the dynamics of 
realized consumption and extrapolation in expectations due to a form of 
belief unanchoring. 
The propensity for beliefs to unanchor is tightly linked to the magnitude of 
general equilibrium amplification, which is large when marginal propensities 
to consume are large, as in heterogeneous-agent models, and when the 
policy regime is less responsive.
I conclude by providing a new rationale that cautions against gradual 
approaches to monetary policy in the form of inertial policy rules and 
to fiscal policy in the form of deferred financing of fiscal deficits, 
both having the potential to inadvertently contribute to longer transmission lags.

\pagebreak
\bibliographystyle{aernoextra}
\bibliography{main}

@article{afrouzi2023overreaction,
  title={Overreaction in expectations: Evidence and theory},
  author={Afrouzi, Hassan and Kwon, Spencer Y and Landier, Augustin and Ma, Yueran and Thesmar, David},
  journal={The Quarterly Journal of Economics},
  volume={138},
  number={3},
  pages={1713--1764},
  year={2023},
  publisher={Oxford University Press}
}

@article{aiyagari1994uninsured,
  title={Uninsured idiosyncratic risk and aggregate saving},
  author={Aiyagari, S Rao},
  journal={The Quarterly Journal of Economics},
  volume={109},
  number={3},
  pages={659--684},
  year={1994},
  publisher={MIT Press}
}

@article{andre2022subjective,
  title={Subjective models of the macroeconomy: Evidence from experts and representative samples},
  author={Andre, Peter and Pizzinelli, Carlo and Roth, Christopher and Wohlfart, Johannes},
  journal={The Review of Economic Studies},
  volume={89},
  number={6},
  pages={2958--2991},
  year={2022},
  publisher={Oxford University Press}
}

@article{angeletos2018forward,
  title={Forward guidance without common knowledge},
  author={Angeletos, George-Marios and Lian, Chen},
  journal={American Economic Review},
  volume={108},
  number={9},
  pages={2477--2512},
  year={2018},
  publisher={American Economic Association 2014 Broadway, Suite 305, Nashville, TN 37203}
}

@article{angeletos2020business,
  title={Business-cycle anatomy},
  author={Angeletos, George-Marios and Collard, Fabrice and Dellas, Harris},
  journal={American Economic Review},
  volume={110},
  number={10},
  pages={3030--3070},
  year={2020},
  publisher={American Economic Association 2014 Broadway, Suite 305, Nashville, TN 37203}
}

@article{angeletos2021imperfect,
  title={Imperfect macroeconomic expectations: Evidence and theory},
  author={Angeletos, George-Marios and Huo, Zhen and Sastry, Karthik A},
  journal={NBER Macroeconomics Annual},
  volume={35},
  number={1},
  pages={1--86},
  year={2021},
  publisher={The University of Chicago Press Chicago, IL}
}

@article{angeletos2021myopia,
  title={Myopia and anchoring},
  author={Angeletos, George-Marios and Huo, Zhen},
  journal={American Economic Review},
  volume={111},
  number={4},
  pages={1166--1200},
  year={2021},
  publisher={American Economic Association 2014 Broadway, Suite 305, Nashville, TN 37203}
}

@techreport{angeletos2023can,
  title={Can Deficits Finance Themselves?},
  author={Angeletos, George-Marios and Lian, Chen and Wolf, Christian K},
  year={2023},
  institution={National Bureau of Economic Research}
}

@techreport{auclert2020micro,
  title={Micro jumps, macro humps: Monetary policy and business cycles in an estimated HANK model},
  author={Auclert, Adrien and Rognlie, Matthew and Straub, Ludwig},
  year={2020},
  institution={National Bureau of Economic Research}
}

@article{auclert2021using,
  title={Using the sequence-space Jacobian to solve and estimate heterogeneous-agent models},
  author={Auclert, Adrien and Bard{\'o}czy, Bence and Rognlie, Matthew and Straub, Ludwig},
  journal={Econometrica},
  volume={89},
  number={5},
  pages={2375--2408},
  year={2021},
  publisher={Wiley Online Library}
}

@article{auclert2024intertemporal,
  title={The Intertemporal Keynesian Cross},
  author={Auclert, Adrien and Rognlie, Matthew and Straub, Ludwig},
  journal={Journal of Political Economy},
  year={2024},
}

@article{azeredo2024optimally,
  title={Optimally imprecise memory and biased forecasts},
  author={Azeredo da Silveira, Rava and Sung, Yeji and Woodford, Michael},
  journal={American Economic Review},
  volume={114},
  number={10},
  pages={3075--3118},
  year={2024},
  publisher={American Economic Association 2014 Broadway, Suite 305, Nashville, TN 37203}
}

@article{bardoczy2023unemployment,
  title={Unemployment insurance in macroeconomic stabilization with imperfect expectations},
  author={Bard{\'o}czy, Bence and Guerreiro, Joao},
  journal={Manuscript, April},
  year={2023}
}

@article{barnichon2019impulse,
  title={Impulse response estimation by smooth local projections},
  author={Barnichon, Regis and Brownlees, Christian},
  journal={Review of Economics and Statistics},
  volume={101},
  number={3},
  pages={522--530},
  year={2019},
  publisher={MIT Press One Rogers Street, Cambridge, MA 02142-1209, USA journals-info~…}
}

@article{barnichon2020identifying,
  title={Identifying modern macro equations with old shocks},
  author={Barnichon, Regis and Mesters, Geert},
  journal={The Quarterly Journal of Economics},
  volume={135},
  number={4},
  pages={2255--2298},
  year={2020},
  publisher={Oxford University Press}
}

@article{bastianello2025expectations,
  title={Expectations and learning from prices},
  author={Bastianello, Francesca and Fontanier, Paul},
  journal={Review of Economic Studies},
  volume={92},
  number={3},
  pages={1341--1374},
  year={2025},
  publisher={Oxford University Press UK}
}

@article{bayer2024shocks,
  title={Shocks, frictions, and inequality in US business cycles},
  author={Bayer, Christian and Born, Benjamin and Luetticke, Ralph},
  journal={American Economic Review},
  volume={114},
  number={5},
  pages={1211--1247},
  year={2024},
  publisher={American Economic Association 2014 Broadway, Suite 305, Nashville, TN 37203}
}

@article{bernanke1997systematic,
  title={Systematic monetary policy and the effects of oil price shocks},
  author={Bernanke, Ben S and Gertler, Mark and Watson, Mark and Sims, Christopher A and Friedman, Benjamin M},
  journal={Brookings papers on economic activity},
  volume={1997},
  number={1},
  pages={91--157},
  year={1997},
  publisher={JSTOR}
}

@article{best2020estimating,
  title={Estimating the elasticity of intertemporal substitution using mortgage notches},
  author={Best, Michael Carlos and Cloyne, James S and Ilzetzki, Ethan and Kleven, Henrik J},
  journal={The Review of Economic Studies},
  volume={87},
  number={2},
  pages={656--690},
  year={2020},
  publisher={Oxford University Press}
}

@article{bewley1986stationary,
  title={Stationary monetary equilibrium with a continuum of independently fluctuating consumers},
  author={Bewley, Truman},
  journal={Contributions to mathematical economics in honor of G{\'e}rard Debreu},
  volume={79},
  year={1986},
  publisher={North-Holland Amsterdam}
}

@article{blanchard1985olg,
  title={Debt, deficits, and finite horizons},
  author={Blanchard, Olivier J},
  journal={Journal of Political Economy},
  volume={93},
  pages={223-247},
  year={1985},
}

@article{blanchard2007real,
  title={Real wage rigidities and the New Keynesian model},
  author={Blanchard, Olivier and Gal{\'\i}, Jordi},
  journal={Journal of money, credit and banking},
  volume={39},
  pages={35--65},
  year={2007},
  publisher={Wiley Online Library}
}

@article{bordalo2018diagnostic,
  title={Diagnostic expectations and credit cycles},
  author={Bordalo, Pedro and Gennaioli, Nicola and Shleifer, Andrei},
  journal={The Journal of Finance},
  volume={73},
  number={1},
  pages={199--227},
  year={2018},
  publisher={Wiley Online Library}
}

@article{bordalo2020overreaction,
  title={Overreaction in macroeconomic expectations},
  author={Bordalo, Pedro and Gennaioli, Nicola and Ma, Yueran and Shleifer, Andrei},
  journal={American Economic Review},
  volume={110},
  number={9},
  pages={2748--2782},
  year={2020},
  publisher={American Economic Association 2014 Broadway, Suite 305, Nashville, TN 37203}
}

@techreport{candia2020communication,
  title={Communication and the beliefs of economic agents},
  author={Candia, Bernardo and Coibion, Olivier and Gorodnichenko, Yuriy},
  year={2020},
  institution={National Bureau of Economic Research}
}

@article{carroll2020sticky,
  title={Sticky expectations and consumption dynamics},
  author={Carroll, Christopher D and Crawley, Edmund and Slacalek, Jiri and Tokuoka, Kiichi and White, Matthew N},
  journal={American economic journal: macroeconomics},
  volume={12},
  number={3},
  pages={40--76},
  year={2020},
  publisher={American Economic Association 2014 Broadway, Suite 305, Nashville, TN 37203-2425}
}

@article{chen2025expectation,
  title={Expectation and Confusion: Evidence and Theory},
  author={Chen, Heng and Liu, Yicheng},
  journal={Available at SSRN 5343753},
  year={2025}
}

@article{chetty2016consumption,
  title={Consumption commitments and habit formation},
  author={Chetty, Raj and Szeidl, Adam},
  journal={Econometrica},
  volume={84},
  number={2},
  pages={855--890},
  year={2016},
  publisher={Wiley Online Library}
}

@article{christiano2005nominal,
  title={Nominal rigidities and the dynamic effects of a shock to monetary policy},
  author={Christiano, Lawrence J and Eichenbaum, Martin and Evans, Charles L},
  journal={Journal of Political Economy},
  volume={113},
  number={1},
  pages={1--45},
  year={2005},
  publisher={The University of Chicago Press}
}

@techreport{christiano2020anchoring,
  title={Anchoring Inflation Expectations},
  author={Christiano, Lawrence and Takahashi, Yuta},
  year={2020},
  institution={Mimeo}
}

@article{christiano2024slow,
  title={Slow Learning},
  author={Christiano, Lawrence J and Eichenbaum, Martin and Johannsen, Benjamin K},
  year={2024}
}

@article{clarida1998monetary,
  title={Monetary policy rules in practice: Some international evidence},
  author={Clarida, Richard and Gal{\i}, Jordi and Gertler, Mark},
  journal={European economic review},
  volume={42},
  number={6},
  pages={1033--1067},
  year={1998},
  publisher={Elsevier}
}

@incollection{crump2023term,
  title={The term structure of expectations},
  author={Crump, Richard K and Eusepi, Stefano and Moench, Emanuel and Preston, Bruce},
  booktitle={Handbook of economic expectations},
  pages={507--540},
  year={2023},
  publisher={Elsevier}
}

@article{del2011fitting,
  title={Fitting observed inflation expectations},
  author={Del Negro, Marco and Eusepi, Stefano},
  journal={Journal of Economic Dynamics and control},
  volume={35},
  number={12},
  pages={2105--2131},
  year={2011},
  publisher={Elsevier}
}

@techreport{desilva2024selective,
  title={Selective Inattention},
  author={De Silva, Tim and Mei, Pierfrancesco},
  year={2024}
}

@article{dynan2000habit,
  title={Habit formation in consumer preferences: Evidence from panel data},
  author={Dynan, Karen E},
  journal={American Economic Review},
  volume={90},
  number={3},
  pages={391--406},
  year={2000},
  publisher={American Economic Association}
}

@article{evans1999learning,
  title={Learning dynamics},
  author={Evans, George W and Honkapohja, Seppo},
  journal={Handbook of macroeconomics},
  volume={1},
  pages={449--542},
  year={1999},
  publisher={Elsevier}
}

@article{eusepi2011expectations,
  title={Expectations, learning, and business cycle fluctuations},
  author={Eusepi, Stefano and Preston, Bruce},
  journal={American Economic Review},
  volume={101},
  number={6},
  pages={2844--2872},
  year={2011},
  publisher={American Economic Association}
}

@article{eusepi2024shortrun,
  title={The Short-run Policy Constraints of Long-run Expectations},
  author={Eusepi, Stefano and Giannoni, Marc and Preston, Bruce},
  journal={Journal of Political Economy},
  year={2024},
}

@article{farmer2024learning,
  title={Learning about the long run},
  author={Farmer, Leland E and Nakamura, Emi and Steinsson, J{\'o}n},
  journal={Journal of Political Economy},
  volume={132},
  number={10},
  pages={000--000},
  year={2024},
  publisher={The University of Chicago Press Chicago, IL}
}

@article{fuhrer2000habit,
  title={Habit formation in consumption and its implications for monetary-policy models},
  author={Fuhrer, Jeffrey C},
  journal={American economic review},
  volume={90},
  number={3},
  pages={367--390},
  year={2000},
  publisher={American Economic Association}
}

@incollection{gabaix2019behavioral,
  title={Behavioral inattention},
  author={Gabaix, Xavier},
  booktitle={Handbook of behavioral economics: Applications and foundations 1},
  volume={2},
  pages={261--343},
  year={2019},
  publisher={Elsevier}
}

@article{gabaix2020behavioral,
  title={A behavioral New Keynesian model},
  author={Gabaix, Xavier},
  journal={American Economic Review},
  volume={110},
  number={8},
  pages={2271--2327},
  year={2020},
  publisher={American Economic Association 2014 Broadway, Suite 305, Nashville, TN 37203}
}

@article{gagliardone2023oil,
  title={Oil prices, monetary policy and inflation surges},
  author={Gagliardone, Luca and Gertler, Mark},
  year={2023},
  publisher={National Bureau of Economic Research Cambridge, MA}
}

@book{gali2015monetary,
  title={Monetary Policy, Inflation, and the Business Cycle},
  author={Gal{\'i}, Jordi},
  year={2015},
  publisher={Princeton University Press}
}

@article{guerreiro2023belief,
  title={Belief disagreement and business cycles},
  author={Guerreiro, Joao},
  journal={Northwestern University manuscript},
  year={2023}
}

@article{hansen1982generalized,
  title={Generalized instrumental variables estimation of nonlinear rational expectations models},
  author={Hansen, Lars Peter and Singleton, Kenneth J},
  journal={Econometrica: Journal of the Econometric Society},
  pages={1269--1286},
  year={1982},
  publisher={JSTOR}
}

@article{havranek2017habit,
  title={Habit formation in consumption: A meta-analysis},
  author={Havranek, Tomas and Rusnak, Marek and Sokolova, Anna},
  journal={European economic review},
  volume={95},
  pages={142--167},
  year={2017},
  publisher={Elsevier}
}

@article{huggett1993risk,
  title={The risk-free rate in heterogeneous-agent incomplete-insurance economies},
  author={Huggett, Mark},
  journal={Journal of economic Dynamics and Control},
  volume={17},
  number={5-6},
  pages={953--969},
  year={1993},
  publisher={Elsevier}
}

@article{imrohorouglu1989cost,
  title={Cost of business cycles with indivisibilities and liquidity constraints},
  author={Imrohoro{\u{g}}lu, Ay{\c{s}}e},
  journal={Journal of Political economy},
  volume={97},
  number={6},
  pages={1364--1383},
  year={1989},
  publisher={The University of Chicago Press}
}

@article{jentsch2019dynamic,
  title={The dynamic effects of personal and corporate income tax changes in the United States: Comment},
  author={Jentsch, Carsten and Lunsford, Kurt G},
  journal={American Economic Review},
  volume={109},
  number={7},
  pages={2655--2678},
  year={2019},
  publisher={American Economic Association 2014 Broadway, Suite 305, Nashville, TN 37203}
}

@article{kamdar2023supplyside,
  title={The effects of news shocks and supply-side beliefs},
  author={Kamdar, Rupal and Ray, Walker},
  year={2023}
}

@article{kanzig2021macroeconomic,
  title={The macroeconomic effects of oil supply news: Evidence from OPEC announcements},
  author={K{\"a}nzig, Diego R},
  journal={American Economic Review},
  volume={111},
  number={4},
  pages={1092--1125},
  year={2021},
  publisher={American Economic Association 2014 Broadway, Suite 305, Nashville, TN 37203}
}

@article{kaplan2022marginal,
  title={The marginal propensity to consume in heterogeneous agent models},
  author={Kaplan, Greg and Violante, Giovanni L},
  journal={Annual Review of Economics},
  volume={14},
  number={1},
  pages={747--775},
  year={2022},
  publisher={Annual Reviews}
}

@incollection{kosar2023expectations,
  title={Expectations data in structural microeconomic models},
  author={Kosar, Gizem and O'Dea, Cormac},
  booktitle={Handbook of Economic Expectations},
  pages={647--675},
  year={2023},
  publisher={Elsevier}
}

@article{kurmann2021revisions,
  title={Revisions in utilization-adjusted TFP and robust identification of news shocks},
  author={Kurmann, Andr{\'e} and Sims, Eric},
  journal={Review of Economics and Statistics},
  volume={103},
  number={2},
  pages={216--235},
  year={2021},
  publisher={Mit Press One Rogers Street, Cambridge, MA 02142-1209, USA journals-info~…}
}

@article{kuvcinskas2022measuring,
  title={Measuring under-and overreaction in expectation formation},
  author={Ku{\v{c}}inskas, Simas and Peters, Florian S},
  journal={Review of Economics and Statistics},
  pages={1--45},
  year={2022},
  publisher={MIT Press One Rogers Street, Cambridge, MA 02142-1209, USA journals-info~…}
}

@article{lewis2022dynamic,
  title={Dynamic Identification Using System Projections and Instrumental Variables},
  author={Lewis, Daniel J and Mertens, Karel},
  year={2022},
  publisher={CEPR Discussion Paper No. DP17153}
}

@article{luo2008consumption,
  title={Consumption dynamics under information processing constraints},
  author={Luo, Yulei},
  journal={Review of Economic dynamics},
  volume={11},
  number={2},
  pages={366--385},
  year={2008},
  publisher={Elsevier}
}

@article{mackowiak2015business,
  title={Business cycle dynamics under rational inattention},
  author={Ma{\'c}kowiak, Bartosz and Wiederholt, Mirko},
  journal={The Review of Economic Studies},
  volume={82},
  number={4},
  pages={1502--1532},
  year={2015},
  publisher={Oxford University Press}
}

@article{mankiw2002sticky,
  title={Sticky information versus sticky prices: a proposal to replace the New Keynesian Phillips curve},
  author={Mankiw, N Gregory and Reis, Ricardo},
  journal={The Quarterly Journal of Economics},
  volume={117},
  number={4},
  pages={1295--1328},
  year={2002},
  publisher={MIT Press}
}

@article{manski2004measuring,
  title={Measuring expectations},
  author={Manski, Charles F},
  journal={Econometrica},
  volume={72},
  number={5},
  pages={1329--1376},
  year={2004},
  publisher={Wiley Online Library}
}

@article{manski2018survey,
  title={Survey measurement of probabilistic macroeconomic expectations: progress and promise},
  author={Manski, Charles F},
  journal={NBER Macroeconomics Annual},
  volume={32},
  number={1},
  pages={411--471},
  year={2018},
  publisher={University of Chicago Press Chicago, IL}
}

@article{mckay2016power,
  title={The power of forward guidance revisited},
  author={McKay, Alisdair and Nakamura, Emi and Steinsson, J{\'o}n},
  journal={American Economic Review},
  volume={106},
  number={10},
  pages={3133--3158},
  year={2016},
  publisher={American Economic Association 2014 Broadway, Suite 305, Nashville, TN 37203}
}

@article{mertens2013dynamic,
  title={The dynamic effects of personal and corporate income tax changes in the United States},
  author={Mertens, Karel and Ravn, Morten O},
  journal={American economic review},
  volume={103},
  number={4},
  pages={1212--1247},
  year={2013},
  publisher={American Economic Association}
}

@article{milani2011expectation,
  title={Expectation shocks and learning as drivers of the business cycle},
  author={Milani, Fabio},
  journal={The Economic Journal},
  volume={121},
  number={552},
  pages={379--401},
  year={2011},
  publisher={Oxford University Press Oxford, UK}
}

@article{molavi2022simple,
  title={Simple models and biased forecasts},
  author={Molavi, Pooya},
  journal={arXiv preprint arXiv:2202.06921},
  year={2022}
}

@article{mountford2009effects,
  title={What are the effects of fiscal policy shocks?},
  author={Mountford, Andrew and Uhlig, Harald},
  journal={Journal of applied econometrics},
  volume={24},
  number={6},
  pages={960--992},
  year={2009},
  publisher={Wiley Online Library}
}

@techreport{nagel2024lean,
  title={Leaning Against Inflation Experiences},
  author={Nagel, Stefan},
  year={2024}
}

@article{newey1994large,
  title={Large sample estimation and hypothesis testing},
  author={Newey, Whitney K and McFadden, Daniel},
  journal={Handbook of econometrics},
  volume={4},
  pages={2111--2245},
  year={1994},
  publisher={Elsevier}
}

@article{piazzesi2009momentum,
  title={Momentum traders in the housing market: Survey evidence and a search model},
  author={Piazzesi, Monika and Schneider, Martin},
  journal={American Economic Review},
  volume={99},
  number={2},
  pages={406--411},
  year={2009},
  publisher={American Economic Association}
}

@article{ramey2011identifying,
  title={Identifying government spending shocks: It's all in the timing},
  author={Ramey, Valerie A},
  journal={The Quarterly Journal of Economics},
  volume={126},
  number={1},
  pages={1--50},
  year={2011},
  publisher={MIT Press}
}

@article{ring2024wealth,
  title={Wealth taxation and household saving: Evidence from assessment discontinuities in Norway},
  author={Ring, Marius Alexander Kalleberg},
  journal={Review of economic studies},
  year={2024}
}

@article{romer2004new,
  title={A new measure of monetary shocks: Derivation and implications},
  author={Romer, Christina D and Romer, David H},
  journal={American economic review},
  volume={94},
  number={4},
  pages={1055--1084},
  year={2004},
  publisher={American Economic Association}
}

@article{rozsypal2023overpersistence,
  title={Overpersistence bias in individual income expectations and its aggregate implications},
  author={Rozsypal, Filip and Schlafmann, Kathrin},
  journal={American Economic Journal: Macroeconomics},
  volume={15},
  number={4},
  pages={331--371},
  year={2023},
  publisher={American Economic Association 2014 Broadway, Suite 305, Nashville, TN 37203-2425}
}

@article{rudebusch2005monetary,
  title={Monetary policy inertia: fact or fiction?},
  author={Rudebusch, Glenn D},
  journal={FRB of San Francisco Working Paper},
  number={2005-19},
  year={2005}
}

@article{sack1998uncertainty,
  title={Uncertainty, learning, and gradual monetary policy},
  author={Sack, Brian P},
  journal={Available at SSRN 121439},
  year={1998}
}

@article{sims2003implications,
  title={Implications of rational inattention},
  author={Sims, Christopher A},
  journal={Journal of monetary Economics},
  volume={50},
  number={3},
  pages={665--690},
  year={2003},
  publisher={Elsevier}
}

@techreport{stock2012disentangling,
  title={Disentangling the Channels of the 2007-2009 Recession},
  author={Stock, James H and Watson, Mark W},
  year={2012},
  institution={National Bureau of Economic Research}
}

@article{williams2003adaptive,
  title={Adaptive learning and business cycles},
  author={Williams, Noah},
  journal={Manuscript, Princeton University},
  year={2003}
}

@article{woodford1999optimal,
  title={Optimal monetary policy inertia},
  author={Woodford, Michael},
  journal={The Manchester School},
  volume={67},
  pages={1--35},
  year={1999},
  publisher={Wiley Online Library}
}

@misc{woodford2001imperfect,
  title={Imperfect common knowledge and the effects of monetary policy},
  author={Woodford, Michael},
  year={2001},
  publisher={National Bureau of Economic Research Cambridge, Mass., USA}
}

@article{woodford2013macroeconomic,
  title={Macroeconomic analysis without the rational expectations hypothesis},
  author={Woodford, Michael},
  journal={Annu. Rev. Econ.},
  volume={5},
  number={1},
  pages={303--346},
  year={2013},
  publisher={Annual Reviews}
}

@article{yaari1965uncertain,
  title={Uncertain lifetime, life insurance, and the theory of the consumer},
  author={Yaari, Menahem E},
  journal={The Review of Economic Studies},
  volume={32},
  number={2},
  pages={137--150},
  year={1965},
  publisher={Wiley-Blackwell}
}

\pagebreak

\appendix
\section{Additional impulse response results}

\subsection{Impulse response comparisons with confidence bands}

\begin{figure}[H]
    {
    \centering

    \begin{subfigure}{0.495\textwidth}
        \centering
        \includegraphics[width=\linewidth]{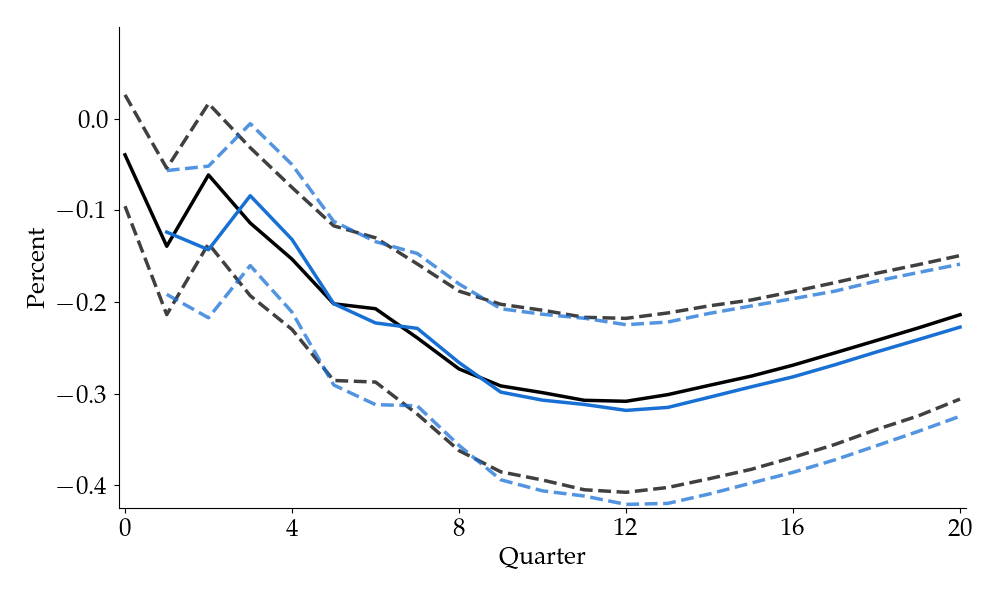}
        \caption{One-quarter ahead Bluechip expectation}
    \end{subfigure}
    \hfill
    \begin{subfigure}{0.495\textwidth}
        \centering
        \includegraphics[width=\linewidth]{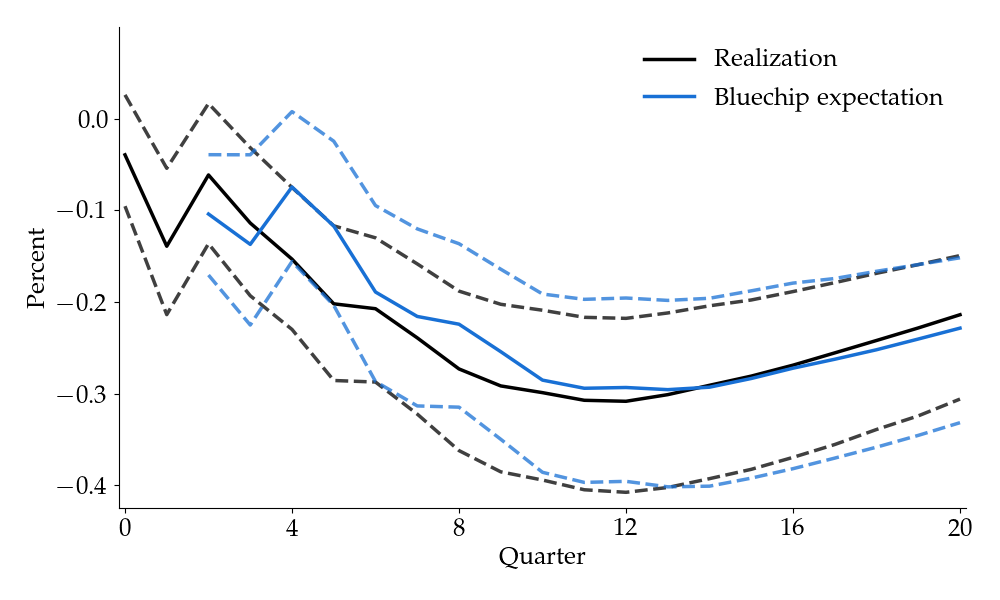}
        \caption{Two-quarters ahead Bluechip expectation}
    \end{subfigure}

    \vspace{1em} 

    \begin{subfigure}{0.495\textwidth}
        \centering
        \includegraphics[width=\linewidth]{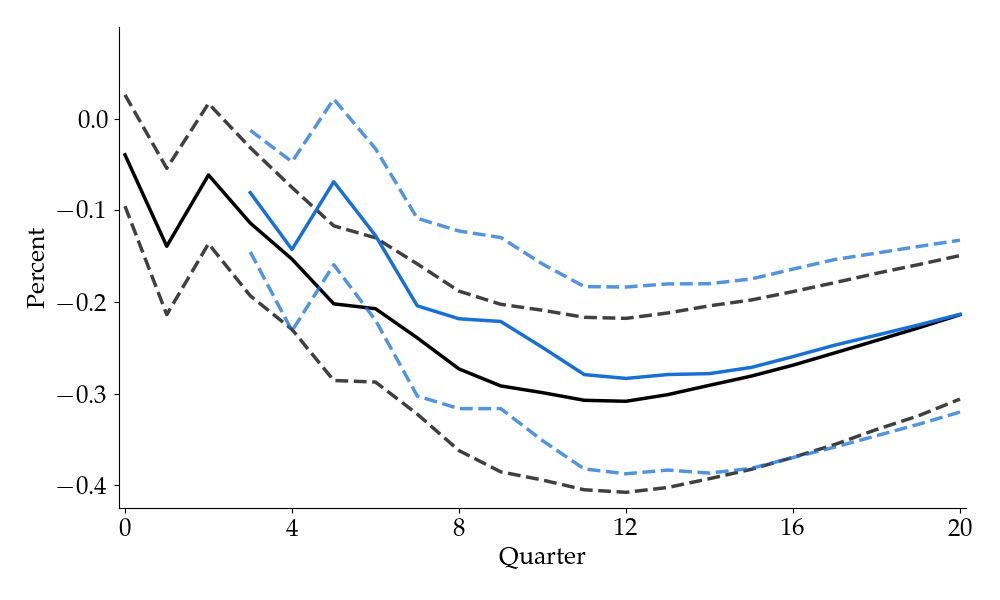}
        \caption{Three-quarters ahead Bluechip expectation}
    \end{subfigure}
    \hfill
    \begin{subfigure}{0.495\textwidth}
        \centering
        \includegraphics[width=\linewidth]{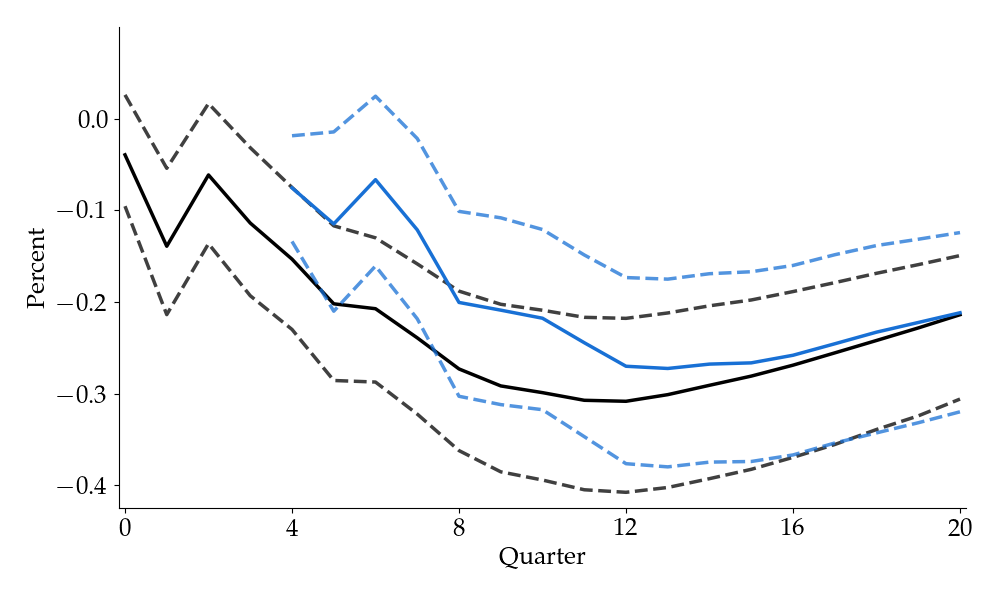}
        \caption{Four-quarters ahead Bluechip expectation}
    \end{subfigure}

    \caption{Real disposable income impulse responses to a \cite{kanzig2021macroeconomic} oil shock}
    }

    \caption*{
    \footnotesize \emph{Note}: each panel contains an impulse response function of realizations (black) 
    and a fixed horizon-$h$ (blue) forecast from Bluechip survey expectations data 
    to a positive \cite{kanzig2021macroeconomic} oil price news shock.
    The dashed lines are 68\% confidence bands produced using moving block bootstrap by
    \cite{jentsch2019dynamic}.
    }
\end{figure}

\begin{figure}[H]
    {
    \centering

    \begin{subfigure}{0.495\textwidth}
        \centering
        \includegraphics[width=\linewidth]{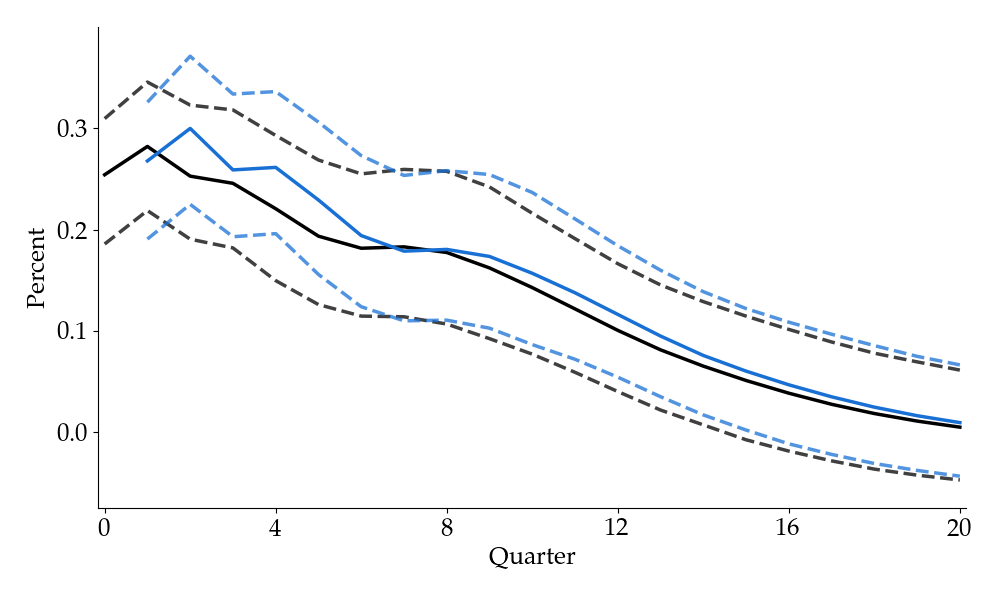}
        \caption{One-quarter ahead Bluechip expectation}
    \end{subfigure}
    \hfill
    \begin{subfigure}{0.495\textwidth}
        \centering
        \includegraphics[width=\linewidth]{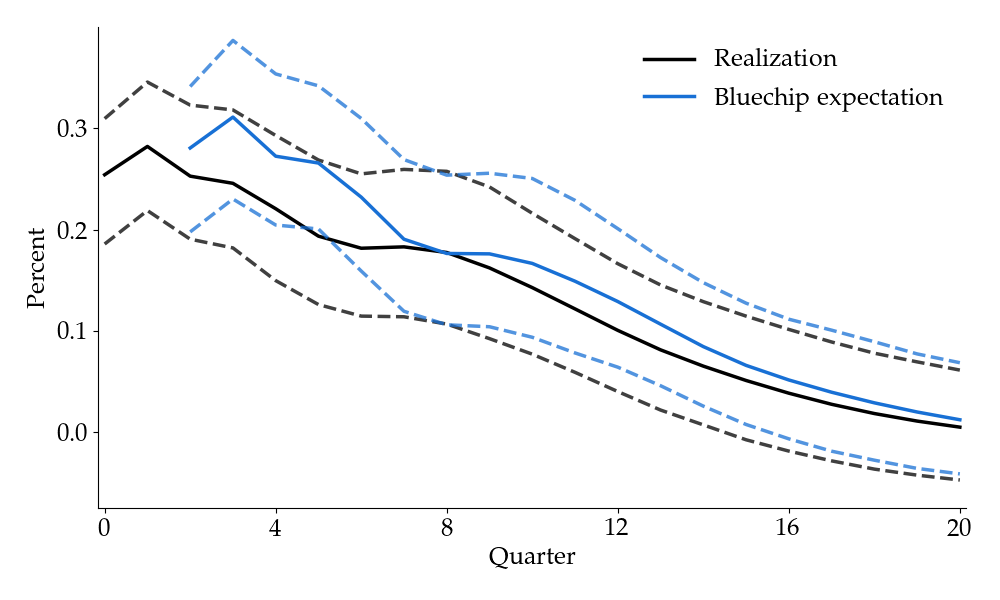}
        \caption{Two-quarters ahead Bluechip expectation}
    \end{subfigure}

    \vspace{1em} 

    \begin{subfigure}{0.495\textwidth}
        \centering
        \includegraphics[width=\linewidth]{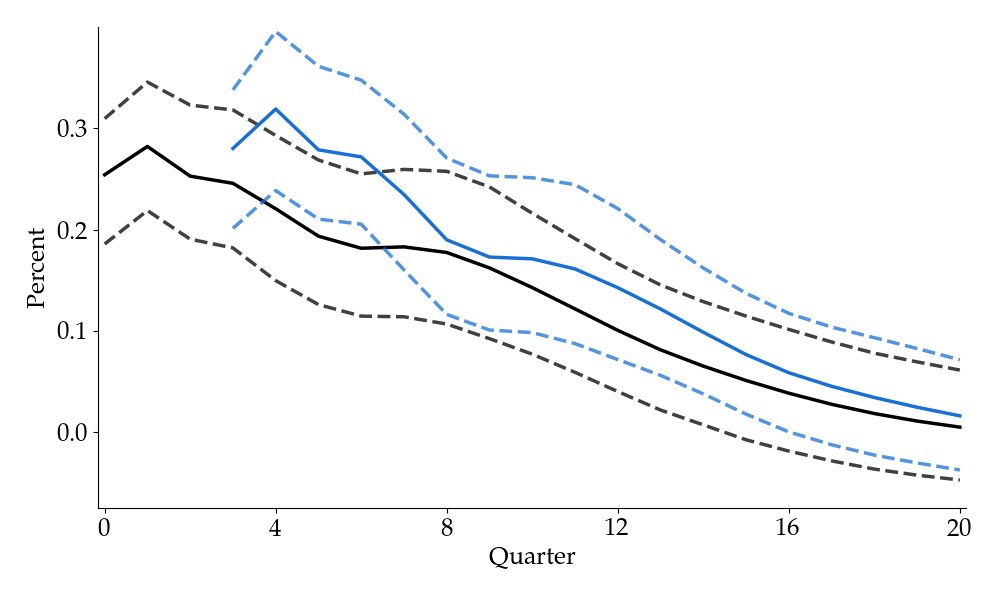}
        \caption{Three-quarters ahead Bluechip expectation}
    \end{subfigure}
    \hfill
    \begin{subfigure}{0.495\textwidth}
        \centering
        \includegraphics[width=\linewidth]{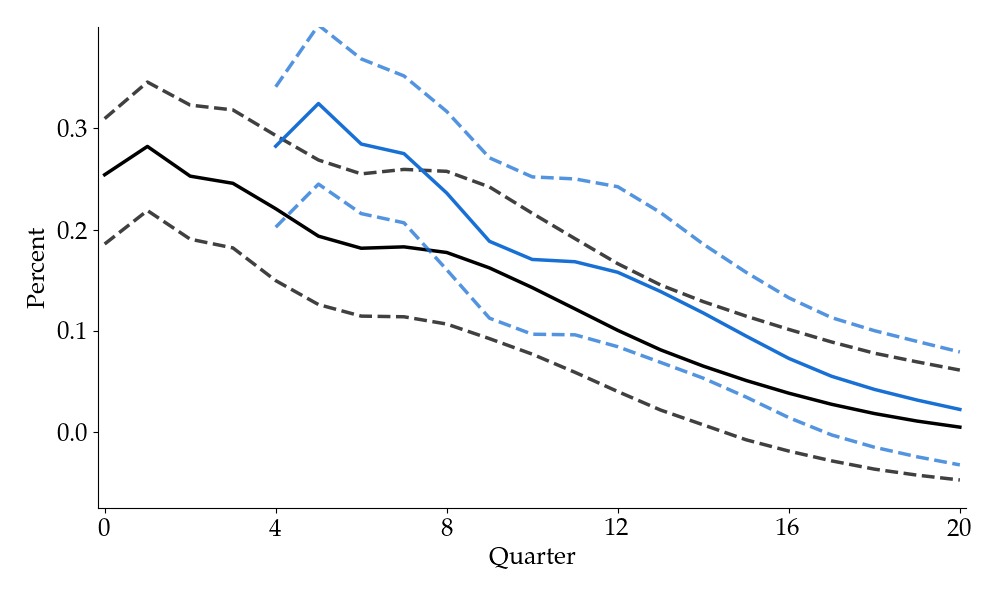}
        \caption{Four-quarters ahead Bluechip expectation}
    \end{subfigure}

    \caption{CPI inflation impulse responses to a \cite{kanzig2021macroeconomic} oil shock}
    }

    \caption*{
    \footnotesize \emph{Note}: each panel contains an impulse response function of realizations (black) 
    and a fixed horizon-$h$ (blue) forecast from Bluechip survey expectations data 
    to a positive \cite{kanzig2021macroeconomic} oil price news shock.
    The dashed lines are 68\% confidence bands produced using moving block bootstrap by
    \cite{jentsch2019dynamic}.
    }
\end{figure}

\begin{figure}[H]
    {
    \centering

    \begin{subfigure}{0.495\textwidth}
        \centering
        \includegraphics[width=\linewidth]{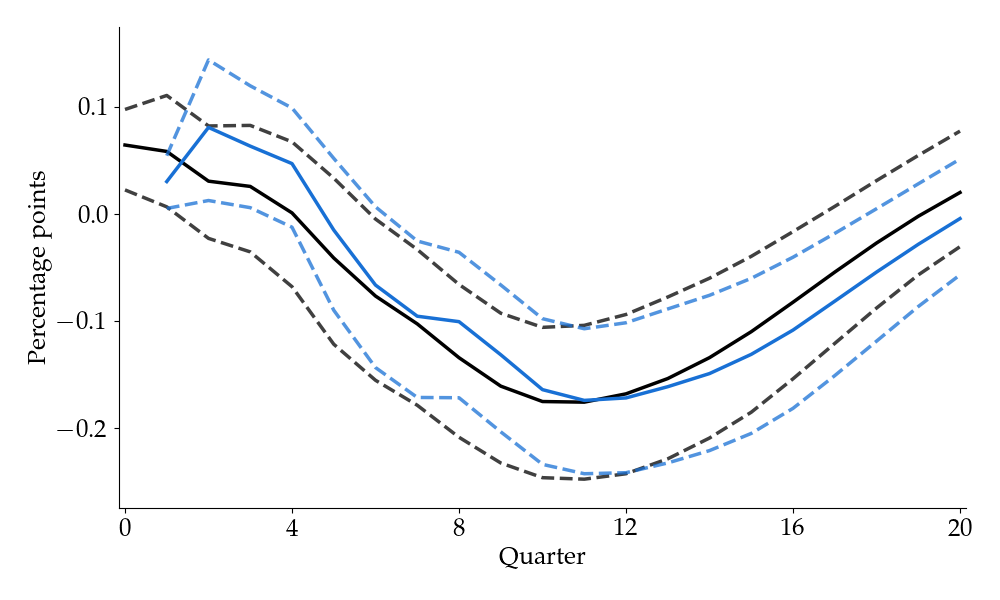}
        \caption{One-quarter ahead Bluechip expectation}
    \end{subfigure}
    \hfill
    \begin{subfigure}{0.495\textwidth}
        \centering
        \includegraphics[width=\linewidth]{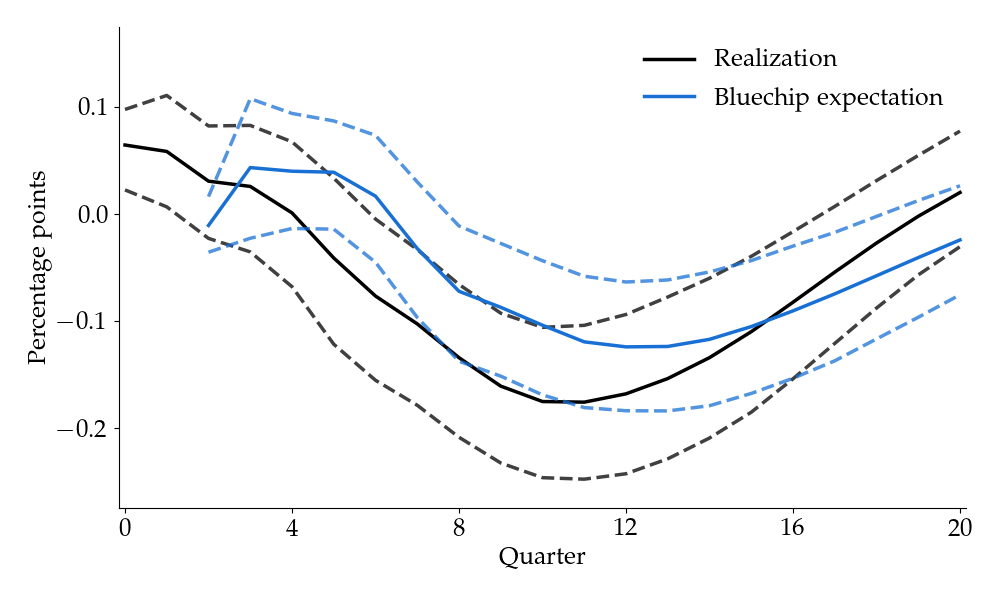}
        \caption{Two-quarters ahead Bluechip expectation}
    \end{subfigure}

    \vspace{1em} 

    \begin{subfigure}{0.495\textwidth}
        \centering
        \includegraphics[width=\linewidth]{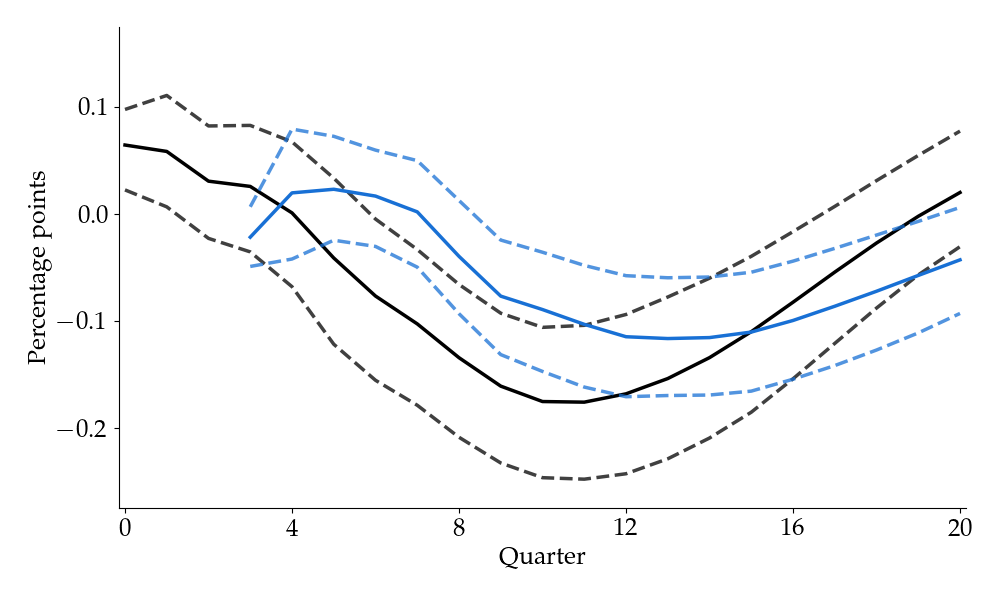}
        \caption{Three-quarters ahead Bluechip expectation}
    \end{subfigure}
    \hfill
    \begin{subfigure}{0.495\textwidth}
        \centering
        \includegraphics[width=\linewidth]{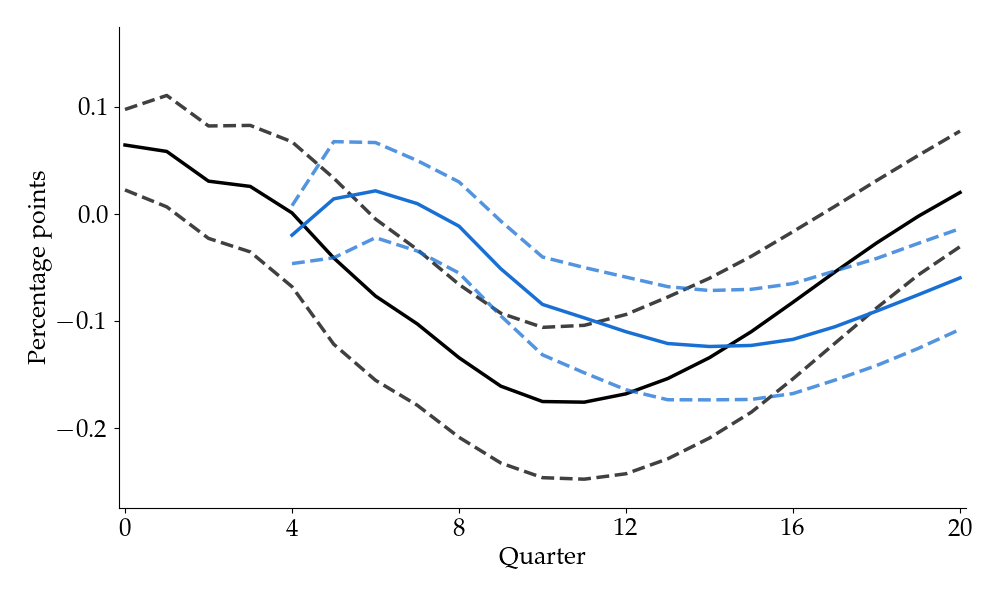}
        \caption{Four-quarters ahead Bluechip expectation}
    \end{subfigure}

    \caption{Nominal federal funds rate impulse responses to a \cite{kanzig2021macroeconomic} oil shock}
    }

    \caption*{
    \footnotesize \emph{Note}: each panel contains an impulse response function of realizations (black) 
    and a fixed horizon-$h$ (blue) forecast from Bluechip survey expectations data 
    to a positive \cite{kanzig2021macroeconomic} oil price news shock.
    The dashed lines are 68\% confidence bands produced using moving block bootstrap by
    \cite{jentsch2019dynamic}.
    }
\end{figure}

\subsection{Other models of expectation formation}\label{asubsec:other_models_exp_formation}

\begin{figure}[H]
    {
    \centering

    \begin{subfigure}{0.495\textwidth}
        \centering
        \includegraphics[width=\linewidth]{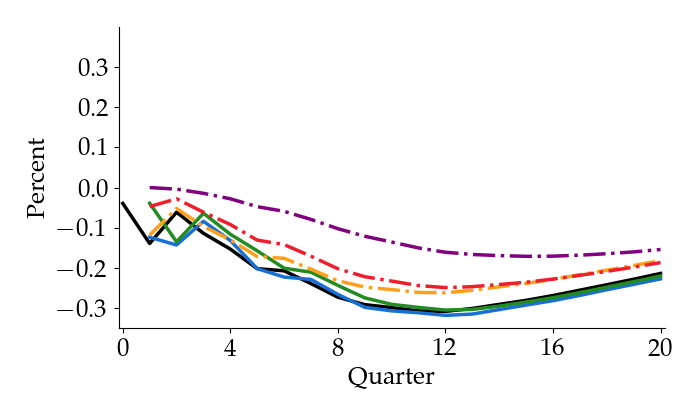}
        \caption{Real disposable income, one-quarter ahead exp.}
    \end{subfigure}
    \hfill
    \begin{subfigure}{0.495\textwidth}
        \centering
        \includegraphics[width=\linewidth]{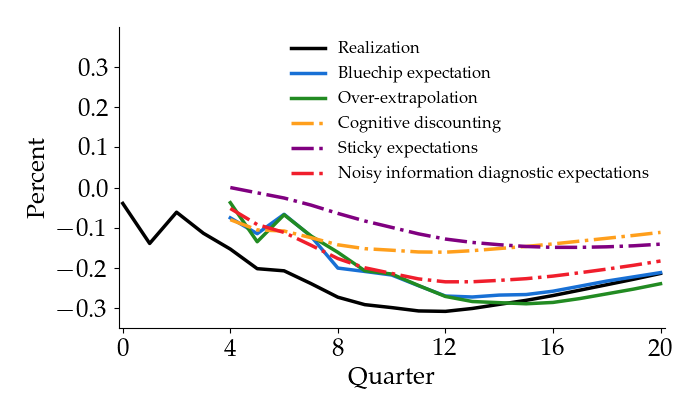}
        \caption{Real disposable income, four-quarters ahead exp.}
    \end{subfigure}

    \vspace{1em} 

    \begin{subfigure}{0.495\textwidth}
        \centering
        \includegraphics[width=\linewidth]{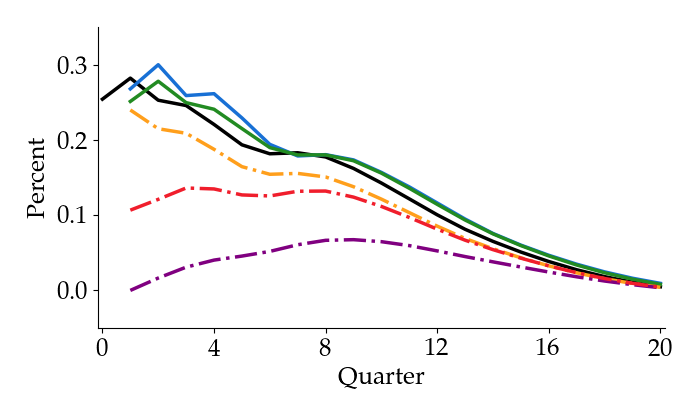}
        \caption{CPI inflation, one-quarter ahead exp.}
    \end{subfigure}
    \hfill
    \begin{subfigure}{0.495\textwidth}
        \centering
        \includegraphics[width=\linewidth]{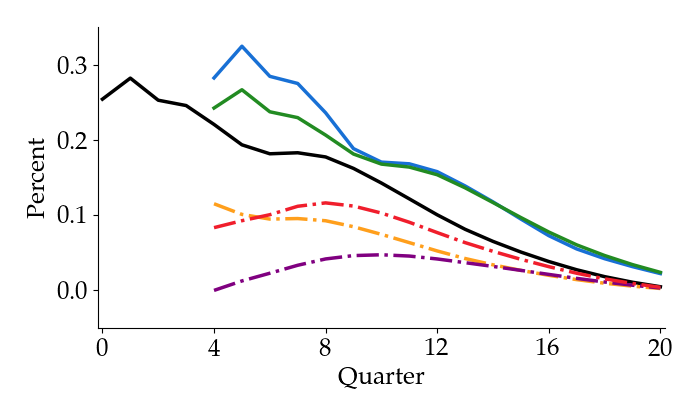}
        \caption{CPI inflation, four-quarters ahead exp.}
    \end{subfigure}

    \vspace{1em} 

    \begin{subfigure}{0.495\textwidth}
        \centering
        \includegraphics[width=\linewidth]{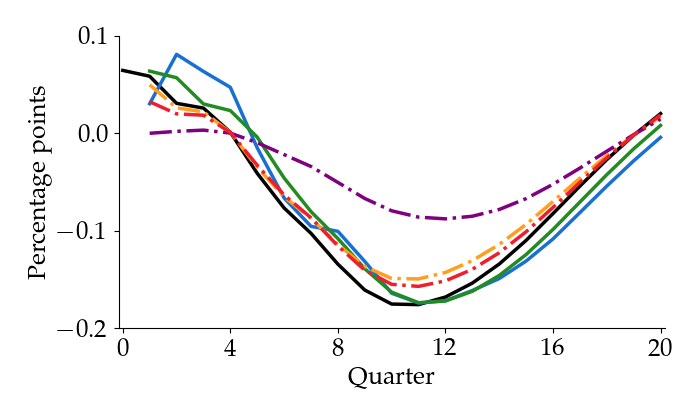}
        \caption{Nominal Fed funds rate, one-quarter ahead exp.}
    \end{subfigure}
    \hfill
    \begin{subfigure}{0.495\textwidth}
        \centering
        \includegraphics[width=\linewidth]{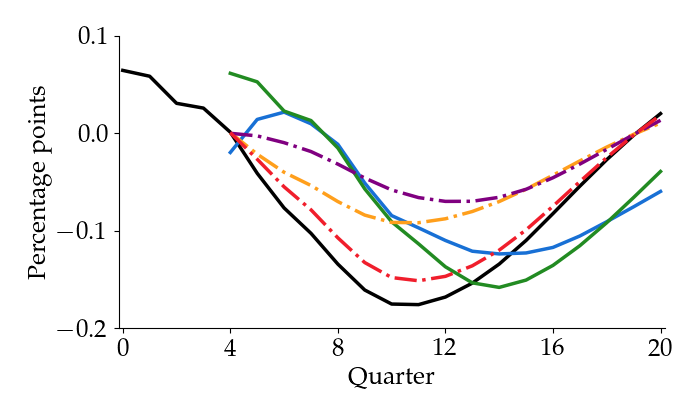}
        \caption{Nominal Fed funds rate, four-quarters ahead exp.}
    \end{subfigure}

    \caption{Expectation data and model impulse response comparisons}
    \label{fig:exp_irfs_model_comp}
    }

    \caption*{
    \footnotesize \emph{Note}: each panel contains impulse response functions of realizations (black) 
    and expectations data or model expectations (color) to a positive \cite{kanzig2021macroeconomic} oil price news shock. 
    The left column contains impulse response functions of one-quarter ahead expectations, 
    and the right column contains analogous responses of four-quarter ahead expectations. 
    }
\end{figure}

Figure \ref{fig:exp_irfs_model_comp} illustrates the difficulty that many existing 
models of expectation formation have matching impulse responses of expectations data. 
The left column plots the impulse response functions of each variables' 
realization, the one-quarter ahead Bluechip survey expectation and the 
same expectation implied by models of expectation formation. 
The right column plots the impulse response functions for the four-quarter ahead expectations. 
The horizons of each expectational impulse response is (vertically) aligned to the period it is 
forecasting.

The over-extrapolation model is the model from Equation (\ref{eq:simple_overextrap}) and 
whose hair plot is displayed in Figure \ref{fig:output_cpi_extrap}. 
As shown earlier, the over-extrapolation model is able to rationalize the 
Bluechip expectations impulse responses across variables, time, and horizons.
We can similarly construct impulse responses implied by a number of other models, 
treating the realized impulse response as the relevant full-information rational 
expectations (FIRE) benchmark.

The impulse response implied by the \cite{gabaix2019behavioral} model of cognitive discounting 
with cognitive discount parameter $\theta$ is
$$
\Psi(E^{\text{CD}}_t[Y_{t+h}], \varepsilon_{t-\ell}) = \theta^h \underbrace{\Psi(Y_{t+h}, \varepsilon_{t-\ell})}_{\text{Full-information rational expectation IR}}
$$
Cognitive discounting implies uniform under-reaction relative to FIRE, where the degree of under-reaction 
increases with the horizon. 
Hence, the expectation impulse response under cognitive discounting (gold dot-dashed) under-reacts 
by more for the four-quarter ahead expectation (right panel) than the one-quarter ahead expectation 
(left panel). 
While we see that this under-reaction is largely consistent with Bluechip expectations of 
real disposable income, it is inconsistent with Bluechip expectations of CPI inflation. 
Figure \ref{fig:exp_irfs_model_comp} uses $\theta = 0.85$ from \cite{gabaix2019behavioral}.

The impulse response implied by the \cite{carroll2020sticky} model of sticky expectations with 
parameter $\theta$ for horizon $h > 0$ is
$$
\Psi(E^{\text{SE}}_t[Y_{t+h}], \varepsilon_{t-\ell}) = (1 - \theta^{\ell + 1}) \underbrace{\Psi(Y_{t+h}, \varepsilon_{t-\ell})}_{\text{Full-information rational expectation IR}}
$$
Sticky expectations implies uniform under-reaction relative to FIRE, where the degree of 
under-reaction decreases with the time elapsed since the initial shock. 
While this model of expectation formation struggles to match the CPI inflation expectations for 
this reason, similarly to cognitive discounting, it also implies too much under-reaction at 
early periods of the impulse response for each horizon. 
In contrast to this model, the Bluechip expectations do not exhibit more pronounced under- or 
over-reaction in early impulse response periods. 
Figure \ref{fig:exp_irfs_model_comp} uses $\theta = 0.935$ estimated in \cite{auclert2020micro}.

The impulse response implied by the \cite{bordalo2020overreaction} model of dispersed 
noisy information and diagnostic expectations with parameters $\theta, \tau$, the diagnosticity 
and signal-to-noise precision ratio, is
$$
\Psi(E^{\text{NIDE}}_t[Y_{t+h}], \varepsilon_{t-\ell}) = 
\begin{cases}
    (1 + \theta)\left( \frac{1}{\tau + 1} \right) \Psi(Y_{t+h}, \varepsilon_t) & \text{ for } \ell = 0 \\
    \left((1 + \theta) \left(\frac{\ell + 1}{\tau + \ell + 1} \right) - \theta \left(\frac{\ell}{\tau + \ell}\right)\right) \Psi(Y_{t+h}, \varepsilon_{t-\ell}) & \text{ for } \ell > 0
\end{cases}
$$
This is obtained by first considering the diagnostic expectation relative to the 
noisy information rational expectation benchmark, denoted $\tilde{\mathbb{E}}_t$
$$
\mathbb{E}^{\text{DE}}_t[Y_{t+h}] = \tilde{\mathbb{E}}_t[Y_{t+h}] + \theta (\tilde{\mathbb{E}}_t[Y_{t+h}] - \tilde{\mathbb{E}}_{t-1}[Y_{t+h}])
$$
All agents receive a signal $s_t$ each period about the exogenous impulse response shock 
$\varepsilon_{t-\ell}$ of the form $s_t = \varepsilon_{t-\ell} + \nu_t$, as in 
the Appendix of \cite{auclert2020micro}. 
Let $\tau$ denote the ratio of (constant) signal precision (inverse standard 
deviation of $\nu_t$) to the precision of $\varepsilon_{t-\ell}$. 
Then the impulse response of the horizon-$h$ noisy information rational expectation can be written as
$$
\Psi(\tilde{\mathbb{E}}_t[Y_{t+h}], \varepsilon_{t-\ell}) = \frac{\ell + 1}{\tau + \ell + 1} \Psi(Y_{t+h}, \varepsilon_{t-\ell})
$$
On the initial shock impact period when $\ell = 0$, past expectations $\mathbb{E}_{t-1}$ are still anchored 
at 0, hence the $\ell=0$ case in the above impulse response of the noisy information, diagnostic expectation. 
However, after $\ell > 0$, the full-information rational expectation of the prior referenced period will 
fully adjust, hence $\mathbb{E}_t = \mathbb{E}_{t-1}$ for $\ell > 0$.

I use estimated values of $\theta, \tau$ from \cite{bordalo2020overreaction} for the 
noisy information, diagnostic expectation of variables plotted in Figure \ref{fig:exp_irfs_model_comp}.
For real disposable income expectations, I use the estimated values in \cite{bordalo2020overreaction} 
for Bluechip real GDP growth expectations. 
They estimate $\theta, \tau$ for CPI inflation and the nominal Federal funds rate expectations 
from the Bluechip, so I use exactly those values for these variables.

While in principle diagnostic expectations can produce over-reaction, due to the diagnosticity parameter 
$\theta$, in \cite{bordalo2020overreaction} the values of $\tau$ are sufficiently large that 
the average expectation does not display over-reaction, as it would with pure diagnostic expectations 
as in \cite{bordalo2018diagnostic}.

\subsection{Extrapolating missing horizons of expectations data}\label{asubsec:extrap}
\begin{table}[t]
    \centering
    \begin{tabular}{ccccc} 
        & & \multicolumn{3}{c}{\textbf{Consumption-Savings Models}} \\
        \toprule
        Extrapolation & Parameter & Perpetual Youth & Standard Incomplete Markets & Rep. agent \\ 
        \midrule
        \midrule
        \multirow{2}{*}{AR(2)} & EIS &  0.08  &   0.09  & 0.00 \\ 
        & MPC &  0.04  &  0.05  & 0.005 \\ 
        \midrule
        \multirow{2}{*}{AR(1)} & EIS &  0.11  &   0.07  & 0.00 \\ 
        & MPC &  0.05  &  0.07  & 0.005 \\ 
        \midrule
        \multirow{2}{*}{AR(1) of AR(1)s} & EIS &  0.06  &   0.07  & 0.00 \\ 
        & MPC &  0.05  &  0.07  & 0.005 \\ 
        \midrule
        \multirow{2}{*}{``Over-extrapolation''} & EIS &  0.01  &   0.10  & 0.00 \\ 
         & MPC &  0.03  &  0.07  & 0.005 \\ 
        \bottomrule
    \end{tabular}
    \caption{Estimated parameters across missing-horizon extrapolation models}
    \label{tab:mod_moments_extrap}

    \caption*{\footnotesize 
    \emph{Note}: 
    Each panel contains of estimated parameters for each consumption-savings model under 
    different extrapolation methods for missing expectations data horizons. 
    The parameter estimates enforce that the steady state assets-to-income ratio 
    is equal to the initial calibration target.
    }
\end{table}
In Section \ref{subsec:irf_income_rates}, I discussed the need to extrapolate the missing horizons 
of expectations data. 
While the Bluechip expectations data has forecasts for a finite number of horizons, 
we need an infinite set of horizons of expectations to evaluate model-implied consumption.
To obtain these missing horizons, I estimate an auxiliary, parametric model on the existing 
horizons and impulse response periods and use it to extrapolate the missing horizons.

Using two-stage least squares where moments in Equation (\ref{eq:shape_restr}) are targeted with 
weights given by the inverse covariance matrix of the sequence of instruments $\{z_{t-\ell}\}$, 
I estimate the following auxiliary models, which result in the parameter 
estimates I report in Table \ref{tab:mod_moments_extrap}. 
I also impose an additional penalty on each model to ensure that the far-horizon 
expectations implied by each model are stationary, that is
$$
\lim_{h \to \infty} \mathbb{E}[F_t[W_{t+h};\boldsymbol{\vartheta}] z_{t-\ell}] = 0
$$
Note that the top panel of Table \ref{tab:mod_moments_extrap} is the baseline extrapolation 
I choose in the main set of results reported in Table \ref{tab:mod_moments} in the main body 
of the paper.

The AR(2) and AR(1) are the standard univariate autoregressive processes, with two and one 
period lags respectively. 
The ``AR(1) of AR(1)s'' is a single lag autoregressive process, whose innovation term is 
also an AR(1) process. 
This functional form choice is motivated by the functional form of the 
equilibrium law of motion of output $Y_t$ in Section \ref{sec:macro_model_inertia}, 
where the belief component law of motion follows a vector equivalent of 
a ``AR(1) of AR(1)s''. 
Finally, the ``over-extrapolation'' model is the one given by Equation (\ref{eq:simple_overextrap}) 
and displayed in Figure \ref{fig:output_cpi_extrap}.


\section{Proofs and derivations}

\subsection{Proof of Proposition \ref{prop:main_prop}}\label{asubsec:main_prop_proof}
Let the persistent shock component $\lambda_t$ and the transitory shock component $\eta_t$ 
each follow AR(1) processes
\begin{align*}
    \lambda_t &= \rho_\lambda \lambda_{t-1} + u_{\lambda, t}\\
    \eta_t &= \rho_\eta \eta_{t-1} + u_{\eta, t}
\end{align*}
The perceived law of motion $\tilde{Y}_t$ and equilibrium law of motion of output $Y_t$ are given by
\begin{align*}
    \tilde{Y}_t &= \lambda_t + \eta_t \\
    Y_t &= \chi (h_\lambda E_{t-1}[\lambda_t] + h_\eta E_{t-1}[\eta_t]) + \lambda_t + \eta_t
\end{align*}
where $h_\lambda := \frac{\rho_\lambda}{1 - \beta \omega \rho_\lambda}$ and $h_\eta$ is analogously defined 
with respect to $\rho_\eta$.
Beliefs about each shock component evolve according to the Kalman update equation
\begin{align*}
    \begin{bmatrix}
        E_t[\lambda_{t+1}] \\
        E_t[\eta_{t+1}]
    \end{bmatrix} 
    &= 
    \begin{bmatrix}
        \rho_\lambda & 0 \\
        0 & \rho_\eta
    \end{bmatrix}
    \begin{bmatrix}
        E_t[\lambda_{t+1}] \\
        E_t[\eta_{t+1}]
    \end{bmatrix} 
    + 
    \begin{bmatrix}
        g_\lambda \\
        g_\eta
    \end{bmatrix} 
    (Y_t - E_{t-1}[Y_t])
\end{align*}
where the subjective expectation $E_{t-1}[Y_t] = E_{t-1}[\lambda_t] + E_{t-1}[\eta_t]$ is the 
conditional expectation of output induced by the perceived law of motion, given 
the history of past output observations $\{Y_{t-\ell}\}_{\ell > 0}$, 
and $g_\lambda, g_\eta$ are the steady state Kalman gains under the perceived law of motion.

Evaluating and re-organizing terms in the belief component law of motion, we obtain
\begin{equation}\label{eq:belief_comp_lom}
    \begin{bmatrix}
        E_t[\lambda_{t+1}] \\
        E_t[\eta_{t+1}]
    \end{bmatrix} 
    = 
    \underbrace{\left(
    \begin{bmatrix}
        \rho_\lambda & 0 \\
        0 & \rho_\eta
    \end{bmatrix}
    + 
    \begin{bmatrix}
        g_\lambda( \chi h_\lambda - 1) & g_\lambda(\chi h_\eta - 1) \\
        g_\eta(\chi h_\lambda - 1) & g_\eta(\chi h_\eta - 1)
    \end{bmatrix}
    \right)
    }_{\text{Let } \mathbf{A} := }
    \begin{bmatrix}
        E_{t-1}[\lambda_{t}] \\
        E_{t-1}[\eta_{t}]
    \end{bmatrix} 
    + 
    \underbrace{\begin{bmatrix}
        g_\lambda & g_\lambda \\
        g_\eta & g_\eta
    \end{bmatrix}
    }_{\text{Let } \mathbf{G}:=} 
    \begin{bmatrix}
        \lambda_t \\
        \eta_t
    \end{bmatrix}
\end{equation}
where I require the eigenvalues of $\mathbf{A}$ to be within the unit circle, such that the 
belief component law of motion is stationary.

\paragraph{Inertia at time-1} for $Y_t$ to exhibit inertia with respect to an innovation to 
a component shock, the net increase in belief feedback at time-1 must exceed the decay from 
the direct effect of the component shock.

Consider a time-0 positive innovation to a component shock 
$e_0 > 0, \text{ where } e_0 \in \{\lambda_0, \eta_0\}$. 
At time-0, only the direct shock effect occurs so $Y_0 = e_0$, 
given beliefs prior to time-0 are zero in steady state. 
At time-1, we have
$$
\begin{bmatrix}
    E_0[\lambda_1] \\
    E_0[\eta_1]
\end{bmatrix} 
= 
\begin{bmatrix}
    g_\lambda \\
    g_\eta
\end{bmatrix} e_0
$$
Evaluating $Y_1$ and solving for the threshold $Y_1 > Y_0$, we obtain the lower bound
\begin{equation}\label{eq:chi_t1_lb}
    \chi > \frac{1 - \rho_e}{h_\lambda g_\lambda + h_\eta g_\eta} > 0
\end{equation}
Let us denote this lower threshold for $\chi$ as $\underline{X}_{e, 0}$. 
Given the range of permissible parameters, where $\beta, \omega, \rho_e \in (0, 1)$, 
it must be that this threshold is strictly positive, i.e. $\underline{X}_{e, 0} > 0$.
Thus, if $\chi > \underline{X}_{e, 0}$, then $Y_t$ will exhibit inertia with the inertial 
peak period $\bar{\ell} \geq 1$.

\paragraph{Regularity conditions on $\mathbf{A}$} 
A necessary condition for $Y_t$ to be increasing with time-$t$ up until an inertial 
peak period $\bar{\ell}$, is for the belief feedback term, 
$\chi (h_\lambda E_{t-1}[\lambda_t] + h_\eta E_{t-1}[\eta_t])$, to be increasing with time-$t$.
I now derive some restrictions on parameters, or equivalently regularity conditions on $\mathbf{A}$,
that ensures that the belief feedback term $\chi (h_\lambda E_{t-1}[\lambda_t] + h_\eta E_{t-1}[\eta_t])$ 
is positive for all time-$t$. 
For times-$t$ leading up to the inertial peak period $\bar{\ell}$ this is required 
for the necessary condition to hold\footnote{This is potentially stronger than necessary 
for bounding the absolute value of the MA coefficients of $Y_t$ for 
times-$t > \bar{\ell}$ to be less than the time-$\bar{\ell}$ coefficient. 
However, another way to justify the strength of these regularity conditions 
for periods after $\bar{\ell}$ is if we desire $Y_t$ to return to its steady 
state at zero without ``over-shooting'' and becoming negative in response to a 
positive component shock. If so, these regularity conditions will ensure this.}.\\

\noindent \textbf{Positive, real eigenvalues of $\mathbf{A}$}\\
For the eigenvalues of $\mathbf{A}$ to be real (note that $\mathbf{A}$ is not 
positive semi-definite), the discriminant of its characteristic polynomial 
must be positive. 
This can be simplified to checking whether the following expression is greater 
than zero
$$
(\rho_\lambda - \rho_\eta)^2 + (\theta_\lambda + \theta_\eta)^2 + 2 (\rho_\lambda - \rho_\eta) (\theta_\eta - \theta_\lambda) > 0
$$
where $\theta_\lambda := g_\lambda (1 - \chi h_\lambda)$ and likewise for $\theta_\eta$. 
Given our definitions of $\chi := (1 - \beta \omega - \beta \omega \sigma \phi)$ and the 
range of permissible parameters, it must be that $\theta_\lambda, \theta_\eta > 0$. 
Hence, simplifying the above expression in terms of lower bound on $\chi$, we obtain
\begin{equation}\label{eq:chi_real_eig_lb}
    \chi \geq \frac{g_\lambda - g_\eta}{h_\lambda g_\lambda - h_\eta g_\eta}
\end{equation}
which ensures the eigenvalues of $\mathbf{A}$ are real.

For the eigenvalues to be positive, we have
\begin{align*}
    w_1 + w_2 &= \rho_\lambda + \rho_\eta - \theta_\lambda - \theta_\eta \\
    w_1 w_2 &= \rho_\lambda \rho_\eta - \rho_\lambda \theta_\eta - \rho_\eta \theta_\lambda
\end{align*}

The following lower bounds on $\chi$ ensure that the eigenvalues will be positive
\begin{equation}\label{eq:chi_pos_eig_lb}
    \chi \geq \frac{g_\lambda - \frac{1}{2} \rho_\lambda}{h_\lambda g_\lambda}, \quad \chi \geq \frac{g_\eta - \frac{1}{2} \rho_\eta}{h_\eta g_\eta}
\end{equation}
with at least one holding as a strict inequality.\\

\noindent \textbf{Effective horizons, gains $\mathbf{h}, \mathbf{g}^\prime$ and eigenvectors of $\mathbf{A}$}\\
Given the $\chi$ lower bound in (\ref{eq:chi_t1_lb}) and $\theta_\lambda, \theta_\eta > 0$, 
the matrix $\mathbf{A}$ will have positive diagonal entries and negative off-diagonal entries. 
Due to this, the eigenvector $v_1$ corresponding to the dominant eigenvalue $w_1 > w_2$ will have 
one positive $v_{11} > 0$ and one negative entry $v_{12} < 0$. 
The non-dominant eigenvector $v_2$ will solely have positive entries, i.e. $v_{21}, v_{22} > 0$.

Consider the following expression, where we can unwind the recursion in Equation (\ref{eq:belief_comp_lom}) 
given the initial component shock $e_0$
$$
\begin{bmatrix}
    E_t[\lambda_{t+1}] \\
    E_t[\eta_{t+1}]
\end{bmatrix} 
= \sum_{\ell=0}^t \mathbf{A}^{t-\ell} \mathbf{g}^\prime \rho_e^\ell e_0
$$
Left-multiplying by $\mathbf{h}$ to compute the contribution of belief feedback and expanding 
the eigendecomposition of $\mathbf{A} = \mathbf{V} \mathbf{W} \mathbf{V}^{-1}$, we obtain
$$
\mathbf{h}
\begin{bmatrix}
    E_t[\lambda_{t+1}] \\
    E_t[\eta_{t+1}]
\end{bmatrix} 
= \mathbf{h} \mathbf{V} \sum_{\ell=0}^t \mathbf{W}^{t-\ell} \mathbf{V}^{-1} \mathbf{g}^\prime \rho_e^\ell e_0
$$
With the conditions
\begin{equation}\label{eq:regularity_eigvec}
    \frac{h_\lambda}{h_\eta} > -\frac{v_{12}}{v_{11}}, \quad \frac{g_\lambda}{g_\eta} < \frac{v_{22}}{v_{21}}
\end{equation}
the above expression constitutes the sum of strictly positive, bilinear forms which itself must be positive, 
hence we will have as desired
$$
\mathbf{h}
\begin{bmatrix}
    E_t[\lambda_{t+1}] \\
    E_t[\eta_{t+1}]
\end{bmatrix} > 0, \quad \forall{t}
$$
Without condition (\ref{eq:regularity_eigvec}), we cannot ensure the belief feedback term 
is positive for any time-$t$.

\paragraph{Inertia at time-$t$} I now proceed with the induction to prove that 
the inertial peak $\bar{\ell}$ is (weakly) increasing in $\chi$, assuming 
conditions (\ref{eq:chi_t1_lb}), (\ref{eq:chi_real_eig_lb}), (\ref{eq:chi_pos_eig_lb}), (\ref{eq:regularity_eigvec}) hold.

Suppose $Y_{t-\ell} > Y_{t-\ell-1}$ for $\ell \in \{0, t\}$, and at time-$t+1$ we have $Y_{t+1} = Y_t$, 
placing the inertial peak $\bar{\ell} = t$.
The equality $Y_{t+1} = Y_t$ can be written as
\begin{equation}\label{eq:ytp1_inertia_cond}
    \chi (h_\lambda \Delta E_t[\lambda_{t+1}] + h_\eta \Delta E_t[\eta_{t+1}]) 
    = (1 - \rho_e) \rho_e^t e_0
\end{equation}
where $\Delta E_t[\lambda_{t+1}] := E_t[\lambda_{t+1}] - E_{t-1}[\lambda_t]$ and likewise for $\eta$.

Our goal is to demonstrate that as $\chi$ increases the left-hand side exceeds the 
right-hand side which is invariant to $\chi$, thus shifting the inertial peak to $\bar{\ell} = t+1$. 
If Equation (\ref{eq:ytp1_inertia_cond}) held with an inequality $<$, then a marginal increase 
in $\chi$ would not shift the peak, hence this result only implies $\bar{\ell}$ is weakly increasing 
in $\chi$.

Differentiating the left-hand side of Equation (\ref{eq:ytp1_inertia_cond}), we obtain
\begin{align*}
    \overbrace{h_\lambda \Delta E_t[\lambda_{t+1}] + h_\eta \Delta E_t[\eta_{t+1}]}^{=\chi^{-1} (1 - \rho_e) \rho_e^t e_0 > 0} + \\
    \chi (h_\lambda \partial_\chi \Delta E_t[\lambda_{t+1}] + h_\eta \partial_\chi \Delta E_t[\eta_{t+1}])
\end{align*}
Given $\chi > 0$ by Equation (\ref{eq:chi_t1_lb}), 
it suffices to verify 
$h_\lambda \partial_\chi \Delta E_t[\lambda_{t+1}] + h_\eta \partial_\chi \Delta E_t[\eta_{t+1}] > 0$.\\

Differentiating the component beliefs in Equation (\ref{eq:belief_comp_lom}) with respect to $\chi$, we obtain
\begin{equation*}
    \begin{bmatrix}
        \partial_\chi E_t[\lambda_{t+1}] \\
        \partial_\chi E_t[\eta_{t+1}]
    \end{bmatrix} 
    = 
    \mathbf{A}
    \begin{bmatrix}
        \partial_\chi E_{t-1}[\lambda_{t}] \\
        \partial_\chi E_{t-1}[\eta_{t}]
    \end{bmatrix} 
    +
    \mathbf{g}^\prime \mathbf{h}
    \begin{bmatrix}
        E_{t-1}[\lambda_{t}] \\
        E_{t-1}[\eta_{t}]
    \end{bmatrix} 
\end{equation*}
Differencing, unwinding the recursion, and left-multiplying again by $\mathbf{h}$ we obtain
$$
\mathbf{h} 
\begin{bmatrix}
    \partial_\chi \Delta E_t[\lambda_{t+1}] \\
    \partial_\chi \Delta E_t[\eta_{t+1}]
\end{bmatrix} 
= 
\mathbf{h} \mathbf{V} \mathbf{W}^{t-1} \mathbf{V}^{-1} \mathbf{g}^\prime \mathbf{h} 
\begin{bmatrix}
    E_0[\lambda_1] \\
    E_0[\eta_1]
\end{bmatrix} 
+ 
\mathbf{h} \mathbf{V} \sum_{\ell=1}^{t-1} \mathbf{W}^{t-1-\ell} \mathbf{V}^{-1} \mathbf{g}^\prime \mathbf{h}
\begin{bmatrix}
    \Delta E_t[\lambda_{t+1}] \\
    \Delta E_t[\eta_{t+1}]
\end{bmatrix}
> 0
$$
Finally, given the $\chi > 0$ by Equation (\ref{eq:chi_t1_lb}) and the above expression, 
we have 
$$
\chi (h_\lambda \partial_\chi \Delta E_t[\lambda_{t+1}] + h_\eta \partial_\chi \Delta E_t[\eta_{t+1}]) > 0
$$
which implies the left-hand side of Equation (\ref{eq:ytp1_inertia_cond}) is strictly increasing 
in $\chi$.



\subsection{Proof of Proposition \ref{prop:rational_comp}}\label{asubsec:rational_comp_proof}
Given $\eta_t$ is i.i.d, the output law of motion under rational learning is given by
\begin{equation}\label{eq:yt_lom_rl}
    Y^R_t = \frac{\chi h_\lambda}{1 - \chi h_\lambda} E_{t-1}[\lambda_t] + \lambda_t + \eta_t
\end{equation}
and the output law of motion under constrained-rational learning is given by
\begin{equation}\label{eq:yt_lom_crl}
    Y^{CR}_t = \chi h_\lambda \underbrace{(1 - \rho_\lambda^2) \sum_{\ell=0}^\infty \rho^\ell a_\ell}_{\text{Denote } \tilde{a}:=} E_{t-1}[\lambda_t] + \lambda_t + \eta_t
\end{equation}
$a_\ell$ denote the MA coefficients of $Y_t^{CR}$ with respect to 
persistent component innovations $u_{\lambda, t-\ell}$.

Given $\eta_t$ is i.i.d, the component belief law of motion for $\lambda_t$ under rational 
learning is given by
\begin{equation}\label{eq:lam_lom_rl}
    E_t[\lambda_{t+1}] = \underbrace{(\rho_\lambda - g_\lambda)}_{\text{Denote } f_\lambda :=} E_{t-1}[\lambda_t] + g_\lambda (\lambda_t + \eta_t)
\end{equation}
and under constrained-rational learning
\begin{equation}\label{eq:lam_lom_crl}
    E_t[\lambda_{t+1}] = \underbrace{(\rho_\lambda - \tilde{g}_\lambda \tilde{a} (1 - \chi h_\lambda) )}_{\text{Denote } \tilde{f}_\lambda :=} E_{t-1}[\lambda_t] + \tilde{g_\lambda} (\lambda_t + \eta_t)
\end{equation}
To simplify Equation (\ref{eq:lam_lom_crl}), compare the Kalman gains $g_\lambda, \tilde{g}_\lambda$ under 
the two different perceived state space models. 
The perceived law of motion under rational learning yields the measurement equation
$$
Y^R_t = \frac{\chi h_\lambda}{1 - \chi h_\lambda} E_{t-1}[\lambda_t] + \lambda_t + \eta_t
$$
Under constrained-rational learning, the measurement equation is
$$
Y^{CR}_t = \tilde{a} \lambda_t + \eta_t
$$
Given $E_{t-1}[\lambda_t]$ is pre-determined in $Y^R_t$ and the state transition 
equations under rational and constrained-rational learning are known to be the same, 
the steady state variance of the one-step ahead prediction errors $p$ is the same.
Applying the steady-state Kalman gain formula under constrained-rational learning
and setting $\sigma^2_\eta = \frac{\tilde{\sigma}_\eta^2}{\tilde{a}}$ obtains $\tilde{g}_\lambda \tilde{a} = g_\lambda$.

Denote the MA representations of $Y^R_t, Y^{CR}_t$ as
\begin{align*}
    Y^R_t &= \sum_{\ell=0}^\infty a^R_\ell u_{\lambda, t-\ell} + b^R_\ell u_{\eta, t-\ell} \\
    Y^{CR}_t &= \sum_{\ell=0}^\infty a^{CR}_\ell u_{\lambda, t-\ell} + b^{CR}_\ell u_{\eta, t-\ell}
\end{align*}

\paragraph{$Y_t$ responses to a transitory innovation $u_{\eta, 0}$}\,

\noindent Iterating Equations (\ref{eq:lam_lom_rl}), (\ref{eq:lam_lom_crl}) forward 
and plugging into (\ref{eq:yt_lom_rl}), (\ref{eq:yt_lom_crl}), we obtain the expressions 
for the MA coefficients with respect to a transitory innovation $t$-periods ago
\begin{align*}
    b^R_t &= \frac{\chi h_\lambda}{1 - \chi h_\lambda} g_\lambda f_\lambda^{t-1} \\
    b^{CR}_t &= \chi h_\lambda g_\lambda \tilde{f}_\lambda^{t-1}
\end{align*}
The inequality $b^R_t > b^{CR}_t$ can be written as
\begin{equation}
    \frac{1}{1 - \chi h_\lambda} \left( \frac{f_\lambda}{\tilde{f}_\lambda} \right)^{t-1} > 1
\end{equation}
Therefore, in the initial period (time-1) in which beliefs respond the inequality 
will hold because $(1 - \chi h_\lambda)^{-1} > 1$, given $1 - \chi h_\lambda \in (0, 1)$ for $\chi > 0$. 
However, as $t \to \infty$, the left-hand side approaches zero because the ratio 
$\frac{f_\lambda}{\tilde{f}_\lambda} \in (0, 1)$. 
This indicates that at some positive period $\bar{t} > 1$ the inequality will no longer hold.

\paragraph{$Y_t$ responses to a persistent innovation $u_{\lambda, 0}$}\,
\noindent Iterating Equations (\ref{eq:lam_lom_rl}), (\ref{eq:lam_lom_crl}) forward 
and plugging into (\ref{eq:yt_lom_rl}), (\ref{eq:yt_lom_crl}), we obtain the expressions 
for the MA coefficients with respect to a persistent innovation $t$-periods ago
\begin{align*}
    a^R_t &= \frac{\chi h_\lambda}{1 - \chi h_\lambda} g_\lambda \sum_{\ell=0}^{t-1} f^{t-1-\ell}_\lambda \rho_\lambda^\ell + \rho_\lambda^t \\
    a^{CR}_t &= \chi h_\lambda g_\lambda \tilde{a} \sum_{\ell=0}^{t-1} \tilde{f}_\lambda^{t-1-\ell} \rho_\lambda^\ell + \rho_\lambda^t
\end{align*}

The expression for $a^{CR}_t$ is implicit, since $\tilde{a}$ contains MA coefficients $\{a_\ell\}_{\ell \geq 0}$. 
Unwinding the implicit expression in $\tilde{a}$ and solving out the following sum
$$
\sum_{\ell=0}^{t-1} f_\lambda^{t-1-\ell} \rho_\lambda^\ell = \frac{\rho_\lambda^t - f_\lambda^t}{\rho_\lambda - f_\lambda}
$$
and likewise for $\tilde{f}_\lambda$, we obtain
\begin{align*}
    a^R_t &= \frac{\chi h_\lambda}{1 - \chi h_\lambda} g_\lambda \left( \frac{\rho_\lambda^t - f_\lambda^t}{\rho_\lambda - f_\lambda} \right) + \rho_\lambda^t\\
    a^{CR}_t &= \chi h_\lambda g_\lambda \left( \frac{\rho_\lambda}{1 - \tilde{f}_\lambda \rho_\lambda} \chi h_\lambda g_\lambda + 1 \right) \left( \frac{\rho_\lambda^t - \tilde{f}_\lambda^t}{\rho_\lambda - \tilde{f}_\lambda} \right) + \rho_\lambda^t
\end{align*}
Setting the inequality $a^R_t > a^{CR}_t$ and simplifying we obtain
$$
\frac{\rho_\lambda^t - f_\lambda^t}{\rho_\lambda^t - \tilde{f}_\lambda^t} > 1 + \frac{\rho_\lambda}{1 - \tilde{f}_\lambda \rho_\lambda} \chi h_\lambda g_\lambda
$$
Given $0 < f_\lambda < \tilde{f}_\lambda < \rho_\lambda$, the left-hand side is positive and increasing in time-$t$. 
Checking the inequality at time-1, I obtain
$$
\frac{1}{1 - \chi h_\lambda} > \frac{\rho_\lambda}{1 - \tilde{f}_\lambda \rho_\lambda} \chi h_\lambda g_\lambda + 1
$$
which further simplifies to $1 > \rho_\lambda^2$, which always holds for the cases we consider $\rho_\lambda \in (0, 1)$.
Therefore, $Y^R_t > Y^{CR}_t$ for all times-$t$.

\end{document}